%% file: main-supp-combined.tex
\documentclass[aps,prb,twocolumn,groupedaddress,superscriptaddress,amsfonts,amssymb,amsmath,citeautoscript,longbibliography,a4paper]{revtex4-2}
\usepackage{layouts}

\input{setup}
\usepackage{graphicx} 
\usepackage{layouts}

\begin{document}
\title{Plasmonic coated scatterers for tunable coherent perfect absorption}

\newcommand{\suppdcmobility}{S2}
\newcommand{\suppmiespheres}{S3}
\newcommand{\suppmiescylinders}{S4}
\newcommand{\suppsigmarealindependence}{S5}
\newcommand{\suppspheresstatic}{S6}
\newcommand{\suppsecondspheresstatic}{S7}
\newcommand{\suppcylindersstatic}{S8}
\newcommand{\supptransferspheres}{S9}
\newcommand{\supptransfercylinders}{S10}
\newcommand{\supplargekspheres}{S11}
\newcommand{\supplargekcylinders}{S12}
\newcommand{\suppsectiondebye}{S13}
\newcommand{\suppdivergence}{S14}
\newcommand{\suppbandwidth}{S15}
\newcommand{\suppcomplexfrequencies}{S16}
\newcommand{\suppnonlocalityspheres}{S17}
\newcommand{\suppnonlocalitycylinders}{S18}
\newcommand{\suppcoupledlattice}{S19}
\newcommand{\suppspherepolarizability}{S20}
\newcommand{\suppfinitetemp}{S21}
\newcommand{\suppmultichannel}{S22}
\newcommand{\supptebandwidth}{S23}

\newcommand{\yaleaffil}{Department of Applied Physics, Yale University, New Haven, Connecticut 06520, USA \\ 
{\textasteriskcentered{} ali.ghorashi@yale.edu}
}
\author{Ali Ghorashi$^{*}$}
\affiliation{\yaleaffil}
\author{Ali H. Alhulaymi}
\affiliation{\yaleaffil}
\author{A. Douglas Stone}
\affiliation{\yaleaffil}

\date{June 2026}
\begin{abstract}
    We derive, in closed-form, the surface conductivity required for coated subwavelength-scale spherical and cylindrical scatterers to perfectly absorb incident coherent light of fixed angular momentum. To address the challenge of synthesizing an incident wave of a fixed angular momentum, we analyze two geometries where this physics can be accessed from the far-field: a single coated sphere suspended above a good conducting surface, and an array of dipole-coupled coated cylindrical scatterers. We show that the required complex surface conductivities necessary for coherent perfect absorption over a large bandwidth in the terahertz may be easily achieved in moderately doped graphene.
\end{abstract}
\maketitle
\paragraph{Coherent perfect absorption}
Coherent perfect absorption (CPA) (also referred to as time-reversed lasing \cite{chong2010coherent,wan2011time}) occurs when a wavefront incident on a scatterer is trapped by destructive interference and completely absorbed, resulting in no outgoing field. In the special case of single channel absorption, this reduces to the well-known critical coupling condition \cite{Yariv2000UniversalWaveguides, Cai2000ObservationSystem}, where the incoming coupling rate is balanced by the internal coupling losses, yielding the interference necessary to have zero outgoing wave in the single channel. 
CPA generalizes the notion of critical coupling to multichannel scattering by mapping the system under time-reversal to a laser at threshold, showing that complete absorption of incoming radiation can still be achieved by utilizing the interchannel interference effects of an adapted wavefront.
Under time-reversal, gain maps to loss, so CPA occurs when a scatterer has an absorption (instead of gain) rate exactly equal to the scattering rate, which is both a rigorous version of the critical coupling condition and defines the required wavefront as the time-reverse of the lasing mode. The rate of absorption is not enhanced at the critical coupling condition, but the ``interference trap" created by this special wavefront guarantees that all of the incident energy is absorbed. Thus, CPA is a means of greatly enhancing the absorption of a weak absorber.
In addition, CPA has been proposed as a means of controlling ``light with light" in the absence of nonlinearities \cite{zhang2012controlling}, and hence as a strategy for sensing \cite{liu2010infrared}. As a general principle, CPA is not restricted to the domain of optics/photonics, and there are analogs of CPA in all linear wave systems of physics (e.g. acoustic and matter waves \cite{song2014acoustic, duan2015theoretical, ma2016acoustic, romero2020perfect}).


CPA/critical coupling to a finite object is a resonance phenomenon: a resonance is a are purely \textit{outgoing} solution of the wave equation at a complex frequency, and CPA is associated with a purely \textit{incoming} solution at the complex conjugate frequency (in the absence of absorption). Adding the right amount of absorption moves this solution to the real axis, making the system a physically realizable steady-state perfectly absorbing sink (for the correct associated wavefront). For dielectric structures, the necessity of resonances at the frequency of interest limits the size of absorbers to the wavelength scale (and above). However, it was realized shortly after the discovery and demonstration of CPA that, since plasmonic structures do support resonances in structures much smaller than the free-space wavelength, it should, in principle, be possible to achieve CPA with subwavelength scatterers. This was first theoretically shown by Noh et al. \cite{noh2012perfect}; they studied nanoscale conducting cylinders and spheres and demonstrated that they could perfectly absorb adapted wavefronts in the visible range if certain values of the dielectric constant could be achieved.  

However, it has been more than a decade since this exciting finding without any successful experimental demonstration of this effect. There are two main difficulties. First, and most problematic, since angular momentum is conserved for cylindrical and spherical structures, the time-reversed incoming wavefronts required for perfect absorption correspond to  spherical/cylindrical waves of definite angular momenta, which are very difficult to generate, even approximately. Second, assuming that such wavefronts could be generated, \cite{noh2012perfect} found that the degree of absorption loss in standard metals is much larger than that needed to achieve critical coupling in the subwavelength limit. Consequently, the original proposal suggested the use of metal-dielectric (core-shell) nanospheres to reach the required physical parameters.

Before addressing the two problems listed above, we emphasize that CPA in the context of the systems studied by \cite{noh2012perfect} is, more precisely, single-channel critical coupling; because of the conservation of angular momentum in a system with spherical/cylindrical symmetry, the system is essentially single-channel. Said differently, CPA in such systems can only be achieved through a single incoming eigenstate of angular momentum, for which the absorption rate equals the scattering loss (quantified by the resonance width in the absence of absorption). In passing, we note that the use of the ``critical coupling" concept in the context of scattering from spheres and cylinders was not widely employed in the Mie scattering and plasmonics literature prior to the work of \cite{noh2012perfect}, as the researchers in those subfields generally assumed that the incident excitation would be a plane wave.  

In this work, we will concentrate on critical coupling and CPA of subwavelength structures under electromagnetic excitation, ``plasmonic CPA", and propose solutions to the two problems described above, based on atomic coating of nanostructures. There has been significant earlier work on this (e.g., \cite{noh2012perfect, pu2012ultrathin, teperik2004radiative, proskurin2021perfect}) for thin-film, cylindrical, and spherical geometries. However, we are considering for the first time the possibility of using two-dimensional coatings (e.g. graphene, \cite{jablan2009plasmonics, chen2011atomically, alu2010plasmonic}) on \textit{isolated} structures to generate the necessary absorption (due to dissipative surface currents) to realize CPA. Previous works on plasmonic CPA were primarily numerical and did not find analytic conditions for achieving CPA. Here we show that the material parameters of the coating required for achieving subwavelength CPA admit closed-form solutions, something which has only been achieved previously for the well-studied thin-film geometry \cite{chong2010coherent, wan2011time, liu2014gate}. In addition, we show that with two tuning parameters (e.g. the chemical potential in the graphene layer and the radius of the scatterer), the required surface impedance may be theoretically achieved at a wide range of frequencies in the terahertz (THz). We then present two setups that are more practical for experiments in which plasmonic CPA can also be observed in the THz frequency range.

Before we present our results on coatings and their application to CPA, we clarify two issues which have generated some confusion in the literature. The CPA solutions found by \cite{noh2012perfect} show that subwavelength CPA is achievable under excitation from the farfield, seemingly in contradiction with the diffraction resolution limit and to works \cite{proskurin2021perfect} which discuss the necessity of generating evanescent waves near the nanostructure to achieve perfect absorption. These issues (primarily the first one) were already discussed and clarified to some extent by a second work by Noh. et al. \cite{noh2013broadband}, but we make some further remarks here. It is indeed not possible to focus from the farfield to a spot in space much smaller that the wavelength of the incident radiation, assuming that there is no matter at that focal spot. However, if there is matter at that spot, and if that matter can resonantly couple to light, as with a plasmonic nanostructure, strong field and energy enhancement in a subwavelength region is possible (at the resonant frequency) \cite{noh2012perfect, noh2013broadband}. Consider the cylindrical case and imagine exciting the system with a cylindrical wave of finite angular momentum, $m$, and fixed frequency, $\omega$. This wave will decay at a distance from the origin $r_c \approx m/k= m\lambda/2 \pi, (k= \omega/c)$ \footnote{Strictly speaking, this is true for large $m$, see e.g. \cite{jacob2006optical}.}, which corresponds to a caustic in the ray limit. If the wave is in resonance with the plasmonic excitations of a cylinder located at the origin, however, it will initially decay for distances $r<r_c$, but it will then excite a scattered field that grows monotonically up to the cylinder's radius, $R$, giving a subwavelength focal spot. This was shown in detail and discussed in a somewhat different language in \cite{noh2013broadband}. Furthermore, note that the \textit{incident} field (field in the absence of the scatterer) has the radial dependence of a Bessel function and can always be written as a superposition of \textit{propagating} plane waves \footnote{See Section \suppmultichannel \space of the Supporting Information for more details}, meaning that the incident excitation of the structure has no evanescent components. The \textit{total} field at CPA, with only incoming flux, does have evanescent components, but this has no bearing on the incident excitation necessary to achieve perfect absorption \footnote{Though perhaps obvious, we stress that the \textit{incident} field, in the terminology of scattering theory, is distinct from the \textit{incoming} field. The incident field is the field in the absence of the scatterer, which must have both incoming and outgoing components.}. 
Hence the possibility of subwavelength CPA from the farfield is not in contradiction with the diffraction limit of focusing and does not require evanescent wave excitation.

The general effect described here is the well-known ``plasmonic field enhancement" induced by a sharp metallic spatial structure; the new feature with plasmonic CPA is that, if just the right degree of absorption is present in the structure, and if it is excited with the correct wavefront, there will be no outgoing wave at all and the incoming flux will be perfectly absorbed. With all of these preliminaries out of the way, we now show how atomically-thin coatings can be used to tune cylinders and spheres of varying radii to satisfy the CPA condition.



\begin{figure}
\centering
      \includegraphics[scale=1]{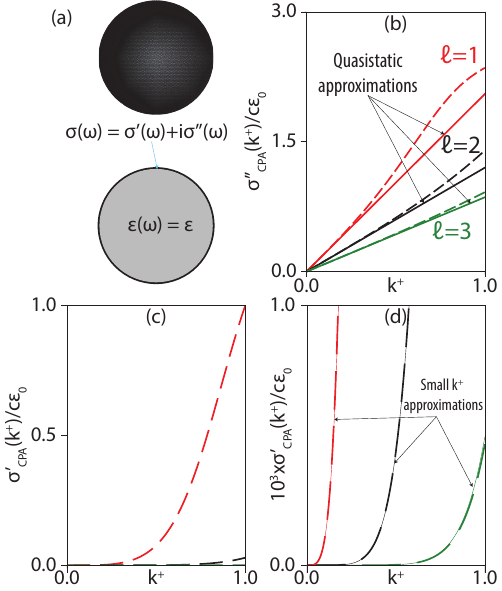}
  \caption{\textbf{Critically coupling to an isolated scatterer with a plasmonic coating}: (a) A spherical scatterer (top) and its 2D cross-section (bottom). The scatterer has a non-dispersive, lossless, bulk permittivity, $\varepsilon$, which is taken to be that of fused silica ($\varepsilon=2.1$ \cite{johnson2001block, malitson1965interspecimen}) for the results in (b) and (c), and a thin coating with surface conductivity, $\sigma(\omega)$. (b),(c),(d) show the imaginary and real parts of the conductivity corresponding to perfect absorption of light with fixed angular momentum, $l$, as a function of the normalized free space wavenumber, $k^+\equiv R\omega/c$. Fig. (d) is a zoomed-in version of (c) to show the behavior of $\sigma'_{CPA}(k^+, l)$ for higher angular momenta in the subwavelength regime. We superimpose analytic approximations for the real and imaginary parts of the CPA conductivity in (d) and (a), respectively. The latter corresponds to a quasistatic approximation that is linear in $k^+$ for all angular momenta.}
  \label{fig: figure 1}
\end{figure}

\begin{figure}
\centering
\includegraphics[scale=1]{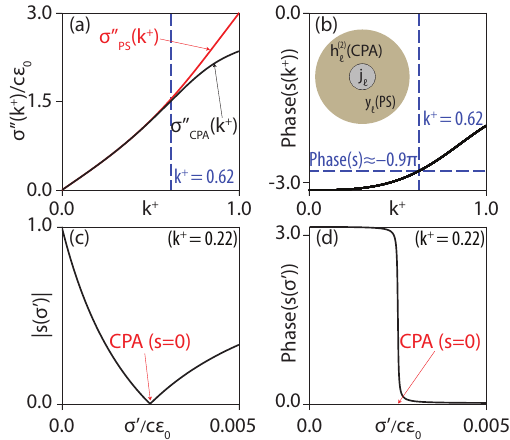}
  \caption{\textbf{CPA vs the lossless passive sink}: (a) comparison of $\sigma''_{PS}(k^+, l)$ and $\sigma''_{CPA}(k^+, l)$ for bulk permittivity $\varepsilon=2.1$ and $l=1$. Note that it would be nonsensical to include an analogous comparison for the real part of the CPA conductivity, since the passive sink conductivity is purely imaginary. (b) Phase of $s_{l=1}$ for a system with surface conductivity $\sigma=i\sigma''_{CPA}(k^+, l=1)$. In the deep subwavelength regime, we get a perfect $\pi$ phase shift, consistent with the agreement of $\sigma''_{CPA}$ and $\sigma''_{PS}$ at small $k^+$ seen in (a). Inset shows a scatterer (grey) in a uniform background (brown). For both CPA and the passive sink, the field inside the scatterer is a Bessel function of the first kind, $j_l$. Outside, however, $s_{l=1}=0$ for CPA and $s_{l=1}=-1$ for a passive sink, so the field goes as $h_l^{(2)}$ (CPA) or $y_l$ (passive sink). In (c) and (d), we fix the conductivity to $\sigma=i\sigma_{PS}''+\sigma'$ and sweep $\sigma'$. We achieve CPA by only changing the real part of the conductivity, consistent with the observation that $\sigma''_{CPA}\approx \sigma''_{PS}$ in the subwavelength regime. Note that (d) immediately implies that absorption pulls the value of the scattering coefficient along the real axis, yielding a sudden jump of the phase from $\pi$ to zero.}
  \label{fig: figure 3}
\end{figure}

\paragraph{Critical coupling by isolated coated scatterers}
We first consider critical coupling of transverse-magnetic (TM) polarized light to an isolated coated sphere suspended in vacuum, an extension of the standard Mie theory \cite{jackson1998classical, bohren2008absorption}, variations of which have been previously addressed in the literature \cite{alu2010plasmonic, christensen2015localized, mostafazadeh2012optical}. As we are primarily interested in the subwavelength, plasmonic, regime, we restrict our study (at first) to the TM polarization, which couples to surface charge oscillations and is thus most easily excited in the subwavelength regime (with a complete study of TM and transverse-electric, TE, polarizations, for both spheres and cylinders, relegated to the Supporting Information (SI) Sections \suppmiespheres\space and \suppmiescylinders). The sphere under consideration has a radius, $R$, a bulk permittivity, $\varepsilon\equiv n^2$ (which we take to be, unless otherwise specified, real-valued and non-dispersive--corresponding to a dielectric with negligible loss), and a dispersive surface conductivity, engendered by an atomically-thin coating, $\sigma(\omega)=\sigma'(\omega)+i\sigma''(\omega)$ (see \cref{fig: figure 1}(a)). The imaginary part of the surface conductivity, $\sigma''(\omega)$, along with the bulk geometry and permittivity, determine the resonance frequencies of the scatterer, while the dissipative real part, $\sigma'(\omega)$, determines the rate of absorption. In practice, the conductivity may be both temporally and spatially dispersive. As such, we will include or omit the frequency/momentum dependence of $\sigma$ depending on the context. In addition to the material parameters, ($R$, $\varepsilon$, and $\sigma$), as noted, the incoming wavefront must be a spherical wave if it is to be perfectly absorbed. In the final sections of this paper, we consider scatterers coupled to conducting surfaces and periodic extended geometries, and show that they can absorb other, more feasible, wavefronts--such as plane waves. 

Returning to the isolated sphere, the scattering problem is as follows. The magnetic field outside of the sphere is given by a linear superposition of an incoming spherical wave (at frequency $\omega$), $h_l^{(2)}(r\omega/c)\mathbf{L}Y_{lm}(\theta, \phi)$, and an outgoing spherical wave, $h_l^{(1)}(r\omega/c)\mathbf{L}Y_{lm}(\theta, \phi)$, with total field given by $\Big[h_l^{(2)}(r\omega/c)+s_lh_l^{(1)}(r\omega/c)\Big]\mathbf{L}Y_{lm}(\theta, \phi)$. $\mathbf{L}\propto \mathbf{r}\times\nabla$ is the angular momentum operator, $h_l^{(1)}$ and $h_l^{(2)}$ are outgoing and incoming spherical Hankel functions and the $Y_{lm}$ are spherical harmonics. $s_l$ is the scattering coefficient (which generalizes to a scattering matrix when rotational symmetry is broken), and we require $|s_l|=0$ for CPA. Note that we use $l$ for the angular momentum in the spherical setting, reserving $m$ for the azimuthal quantum number. When referring to cylinders, in subsequent sections, we will use $m$ for the angular momentum.  

To find the condition on $\sigma(\omega)$ to achieve $|s_l|=0$, one may approach the problem in two different ways. The first way is to consider the coating to have some finite thickness, $d$, (taking $d\rightarrow 0$ in the end) and an effective \textit{thickness-dependent} permittivity, $\varepsilon_{coating}=1+i\sigma(\omega)/\varepsilon_0\omega d$. Using the transfer matrix formalism for a nanoshell \cite{teperik2004radiative, noh2012perfect, mostafazadeh2012optical} then immediately yields an expression for $s_l$ (see SI Sections \supptransfercylinders, \supptransferspheres), which may be solved for $s_l=0$. This method, though laborious, gives insight as to the validity of our results beyond the thin coating limit (which could be of interest, for instance, for the study of multilayered coatings). 

The second method, which is more straightforward, is to directly impose boundary conditions on the tangential components of the magnetic and electric fields at $r=R$, considering the effect of the coating as a surface current source (see, SI Sections \suppmiespheres \space and \suppmiescylinders \space for details). Both methods yield the same result for $\sigma_{CPA}$, the surface conductivity required to engender critical coupling, as:
\begin{equation}
  \frac{\sigma_{CPA}(k^+,l)}{c\varepsilon_0}=\frac{ik^+[\varepsilon j_l(k^-)[k^+h_l^{(2)}(k^+)]'-h_l^{(2)}(k^+)[k^-j_l(k^-)]']}{[k^-j_l(k^-)]'[k^+h_l^{(2)}(k^+)]'},
  \label{eq: Equation 1}
\end{equation}
where $c$ is the speed of light, $\varepsilon_0$ is the permittivity of vacuum, and the primes indicate derivatives with respect to the arguments of the Bessel and Hankel functions. In addition, we have introduced $k^+\equiv (R\omega/c)$ and $k^-\equiv nk^+$ for convenience. Although obvious from \cref{eq: Equation 1}, we stress that $\sigma_{CPA}$ depends on $R$ and $\omega$ only through their dimensionless products, $k^+, k^-$--meaning \cref{eq: Equation 1} alone places no restrictions on the size of the scatterer or the frequencies at which CPA may be achieved. These may only be found with a concrete material model for $\sigma(\omega)$ (e.g. a Drude dispersion). We discuss such conductivity models in the next section  when we consider graphene coatings. We stress that although \cref{eq: Equation 1} is a closed-form solution for $s_l=0$, closed-form solutions may be found for $s_l=\tilde{s}$, for $\tilde{s}$ taking \textit{any} value. This is simply because $s_l=\tilde{s}$ is equivalent to an equation that is \textit{linear} in the conductivity (as can be seen from the form of $s_l$ reported in SI Sections \suppmiespheres \space and \suppmiescylinders). Note that an analogous statement cannot be made for a bulk scatterer. That is, $s_l=\tilde{s}$ is \textit{not} equivalent to an equation that is linear in $\varepsilon$. 

In addition, we note the striking result that $\sigma^\prime_{CPA} (k^+, l)\equiv \Re \sigma_{CPA}(k^+, l)$ is completely independent of $\varepsilon$, the bulk permittivity. This may be verified by taking the real part of \cref{eq: Equation 1}, which gives the following:
\begin{equation}
    \sigma'_{CPA}(k^+, l)=c\varepsilon_0 k^+\Bigg[\frac{j_l(k^+)(k^+y_l(k^+))'-y_l(k^+)(k^+j_l(k^+))'}{(k^+j_l(k^+))'^2+(k^+y_l(k^+))'^2} \Bigg],
\end{equation}
where $y_l$ is the spherical Bessel function of the second kind. The notion that $\Re \sigma_{CPA}(k^+, l)$ is  independent of $\varepsilon$ for \textit{all} values of $k^+$ is somewhat surprising because the underlying physics is quite different in character in the $k^+\ll 1$ and $k^+\gg 1$ regimes. For small $k^+$ (our main focus here), the resonant process is plasmonic excitation, a quasistatic process engendered by the coating, while for large $k^+$ the excitation is a bulk dielectric mode, with the coating serving as a damping source to achieve critical coupling (similar to a Salisbury screen). We note, however, that the fact that $\Re \sigma_{CPA}(k^+, l)$ is independent of $\varepsilon$ has a simple and intuitive explanation. If $\varepsilon$ is real, then one can show that the requirements of perfect absorption and energy conservation alone (both of which require knowledge only of the fields outside the scatterer) immediately determine $\sigma'_{CPA}$ (see SI Section \suppsigmarealindependence \space for further details). 

 Furthermore, we note that the imaginary part of the CPA conductivity, $\sigma''_{CPA}(k^+, l)\equiv \Im\sigma_{CPA}(k^+, l)$, is \textit{not} independent of $\varepsilon$. In fact, due to the zeros of the derivative of the Riccati-Bessel function term, $[k^-j_l(k^-)]'$, in the denominator of \cref{eq: Equation 1}, $\sigma''_{CPA}(k^+, l)$ has an infinite number of divergences (see SI Section \suppdivergence), implying a high sensitivity of $\sigma''_{CPA}(k^+, l)$ to variations of $\varepsilon$ on the real axis, with $k^+$ fixed. These divergences further imply that $\sigma_{CPA}''(k^+, l)$ can be tuned to \textit{any} value by varying the bulk permittivity, a property to which we will return when we study the effect of deviations of the surface conductivity from its CPA value.

In summary, \cref{eq: Equation 1}, which is exact and applies to all size regimes, tells us that for a given $k^+$, if we can find a coating for which the conductivity satisfies $\sigma(\omega)=\sigma_{CPA}(k^+,l)$, for some $\omega$, then a spherical scatterer of radius $R=ck^+/\omega$ will perfectly absorb an incoming spherical wave with angular momentum $l$. In the rest of this paper, we will denote the $\omega$ for which $\sigma(\omega)=\sigma_{CPA}(k^+, l)$ by $\omega_{CPA}$ to differentiate it from generic frequencies, which we denote simply by $\omega$. In addition, we define $R_{CPA}=ck^+/\omega_{CPA}$ \footnote{We will usually omit dependencies of $\omega_{CPA}$ and $R_{CPA}$ on the angular momentum/$k^+$ since in all cases such dependencies will be obvious from the context. In all figures showing $\omega_{CPA}$ and $R_{CPA}$ the angular momentum under consideration will be $l=1$.}. 

Before discussing candidate coatings to  achieve $\sigma_{CPA}$, we first discuss the behavior of $\sigma_{CPA}$ in the limiting regimes $k^+\ll1$ and $k^+\gg 1$. For small $k^+$, the value of $\sigma''_{CPA}$ may be well described quasistatically by, for example, finding the poles of the static multipolar polarizability of a coated system (SI Sections \suppspheresstatic \space, \suppsecondspheresstatic \space and \suppcylindersstatic). Such a quasistatic description yields a linear (in $k^+$) approximation to $\sigma''_{CPA}(k^+, l)$ (for all $l$) that agrees with the fully-retarded value of $\sigma''_{CPA}$ for $k^+\lesssim0.5$ (see \cref{fig: figure 1}(b)). Since a static approximation does not take into account radiative coupling, the behavior of $\sigma'_{CPA}(k^+, l)$ for small $k^+$ can only be obtained through a Taylor expansion of \cref{eq: Equation 1}, which yields $\sigma'_{CPA}(k^+, l)\propto (k^+)^{(2l+2)}$ (see SI Section \suppsigmarealindependence \space for explicit expressions). The weaker dissipation required with larger $l$ is expected, since it is more difficult for the incident wave to excite the subwavelength sphere (the origin of the weaker radiative coupling).

In \cref{fig: figure 1}(c, d), we show the behavior of $\sigma'_{CPA}(k^+, l)$ (as well as the $(k^+)^{(2l+2)}$ small-$k^+$ approximations) in the subwavelength regime. Note that $\sigma'_{CPA}(k^+, 1)\gg\sigma'_{CPA}(k^+, 2),\sigma'_{CPA}(k^+, 3)$, consistent with the understanding that radiative coupling is strongest in the dipolar ($l=1$) channel.  

 Before moving on to the $k^+\gg 1$, ``ray limit" regime, we note that a different approximation to $\sigma_{CPA}(k^+, l)$ for $k^+\ll 1$ can yield interesting insights. As noted in \cite{noh2013broadband}, the maximum field enhancement at the boundary of a subwavelength scatterer occurs when the scattering coefficient $s_l=-1$. This corresponds to maximizing the contribution of the divergent part \footnote{i.e. the part proportional to the Bessel function of the second kind} of the electromagnetic field at the origin. 
 As we show in the SI (Section \suppcoupledlattice), a deeply subwavelength scatterer with surface conductivity $\sigma''_{CPA}$ (the CPA conductivity but without the dissipative real part) yields a perfect $\pi$ phase shift of the outgoing wavefront (and, therefore, $s_l=-1$). Thus, in lieu of the quasistatic approximation, we may approximate $\sigma''_{CPA}(k^+\ll 1, l)$ with an expression for the conductivity which would give us $s_l=-1$. This amounts to neglecting the contributions from the non-diverging Bessel functions in \cref{eq: Equation 1} and gives us: 
\begin{equation} 
\frac{\sigma_{CPA}''(k^+,l )}{c\varepsilon_0}\approx\frac{\sigma_{PS}''(k^+, l)}{c\varepsilon_0}=k^+\Bigg[\frac{\varepsilon j_l(k^-)}{[k^-j_l(k^-)]'}- \frac{y_l(k^+)}{[k^+y_l(k^+)]
'}\Bigg],
\end{equation}
where we have introduced the lossless passive sink conductivity (in the terminology of \cite{noh2013broadband}), $\sigma_{PS}(k^+, l)=i\sigma''_{PS}(k^+, l)$, which yields a perfect $\pi$ phase shift of the outgoing wave (for any $k^+$). Note that this implies that if we imagine starting with a system with $\sigma=\sigma_{PS}(k^+, l)$, for sufficiently small $k^+$, then we can reach CPA simply by adding a small amount of dissipation into the coating, without having to further tune the imaginary part of the conductivity. 
The approximate identity of $\sigma_{CPA}''(k^+,l )$ with $\sigma_{PS}''(k^+, l)$ for small $k^+$ is shown in \cref{fig: figure 3}(a) (for $l=1$). Due to this identity, the scattering coefficient has an almost perfect $\pi$ phase shift until $k^+\approx 0.62$ (at which point it reaches 90\% of a perfect phase shift) as shown in \cref{fig: figure 3}(b).

Now we consider the behavior of $|s_{l=1}|$ for $k^+ = 0.22$, well into the regime where the above identity holds very well.  Starting from the lossless $\sigma=\sigma_{PS}(k^+, l=1)$, we vary the dissipation $\sigma'$ from zero through the value needed to achieve CPA at $k^+=0.22$. As shown in \cref{fig: figure 3}(c), $|s_{l=1}|$ decreases monotonically until we reach the critical coupling value, $\sigma'_{CPA}(k^+, l=1)$, after which $|s_{l=1}|$ increases again from zero--although it never reaches unity due to the unavoidable dissipation when $\sigma'>0$. Since $s_{l=1}$ is real to a good approximation here, it simply evolves from negative real values and passes through zero with an attendant phase jump from $\phi = \pi$ to zero (\cref{fig: figure 3}(d)). In keeping with the standard picture of CPA \cite{chong2010coherent}, for small dissipation this system is undercoupled and there is a zero of $s$ in the upper half of the complex $k^+$ plane; as the dissipation increases, the zero crosses the real axis and passes into the lower half plane.

Note that in the subwavelength regime, $\sigma''_{PS}(k^+, l)$ gives an even better approximation to $\sigma_{CPA}''(k^+, l)$ than the linear quasistatic approximation, as can be seen by comparing \cref{fig: figure 3}(a) with \cref{fig: figure 1}(a). 

We now move to the semiclassical limit, $k^+\gg 1$. There are two relevant regimes for large $k^+$. First we consider $k^+$ and $k^- \gg l $; in this limit the asymptotic behavior of the Bessel functions yields the following forms for $\sigma_{CPA}(k^+, l)$ (see Section \supplargekspheres \space and \supplargekcylinders\space of the SI):
\begin{equation}
\begin{split}
    \frac{\sigma_{CPA}(k^+\gg1, l)}{c\varepsilon_0}\approx 1 + i\sqrt{\varepsilon}\tan\Bigg[k^--\Bigg(l+\frac{1}{2}\mp\frac{1}{2}\Bigg)\frac{\pi}{2}\Bigg]
     \end{split}
     \label{equation: asymptotic cpa},
\end{equation}
where the signs correspond to the TM (top) and TE (bottom) channels, respectively. In this limit, the resonances are the Fabry-Perot type resonances of the dielectric sphere, with the coating providing the dissipation needed for critical coupling. That dissipation, given by the real part of \cref{{equation: asymptotic cpa}}, is just $c\varepsilon_0$, exactly one half of the planar surface conductivity \cite{baranov2017coherent, fan2014tunable, fan2015tunable} required for CPA of two counter-propagating beams in a thin film geometry. In \cref{fig: figure 5}(a, b), we show, for large $k^+$ (and small $l$), both the asymptotic and exact CPA conductivities for the TM and TE channels, respectively. Note that since we are in the Fabry-Perot limit of equally spaced resonances, the imaginary part of the CPA conductivity given by \cref{equation: asymptotic cpa} has the property that $\sigma''_{CPA}(k^++\frac{\pi}{2\sqrt{\varepsilon}}, l+1)=\sigma_{CPA}(k^+, l)$, which results in a shift between the even and odd angular momentum channels. In addition, $\sigma''_{CPA}(k^+, l)$  is periodic in $k^-$ with periodicity $\Delta k^-=\pi$. Lastly, we note that the aforementioned divergent behavior of $\sigma''_{CPA}$ at discrete values of $k^-$ is obvious in this limit. 

The second semiclassical of interest is the regime where $k^- = nk^+>l > k^+ \gg 1$; this is the limit of whispering gallery resonances in which rays are trapped by total internal reflection (but the fields leak out evanescently due to the curvature of the surface). In this limit of high-Q resonances we expect the CPA condition to correspond to much lower dissipation. Using the Debye approximation for the Bessel functions, one indeed finds $\sigma'_{CPA}\ll c\varepsilon_0$. In addition, $\sigma''_{CPA}(k^+, l)$ becomes, just as in the Fabry-Perot limit, oscillatory. However, in the whispering gallery limit, the spacing is related to rays circulating around the circumference (as opposed to bouncing along the diameter). That is, we have, in the whispering gallery limit, $\sigma''_{CPA}(k^++\frac{1}{\sqrt{\varepsilon}}, l+1)\approx \sigma''_{CPA}(k^+, l)$ (see SI Section \suppsectiondebye \footnote{The derivation in the SI is for cylinders, for simplicity, but the exact same behavior applies for spheres.}). 

\begin{figure}
\centering
      \includegraphics[scale=1]{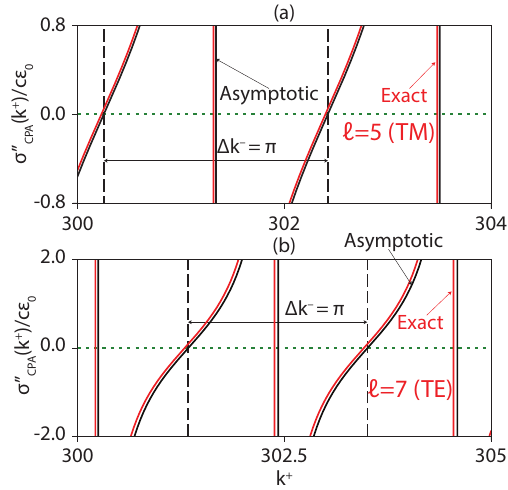}
  \caption{\textbf{CPA for large coated spheres.} We show here the behavior of the CPA conductivity in the Fabry-Perot limit. (a) Imaginary part of the CPA surface conductivity for the TM channel, for $l=5$ and $k^+\gg1$. (b) Same as (a) but for TE modes and $l=7$. For both (a) and (b), the corresponding real part of the CPA conductivity (not plotted) is $\sigma'_{CPA}\approx c\varepsilon_0$. Asymptotic curves are given by \cref{equation: asymptotic cpa}. Vertical dashed lines indicate points at which $\sigma''_{CPA}(k^+)=0$. At these points, the CPA conductivity is exactly one half of the value required for CPA of a two-dimensional resistive sheet at normal incidence. For both (a) and (b), the permittivity of the system is $\varepsilon=2.1$ (in accordance with \cref{fig: figure 1}). }
  \label{fig: figure 5}
\end{figure}

\begin{figure*}
\centering
      \includegraphics[scale=1]{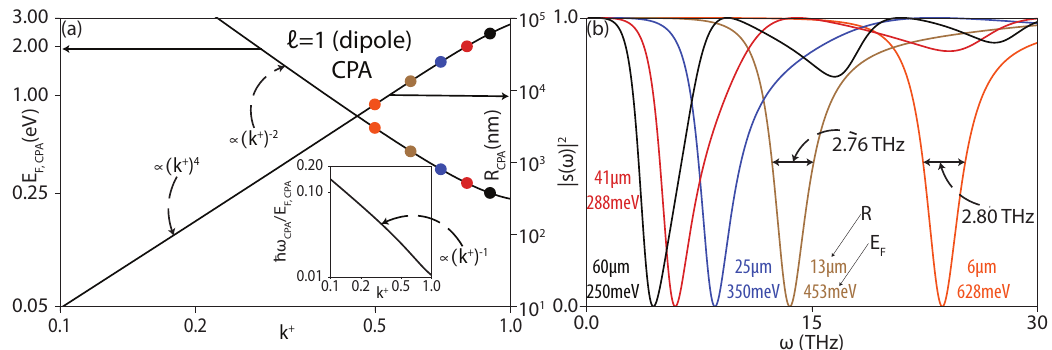}
  \caption{\textbf{Physical parameters for CPA for a graphene-coated sphere:} (a) We show $E_{F, CPA}(k^+, l)$ (left vertical axis) and $R_{CPA}(k^+, l)$ (right vertical axis) in the subwavelength range of $k^+$ for the dipolar ($l=1$) channel. In both cases, the scale is logarithmic for both the horizontal and vertical axes in order to clearly show the scaling with $k^+$. Inset shows the ratio $\hbar\omega_{CPA}/E_{F, CPA}$ as a function of $k^+$, validating the use of the intraband approximation for the graphene conductivity (see main text). (b) $|s(\omega)|^2\equiv |s_{l=1}(\omega)|^2$ for graphene-coated systems for $k^+=0.5, 0.6, 0.7, 0.8, 0.9$ in orange, brown, blue, red and black, respectively. The coatings have radii $R =R_{CPA}(k^+, l=1)$ and are doped at $E_F=E_{F, CPA}(k^+, l=1)$ for each $k^+$ (see color-coded spheres in (a)). CPA occurs at points at which $|s|^2=0$ (at frequencies $\omega_{CPA}(k^+, l=1)$). Color-coded labels show the value for $R_{CPA}(k^+, l=1)$ and $E_{F, CPA}(k^+, l=1)$ and we indicate the full width at half maximum of absorption for two values of $k^+$ with double-sided arrows.}
  \label{fig: figure 7}
\end{figure*}

\paragraph{Graphene as a tunable platform}

We now turn to candidate two-dimensional materials that may be used as plasmonic coatings for subwavelength critical coupling. As stated previously, in order to achieve perfect absorption at a given $k^+$, we require a coated scatterer for which  $\sigma(\omega_{CPA})=\sigma_{CPA}(k^+, l)$ for some $\omega_{CPA}$. Once $\omega_{CPA}$ has been found, the radius of the scatterer must be fixed to $R_{CPA}=k^+c/\omega_{CPA}$. Here, we consider doped graphene and show that for appropriate levels of doping, a graphene coating can indeed achieve $\sigma(\omega_{CPA})=\sigma_{CPA}(k^+)$ for subwavelength values of $k^+$ (for the dipolar, $l=1$, channel and for $\omega_{CPA}$ in the ThZ regime). This then means that a graphene-coated scatterer of suitable radius can absorb coherent light with $l=1$. We do not consider $l>1$ simply because the real part of the CPA conductivity (see \cref{fig: figure 1}(c, d)) is too small to be achievable with graphene (with reasonable physical parameters).

Graphene-coated nanospheres have been studied in the context of their plasmonic properties \cite{christensen2015localized} and as platforms for tailored nonlinearities \cite{smirnova2014second, wang2022detecting}. However, their capacity to engender tunable subwavelength CPA has not been explored, to our knowledge. This may stem from the fact that undoped graphene has an absorbance of only $\approx 2.3\%$ \cite{grigorenko2012graphene}, so, in the absence of patterning or doping, the conductivity of graphene is simply too small in magnitude to sustain perfect absorption, requiring $\approx88$ layers for perfect absorption in a thin film geometry \cite{liu2014gate}. Thus, from the outset, we know that we must restrict ourselves to doped graphene, not only to enhance absorption, but also because we require electronic free carriers if we are to have subwavelength plasmonic resonances. With these preliminaries out of the way, we describe our model for graphene's conductivity, $\sigma(\omega)$, and report our results for the frequencies and radii for which one may expect perfect absorption by a graphene-coated nanosphere. 

For frequencies well below the Fermi energy, $E_F$, the conductivity of doped graphene may be approximated by a Drude form \cite{wunsch2006dynamical, jablan2009plasmonics}:
\begin{equation}
    \sigma(\omega)=\frac{ie^2E_F}{\pi\hbar^2(\omega+i/\tau)}
\end{equation}
In such an approximation, doping graphene (changing $E_F$) rescales both the real and imaginary parts of the conductivity by the same amount. Furthermore, tuning the frequency appropriately allows one to choose any desired ratio, $\sigma''(\omega)/\sigma'(\omega)=\tau\omega$, where $\tau$ is the electronic scattering time. Thus, to achieve CPA, one has to dope graphene to a Fermi energy, $E_{F, CPA}$, and excite the system at frequency, $\omega_{CPA}$, given by:
\begin{equation}
\begin{split}
    &E_{F, CPA}(k^+, l)= \frac{\hslash |{\sigma'_{ CPA}(k^+, l)+i\sigma''_{CPA}}(k^+, l)|^2}{4c\alpha\varepsilon_0\sigma'_{CPA}(k^+, l)\tau}, 
\\& \omega_{CPA}(k^+, l)=\frac{1}{\tau}\Bigg[\frac{\sigma''_{CPA}(k^+, l)}{\sigma'_{CPA}(k^+, l)} \Bigg],
\end{split}
\end{equation}
where in our calculations we used (in all figures) $\tau=640\text{fs}$, which may be motivated by using experimentally observed values for the DC mobility \cite{novoselov2004electric} (Section \suppdcmobility \space of the Supporting Information). $\alpha$ is the fine structure constant and $\hbar$ is Planck's constant. Of course, the radius of the scatterer is no longer an independent parameter and must be set to $R_{CPA}=k^+c/\omega_{CPA}$. In \cref{fig: figure 7}(a), we plot $E_{F, CPA}$ and $R_{CPA}$ on log-log scales. Observe that $E_{F, CPA}(k^+)$ scales as $(k^+)^{-2}$ and $R_{CPA}(k^+)$ scales as $(k^+)^4$ for $l=1$. For $k^+<0.5$, the required Fermi energies are too high for experimental realization. However, for $k^+>0.5$, subwavelength CPA is possible for reasonable levels of doping, on the order of hundreds of meV. 

The discussion above yields the physical parameters for CPA, but it tells us nothing about the linewidth of plasmonic CPA resonances, and, indeed, this has not been addressed analytically in previous work. To address this, in \cref{fig: figure 7}(b), we show the behavior of $|s_{l=1}|^2$ for graphene-coated dielectric spheres for $k^+=0.5, 0.6, 0.7, 0.8$ and $0.9$. In each case, we fix the Fermi energy to $E_F=E_{F, CPA}(k^+, l=1)$, the radius to $R=R_{CPA}(k^+, l=1)$ and vary the frequency over a range including $\omega_{CPA}(k^+, l=1)$. As seen in \cref{fig: figure 7}(b), we achieve CPA at the discrete frequencies given by $\omega_{CPA}(k^+, l=1)$. In addition, we clearly see that the CPA bandwidth is almost independent of $k^+$. Indeed, for a Drude material, the CPA bandwidth for small $k^+$ is exactly given by the electronic scattering rate of the coating (SI Section \suppbandwidth). Thus, the half-width at half maximum of $|s_{l=1}|^2$ is set by $1/\tau\approx 1.56 \text{THz}$.

We close this section by noting that graphene can also, in principle, support subwavelength CPA in the TE channel \cite{he2013analysis, jablan2011transverse, soleimani2025anisotropic}. Subwavelength CPA in the TE channel can only be achieved at frequencies, $\omega_{CPA}$, for which $\sigma''(\omega_{CPA})<0$. This is not possible in the Drude approximation, which takes into account only the free carrier contribution to the conductivity of a metal. Luckily, the low energy physics of graphene yields both Drude and interband contributions to its electronic conductivity, and, for frequencies close to the onset of electronic interband transitions (i.e. $\hbar\omega\approx 2E_F$), the interband contribution can overwhelm the Drude contribution, yielding $\sigma''(\omega\approx 2E_F)<0$. Unfortunately, however, the magnitude of $\sigma''(\omega\approx 2E_F)$ is simply too small to achieve $\sigma''_{CPA}(k^+, l)$ for any non-negligible bandwidth, as we discuss further in the SI (Section \supptebandwidth). The SI also includes calculations for graphene-coated dielectric cylinders, for completeness (SI Section \suppcylindersstatic).

\begin{figure*}
\centering
      \includegraphics[scale=1]{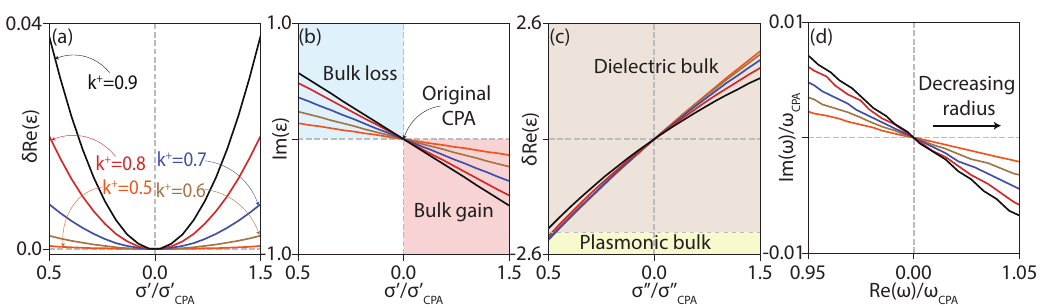}
  \caption{\textbf{Effect of deviations from the CPA conductivity for $l=1$:} (a, b) $\sigma'$ is detuned from $\sigma'_{\text{CPA}}(k^+, l)$, requiring a change in the bulk permittivity to bring the system back to CPA at real frequencies. (a) ((b)) shows the required change in the real (imaginary) part of the permittivity. For $\sigma'>\sigma'_{CPA}$, we require $\Im\varepsilon<0$ (corresponding to a gain medium in the bulk--red shaded region in (b)) and for $\sigma'<\sigma'_{CPA}$ we require  $\Im\varepsilon>0$, corresponding to a lossy bulk (blue shaded region in (b)). (c) Same as (a) but we instead detune $\sigma''$ from $\sigma''_{CPA}$. As discussed in the main text, the imaginary part of the permittivity (not shown) does not need to be changed in this case. We delineate in (c) the phase region corresponding to a dielectric bulk ($\varepsilon>0$) and the region corresponding to a plasmonic bulk ($\varepsilon<0$) (d) CPA solutions at complex frequencies as the radius is detuned from $R_{\text{CPA}}$ (for $0.9<\delta_R/R_{CPA}<1.1$). }
  \label{fig: figure 6}
\end{figure*}

\paragraph{CPA solutions when $\sigma\neq \sigma_{CPA}(k^+, l)$} If the surface conductivity does not match its required value for CPA, the zero of the scattering coefficient, $s_l$, will no longer be on the real frequency axis but will still occur elsewhere in the complex frequency plane. In order to achieve CPA, one can imagine tuning the bulk permittivity to bring the CPA zero back on the real axis. Or, alternatively, one may be able to access the CPA zero with complex frequency excitation \cite{kim2025complex}. In the remainder of this section, we primarily focus on the first method, finding the change in permittivity required to achieve CPA at real-valued frequencies. For completeness, however, we also find the location of the CPA zeros in the complex frequency plane when the system is not critically coupled. 

In \cref{fig: figure 6}(a-c), we consider the first method. As shown in SI Section \suppcomplexfrequencies, if $\sigma'$ ($\sigma''$) is varied from its CPA value, one must change $\Im(\varepsilon)$ ($\Re(\varepsilon)$) by an amount proportional to the change in $\sigma'$ ($\sigma''$) in order to bring the system back to critical coupling. This is clearly verified in \cref{fig: figure 6}(b, c). Surprisingly, however, we see in \cref{fig: figure 6}(a) that if the dissipation in the coating is altered, $\Re(\varepsilon)$ (not just $\Im(\varepsilon)$) must also be tuned--and that the amount by which it must be tuned is \textit{quadratic} in the change of $\sigma'$. Even more surprisingly, if $\sigma''$ is varied, the corresponding change in $\Im(\varepsilon)$ (not shown) is identically zero. This last fact actually follows immediately from our previous observation that $\sigma''_{CPA}$ can be tailored to any value by moving $\varepsilon$ on the real line. Said differently, $\Im\varepsilon$ need not be tuned if $\sigma''_{CPA}$ is varied. 

Turning now to the second method, we track the CPA zeros in the complex $\omega$ plane as the system parameters are varied. To do so properly, we must account for the dispersion of the conductivity of the coating. In keeping with our focus on plasmonic coatings in this paper, we consider a system coated with a Drude material with system parameters initially tuned to CPA. We now choose to alter the radius of the scatterer slightly from $R_{CPA}$ ($R\rightarrow R_{CPA}+\delta_R$), which moves the CPA zero to a complex frequency, $\omega$, given by (SI Section \suppcomplexfrequencies):
\begin{equation}
    \frac{\omega}{\omega_{CPA}}=\Bigg[1-\frac{\delta_R}{2R_{CPA}}\Bigg]+i\frac{(k^+)^3}{2+\varepsilon}\frac{\delta_R}{R_{CPA}}
    \label{Equation: complex omega off cpa}
\end{equation}
As shown in the SI, the shift in the real part of the frequency is governed entirely by the dispersion of the surface conductivity--not by the geometry under consideration. In order to build some physical intuition regarding \cref{Equation: complex omega off cpa}, we note that, clearly, $\Re(R\omega/c)> k^+$ if $\delta_{R}>0$ and $\Re(R\omega/c)< k^+$ if $\delta_R<0$. This simple observation immediately implies that the CPA zero must move up in the complex plane if the radius is increased (as we will have more radiative coupling) and move down if the radius is decreased, consistent with \cref{Equation: complex omega off cpa}. Furthermore, this behavior is exactly what we observe in \cref{fig: figure 6}(d). 

\paragraph{Moving beyond the isolated scatterer regime} The previous discussion always assumed an isolated sphere (or cylinder) illuminated by a spherical/cylindrical wavefront with harmonic time dependence and fixed angular momentum. As noted, such wavefronts are difficult to synthesize in a lab, so we now consider systems that can display the same physics but with more practical incident wavefronts. In both of the cases considered below, the key property will be the polarizability, $\alpha(\omega)$ \footnote{We do not add Cartesian subscripts to the polarizability, since in all cases considered below the polarizability will either be isotropic (as in the case of a sphere) or the direction of the polarization of the incident light will clarify the relevant component of the polarizability tensor.}, of the coated spherical or cylindrical scatterer. Although emphasized below, we stress that the surface conductivity of a coated scatterer can be written in terms of its polarizability in closed form. Thus, if the polarizability required for perfect absorption, $\alpha_{CPA}$, can be evaluated in closed form, the attendant CPA conductivity, $\sigma_{CPA}$, can also be written in closed form (\textit{without} the need for brute-force numerical searches for the CPA zero).
\paragraph{Scatterer above a PEC substrate}
In the next two sections, we address the issue of creating the coherent wavefront required for perfect absorption by a coated scatterer. 
To circumvent the challenge of synthesizing a spherical wave, we consider, in this section, a modified geometry--a scatterer suspended over a perfectly conducting substrate (see \cref{fig: figure 9}(a)). This geometry was previously analyzed \cite{proskurin2021perfect} for the case of an uncoated scatterer, and we build on their results. For this geometry the CPA wavefront is obviously not a spherical wave, nor is it in general a $2\pi$ solid angle truncation of a spherical wave. For the case of a dipolar emitter/absorber the authors of \cite{proskurin2021perfect} were able find a solution for the polarizability which leads to CPA (given by Equation. (6) of \cite{proskurin2021perfect}) as well as the required incident wavefront (given by Equation. (3) of \cite{proskurin2021perfect}). To apply their results to coated scatterers, we relate this CPA polarizability to the retarded polarizability of a coated sphere, expressed in terms of the $l=1$ scattering coefficient, $s_{l=1}$. This we do in closed form in the SI (Section \suppspherepolarizability). For convenience, we reproduce, below, the result derived in detail there: 
\begin{equation}
    s_{l=1}(\omega)=1+\frac{i\alpha(\omega)}{3\pi }\Bigg(\frac{\omega}{c}\Bigg)^3
\end{equation}

Since  $s_{l=1}$ is known analytically in term of the surface conductivity of the coating (and we can analytically invert the relation between $s_{l=1}$ and the conductivity in closed-form to obtain the conductivity required to give \textit{any} value for $s_{l=1}$--as mentioned before), we are able to immediately obtain a closed form expression for $\sigma_{CPA}(k^+)$ for this geometry \footnote{We, of course, do not have angular momentum as a good quantum number in this geometry, so $\sigma_{CPA}$ does not have an angular momentum index.}. 

In \cite{proskurin2021perfect}, only scatterers composed of bulk plasmonic materials (e.g. silver) were considered, yielding CPA only at discrete values of $k^+$ for a given value of the PEC-substrate distance, $h$. We find that a doped coated scatterer yields significantly better tunability. In particular, in  \cref{fig: figure 9} we show that a scatterer coated with graphene can be appropriately tuned (by varying the Fermi energy and the radius) to achieve CPA at any vertical separation.    
Second, we note the asymptotic behavior of $E_{F, CPA}$ and $R_{CPA}$ in the limit $h\gg \lambda$. In this regime, both quantities approach their values for an isolated single sphere, given by \cref{fig: figure 7}.

\begin{figure}
\centering
\includegraphics[scale=1]{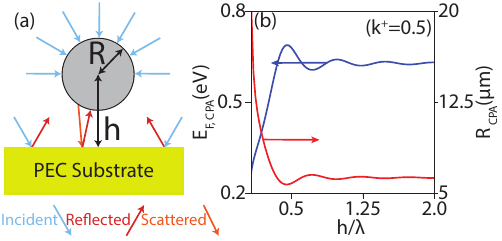}
  \caption{\textbf{CPA for a scatterer above a PEC:} (a) Physical setup of the scattering problem, showing the excitation of the coated scatterer by an incident beam pattern (blue arrows). The response of the scatterer is determined by the incident field and all fields reflected from the PEC substrate (including reflections of the fields scattered from the coated scatterer). (b) $E_{F, CPA}$ and $R_{CPA}$ for a coated spherical scatterer above a PEC substrate as a function of the normalized distance, $h/\lambda$, from the center of the scatterer to the substrate.}
  \label{fig: figure 9}
\end{figure}

\paragraph{Lattice of coated cylindrical scatterers}

\begin{figure}
\centering
      \includegraphics[scale=1]{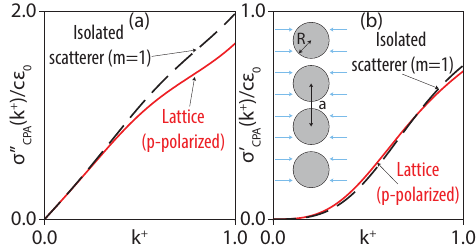}
  \caption{\textbf{CPA from coated metamaterial arrays}:  Coated cylinders, each with a radius, $R$, are arranged in a one-dimensional lattice with periodicity, $a$ (see inset of (b)). The incident field is composed of two counter-propagating plane waves (denoted by light blue arrows), and we take the field to be p-polarized (i.e. electric field polarized along the direction of periodicity). Black and red curves in (a) ((b)) show $\sigma''_{CPA}(k^+)$ ($\sigma'_{CPA}(k^+)$) for isolated  cylindrical scatterers (for $m=1$ and TM polarization) and for the one-dimensional lattice, respectively. For both the isolated and the lattice cases, the bulk permittivity of the cylinders is set to unity (i.e. $\varepsilon=1$). Furthermore, for the lattice case, we take $\lambda/a\approx 2$, which means that all but the lowest diffraction order is evanescent and does not contribute in the far-field.}
  \label{fig: figure 2}
\end{figure}

We now move to the last geometry we analyze in this paper, with the aim of achieving coherent absorption of \textit{plane waves} by a lattice array of critically coupled coated cylinders. A plane wave, of course, is an extended state, and, as such, it cannot be fully absorbed by a single \textit{isolated} nanostructure. Thus, in this section, we instead study periodic arrays of nanostructures (sticking to one-dimensional arrays for simplicity). Such systems have been widely investigated in the literature  \cite{cioci2024tunable, kou2016discrete} for their rich interplay between lattice and geometric resonances \cite{gomez2006extraordinary, babicheva2018metasurfaces}. More relevantly, in the context of tunable absorption, graphene nanodisks \cite{christensen2014classical} were predicted and experimentally demonstrated to yield strong plasmon-enhanced optical absorption \cite{thongrattanasiri2012complete, fang2014active}. In the derivations below, we will deviate somewhat from these works in the sense that we will primarily focus on exploiting the resonances of the individual scatterers in the array to engender tunable CPA (i.e. we will operate in a wavelength regime where lattice resonances do not play a role). Furthermore, our goal will be to, as closely as possible, relate the CPA conductivity of the lattice array to that of the isolated scatterer. Lastly, we will focus on a system in which the polarizability of the isolated scatterer can be written in closed form, and we will try, as best as possible, to uncover all of the relevant physics without having to resort to tedious lattice sums. 

There are three main ingredients involved in extending our formalism to lattice arrays. The first is the mutual interaction between the scatterers (which we take to be infinitely long cylinders) mediated by the appropriate Green's function \cite{morse1946methods} summed over all (but one) lattice sites \cite{garcia2007colloquium, evlyukhin2010optical, babicheva2018metasurfaces, yurkin2007discrete, gomez2006extraordinary, awan2024tunable}. This tells us the field at a given lattice site if the dipole moments at all other lattice sites is known. The second ingredient is the reflection coefficient that is also determined by a lattice-summed Green's function (over \textit{all} lattice sites this time). The last necessary ingredient is the polarizability of each scatterer, which is the only single particle scattering quantity required and can be related to the Mie scattering coefficient, $s_{m=1}$ (see SI Section \suppcoupledlattice). In the following, we neglect the influence of finite size effects of the cylinders, since we expect these to be negligible \cite{zundel2019finite}, and we consider systems for which the wavelength, $\lambda$, is larger than the lattice period, $a$. This is because $\lambda>a$ ensures that the scattering problem may be considered as effectively two-port, since higher diffraction orders will not couple to the far-field in this case \cite{baldacci2015interferometric}. 

To motivate the connection between plane wave absorption and the cylindrical/spherical wave absorption studied up to this point, we note that the total field for an array of \textit{noninteracting} cylinders (placed at lattice sites $(x_n, y_n) = (na, 0)$ with $n$ an integer) may be written using the Jacobi-Anger identity in combination with the scattering coefficients for cylindrical waves: 
\begin{equation}
    H_z = e^{i\omega y/c}+\frac{1}{2}\sum_{n=-\infty}^\infty\sum_{m=-\infty}^\infty \Bigg[(s_m-1)H^{(1)}_m(\omega r_n/c)e^{im\theta_n} \Bigg],
    \label{Equation: Scattering off noninteracting cylinder array}
\end{equation}
where $r_n\equiv \sqrt{(x-x_n)^2+(y-y_n)^2}$ and $\tan(\theta_n)=(y-y_n)/(x-x_n)$. We have chosen the field to be p-polarized (magnetic field, $\mathbf{H}\equiv H_z\mathbf{e}_z$, along the axis of each cylinder), so that the array can support subwavelength plasmonic resonances. Note that the first term in \cref{Equation: Scattering off noninteracting cylinder array} corresponds to the incident wavefront and the second term corresponds to the sum of outgoing scattered waves (cylindrical Hankel functions of the first kind) from each cylinder.  This expression may be simplified by noting that in the long wavelength limit $s_m\approx 1$ for all but the dipole term, meaning that higher multipole scattering can be reasonably neglected. Thus, \cref{Equation: Scattering off noninteracting cylinder array} may be considerably simplified as follows:
\begin{equation} 
H_z \approx  e^{i\omega y/c}+i(s_1-1)\sum_{n=-\infty}^\infty H_1^{(1)}(\omega r_n/c) \sin(\theta_n)
\end{equation}
Furthermore, we note that we have $H_1^{(1)}(\omega r_n/c)\sin(\theta_n)=-\frac{c}{\omega}\partial_y H_0^{(1)}(\omega r_n/c)$; in addition, $H_0^{(1)}$ is trivially related to the Green's function of the Helmholtz equation in two dimensions, meaning that we may easily write it in the Fourier domain and express the sum over lattice sites, to find the far-field behavior, explicitly as follows: 
\begin{equation}
\begin{split}
    &H_z = e^{i\omega y/c}-\frac{i(s_1-1)}{\omega/c}\partial_y\Bigg[\sum_{n=-\infty}^\infty  H_0^{(1)}(\omega r_n/c)\Bigg] = 
    \\&e^{i\omega y/c}\pm\frac{2(s_1-1)}{(a\omega/c)}e^{i\omega |y|/c},
\end{split}
\end{equation}
where the plus sign is taken for $y>0$ and the minus sign is taken for $y<0$. In order to have CPA, we need zero outgoing flux both below and above the lattice. Therefore, clearly, the incident wavefront cannot be a \textit{single} plane wave, and it must, instead, be a linear combination of two counter-propagating plane waves, for which the total field can be written as follows: 
\begin{equation}
H_z = e^{i\omega y/c}-e^{-i\omega y/c}\pm \frac{4(s_1-1)}{a\omega/c}e^{\pm i\omega|y|/c}
\label{Equation: Counterpropagating CPA}
\end{equation}
We stress that for this geometry CPA is thus \textit{not} equivalent to single-channel critical coupling condition and \textit{multichannel} coherent excitation is required to achieve CPA. In \cref{Equation: Counterpropagating CPA}, the first two terms correspond to the incident field, and the last term corresponds to the sum of all fields scattered by the lattice. Furthermore, from \cref{Equation: Counterpropagating CPA}, we immediately see that we must require $4(s_1-1)=-a\omega/c$ for perfect absorption of the counter-propagating plane waves. 

Up to this point, we have neglected the interaction between the individual cylinders. The full derivation of the interacting scattering problem is given in SI Section \suppcoupledlattice. The same basic procedure as the above is followed. Namely, the lattice-summed Green's function, together with the condition $|r+t|=0$ \footnote{Note that the requirement $|r+t|=0$ is an exact condition. Some papers, e.g. \cite{fan2015tunable}, have cited the more relaxed condition, $|r|=|t|$. However, the second relation in Equation. (2) of \cite{fan2015tunable} is incorrect, as may be readily checked.} determines the required polarizability of each isolated cylinder which then determines $s_{m=1}$ and the required conductivity. 

In \cref{fig: figure 2}, we show $\sigma_{CPA}(k^+)$ for a lattice of coated cylinders with $\lambda/a\approx 2$ (well into the two-port scattering regime). We compare our result with the CPA conductivity of an isolated cylinder in the dipolar TM channel, $\sigma_{CPA}(k^+, m=1)$. Importantly, even with the mutual dipolar interactions between the cylinders in the lattice array, the conductivity of the required coating does not differ substantially from that of a single isolated cylinder. Hence, we are justified in considering this geometry as allowing measurement of critically-coupled cylinders via plane wave excitation. Note that our model is two-dimensional (as the cylinders are taken to be infintely long), but this could be approximately realized by finite cylinders on a substrate excited parallel to the substrate surface. 

\paragraph{Discussion}
We begin by listing extensions of the work reported here that are either addressed at length in the SI or could be potentially fruitful, in our minds, directions for future research. First, and perhaps of most interest, we note that one may straightforwardly extend our results to coated systems with nonlocal response functions \cite{raza2013nonlocal, david2011spatial, christensen2014nonlocal, christensen2014classical, mcarthur2017coherent}. For such systems, $\sigma_{CPA}(k^+, l)$ is altered for the TM channel (which couples to charge density oscillations) but $\sigma_{CPA}(k^+, l)$ is unaltered for the TE channel (see SI Sections \suppnonlocalityspheres \space and \suppnonlocalitycylinders \space for derivations concerning spheres and cylinders with nonlocal response functions, respectively). In addition, although the results reported in this article have focused primarily on results obtained in the zero temperature limit, finite temperature effects may be included relatively easily. Finite temperature does not change $\omega_{CPA}$ or $R_{CPA}$ (for any of the geometries we considered in this work) but does change $E_{F, CPA}$. We omit here a lengthy discussion of finite temperature effects, but we refer the interested reader to Section \suppfinitetemp \space of the SI.
Lastly, we note that although we assumed, in this paper, that the effect of the curvature of a scatterer on the  electronic response of its coating should be minimal, one should, in principle, include the effect of strain \cite{castro2009electronic, manes2007symmetry}; indeed, the effect of strain on the critical coupling condition could yield interesting insights and perhaps allow one to achieve highly sensitive mechanically-tunable CPA. 

We consider the first part of this article as a contribution to the theory of Mie scattering, which had not focused on the issue of critical coupling, until the works of \cite{noh2012perfect,noh2013broadband}.  We have been able to find  general closed form solutions for the critical coupling conditions for coated spheres and cylinders which span from plasmonic, subwavelength, resonances to the semiclassical regime where they apply to both Fabry-Perot and whispering gallery resonances. In addition, we do find a geometry where CPA by cylindrical scatterers can be exhibited with plane wave illumination and another geometry where a sphere can perfectly absorb incident light (that is not of fixed angular momentum) when interacting with a conducting surface. We believe our work opens up several directions for experiments studying critical coupling in coated micro and nanostructures, especially because of the high degree of tunability achievable with coated scatterers.

Lastly, we note that the closed-form expressions we have derive here and in the SI are for the exact underlying material properties of the coatings--i.e. the CPA   conductivity. In this sense, our work is distinct from other works that have found closed form CPA solutions for, e.g., polarizabilities of perfectly absorbing dipole scatterers \cite{proskurin2021perfect, proskurin2022coherent}. This is because a polarizability is not a fundamental property of a material itself--it is determined by both the material and its geometry. Furthermore, the polarizability of a bulk (not coated) scatterer, cannot be inverted, in closed form, to determine the attendant permittivity of the bulk material. In this paper, we have emphasized that such an inversion is indeed possible (and is, in fact, mathematically simple) for the case of a coated scatterer. 

There are, of course, other special cases in which CPA may be found in closed form. An example of a platform in which we may know exactly what the parameters for CPA should be is an epsilon-near-zero (ENZ) material \cite{luo2018coherent, yan2023controlling}. However, this may be considered a major edge case, since the ENZ condition essentially nullifies any spatial variation in the electromagnetic fields, and is thus, somewhat by construction, amenable to closed-form solutions. 

\paragraph{Data availability} Our code is publicly available on Github and includes all code to make the figures in the main text and the supplement of this paper \cite{Coherent-Perfect-Absorption}.

\paragraph{Acknowledgements}
The authors thank Hui Cao and her group, Andrea Alu, Nazar Pyvovar, Hao He, Heeso Noh, Ognjen Ilic, Owen Miller and Thomas Christensen for useful discussions throughout the duration of this project. This work was supported by the Simons Collaboration on Extreme Wave Phenomena Based on Symmetries (award no. SFI-MPS-EWP-00008530-09).
\clearpage

\let\oldaddcontentsline\addcontentsline
\renewcommand{\addcontentsline}[3]{}
\bibliography{main}
\let\addcontentsline\oldaddcontentsline

\bibliographystyle{supp/apsrev4-2-longbib}
\clearpage
\setcounter{page}{1}
\setcounter{equation}{0}
\setcounter{figure}{0}
\vspace*{-1.5em}
\onecolumngrid
\renewcommand*{\arraystretch}{1} 

\hyphenation{}
\renewcommand{\thepage}{S\arabic{page}}
\setcounter{page}{1}
\newcommand{\equationcpa}{1}
\newcommand{\equationpassivesink}{3}
\newcommand{\equationdrudegraphene}{5}
\newcommand{\equationefomega}{6}
\newcommand{\figurecpaconductivity}{1}

\makeatletter
\def\ps@titlepage{%
  \def\@oddhead{\hfill\thepage}%
  \def\@evenhead{\thepage\hfill}%
  \def\@oddfoot{}%
  \def\@evenfoot{}%
}
\makeatother
\thispagestyle{titlepage}

\onecolumngrid

\begin{center}
    {\Large SUPPORTING INFORMATION}\\[0.75cm]
    {\Large \textbf{Plasmonic coated scatterers for tunable coherent perfect absorption}}\\[0.5cm]
    {\large Ali~Ghorashi$^{\textasteriskcentered{}}$,$^{1}$ Ali H. Alhulaymi,$^{1}$ 
    A. Douglas Stone$^{1}$
    }\\[0.25cm]
    {$^{1}$Department of Applied Physics, Yale University, New Haven, Connecticut 06520, USA\\}
{\textasteriskcentered{} ali.ghorashi@yale.edu}
\end{center}
\renewcommand{\theequation}{S\arabic{equation}} 
\renewcommand{\thefigure}{S\arabic{figure}} 
\renewcommand{\thetable}{S\arabic{table}} 
\renewcommand{\thesection}{S\arabic{section}}
\renewcommand{\thesubsection}{\Alph{subsection}}
\renewcommand{\thesubsubsection}{\roman{subsubsection}}
\maketitle

\makeatletter
    \DeclareRobustCommand*{\deactivateaddvspace}{\let\addvspace\@gobble} 
    \DeclareRobustCommand*{\deactivatetocsubsections}{
    \def\l@subsection##1##2{}    
    \def\l@subsubsection##1##2{} 
    }
\makeatother

\vspace*{-1.5em}
\noindent{\small\textbf{\textsf{CONTENTS}}}\\ 
\twocolumngrid
\begingroup
    \let\bfseries\relax 
    \deactivateaddvspace 
    \deactivatetocsubsections 
    \makeatletter\@starttoc{toc}\makeatother 
\endgroup
\onecolumngrid
\setlength{\parindent}{0em}
\setlength{\parskip}{.5em}
\section{Summary of the supporting information}
In this section, we provide a brief overview of the Supporting Information, detailing the key derivations and results (with the relevant section and equation numbers). In \cref{Section: dc decay time}, we calculate, using tabulated mobility measurements, the electronic scattering time of conduction electrons in graphene--a parameter that we used in the  Drude model calculations of the main text. In \cref{section: sphere exact cpa}, we derive closed-form expressions (\cref{EQUATION: TM SCATTERING FOR SPHERES} and \cref{EQUATION: TE SCATTERING FOR SPHERES}) for the CPA conductivity, $\sigma_{CPA}(k^+, l)$, of coated spheres (for arbitrary $k^+$ and $l$). Similarly, in \cref{section: cylinder exact cpa}, we find closed-form expressions (\cref{EQUATION: TM SCATTERING FOR CYLINDERS}, \cref{EQUATION: TE SCATTERING FOR CYLINDERS}) for $\sigma_{CPA}(k^+, m)$ for cylinders (again for arbitrary $k^+$ and $m$). 

In \cref{Section: Real sigma independence}, we prove the assertion that we made in the main text that the real part of $\sigma_{CPA}(k^+, l/m)$ is independent of the bulk permittivity, $\varepsilon$, as long as $\varepsilon$ is real. This result holds for both TM and TE channels and for both cylinders and spheres, and we show that it follows simply from energy conservation. 

We present alternative derivations (using a transfer matrix formalism specialized to the case of a thin film) of $\sigma_{CPA}(k^+, l/m)$ for spheres and cylinders in \cref{section: transfer matrix formalism for thin film spheres} and \cref{section: transfer matrix formalism for thin film cylinders}, respectively. 

In \cref{section: nonlocal cpa conductivity for spheres} and \cref{section: nonlocal cpa conductivity for cylinders}, we calculate generalizations of $\sigma_{CPA}(k^+, l/m)$ for spheres and cylinders, respectively, for cases in which the coatings must be described through spatially nonlocal response functions. The generalized expressions for $\sigma_{CPA}(k^+, l/m)$ are given by \cref{Equation: Nonlocal sigma cpa spheres} and \cref{Equation: nonlocal sigma cpa for cylinders} for spheres and cylinders, respectively. 

In \cref{section: quasistatic 1} and \cref{section: quasistatic 2}, we find the quasistatic approximation of $\sigma_{CPA}(k^+, l)$ for spheres (see \cref{Equation: Qausistatic conductivity for TM modes in spheres} for the final result; for cylinders, see \cref{section: quasistatic 3}). To give some idea of the range of $k^+$ for which the quasistatic $\sigma_{CPA}$ should be valid, we calculate explicit numerical values at which $\sigma''_{CPA}$ diverges for both spheres and cylinders in \cref{section: first divergence}. Such divergences are not captured in the quasistatic approximation, and, therefore, the first divergence indicates the onset of  deviations from the region of $k^+$ for which the quasistatic approximation is valid.  

In \cref{section: large k cpa spheres}, and \cref{section: large k cpa cylinders}
we investigate the opposite limit (i.e. $k^+\gg1$) and derive the key results \cref{Equation: Large k sphere TM}, \cref{Equation: Large k sphere TE}, \cref{Equation: Large k cylinder TM}, and \cref{Equation: Large k cylinder TE} which demonstrate the oscillatory behavior of $\sigma''_{CPA}(k^+)$ discussed in the main text. This is what we refer to as the ``Fabry-Perot" limit (here and in the main text). There is another ray-optics limit, in which the angular momentum number is very large ($m\gg1$ for cylinders and $l\gg 1$ for spheres) and $m/l\approx k^-$. This, the ``whispering-gallery" limit, is investigated in \cref{section: Whispering gallery modes}. 

In \cref{section: cpa bandwidth}, we calculate the CPA bandwidth for spherical systems (for the dipolar channel). Crucially, we show in \cref{section: cpa bandwidth} that the bandwidth is intimately related to the electronic scattering rate. In \cref{Section: complex frequencies}, we investigate the effect of being slightly off the CPA condition. In particular, we find CPA zeros in the complex plane, and we find the change in bulk permittivity, $\varepsilon$, required to bring CPA zeros back on the real line if $\sigma\neq \sigma_{CPA}$. 

In \cref{section: Dipole lattice model}, we generalize our results to a lattice of coupled coated cylinders, which sets the stage for the result, shown in the main text, that symmetric illumination of a lattice of coated dipole-coupled cylinders, can yield CPA.

In \cref{section: sphere polarizability from mie}, we calculate the retarded polarizability of a spherical scatterer in terms of the TM scattering coefficients. This result was used in the main text for the calculation of the CPA parameters necessary in the case of a spherical scatterer coupled to a PEC substrate. 

In \cref{section: Finite temperature effects}, we consider the effects of finite temperature, showing that, for a graphene coating, $E_{F, CPA}$ becomes a function of the temperature but $\omega_{CPA}$ does not.

In \cref{Section: Single scatterer plane wave cpa}, we revisit the case of a single scatterer. In particular, we show that although the spherical wave (cylindrical wave) CPA we considered at the outset of the paper may be considered to be single-channel critical coupling, the problem may equivalently be considered as perfect absorption of the flux incident from a linear combination of an infinite number of plane waves. 

In \cref{section: TE bandwidth} we briefly discuss the prospect (or lack thereof) of using graphene coatings for CPA of TE polarized light. In this case, CPA can only be achieved through strong interband optical response. Thus, unlike TM-polarized CPA, the bandwidth for TE-polarized CPA is not set by the electronic scattering time.

\textbf{Notational point}: Unless otherwise specified, the notation here is consistent with the main text. Since many of the quantities defined here and in the main text (e.g. $\omega_{CPA}, R_{CPA}$) depend on \textit{many} other quantities (e.g. $k^+, l/m$), we may keep some dependencies implicit (but only if the surrounding context makes clear all relevant dependencies). 

\section{Electronic scattering time of graphene in the DC limit as obtained from mobility measurements}
\label{Section: dc decay time}
In the main text, we used a constant value of $\tau=6.4\times 10^{-13}\text{s}$ for the Drude scattering time, $\tau$, of graphene coatings. In this section, we motivate this choice. 

In particular, we use the mobility measurements reported in \cite{novoselov2004electric} in order to find the DC scattering time in graphene. As stated in \cite{novoselov2004electric}, the mobility, $\mu$, and the DC conductivity, $\sigma\equiv \sigma(\omega\rightarrow 0)$, are related as follows: 
\begin{equation}
\sigma\equiv \lim_{\omega\rightarrow 0} \frac{ie^2E_F}{\pi\hbar^2(\omega+i/\tau)}=n e\mu \rightarrow \frac{e^2E_F\tau}{\pi\hbar^2}=ne\mu, 
\end{equation}
where we have used the Drude conductivity of graphene \cite{jablan2009plasmonics}, written in terms of the Fermi energy, $E_F$. We note that the electronic density, $n$, may be written in terms of the Fermi momentum, $k_F$, and thus also in terms of the Fermi energy. Explicitly, $n=g_sg_v\pi k_F^2/(4\pi^2)$, where $g_s=g_v=2$ are the spin and valley degeneracies, respectively, and $(4\pi^2)/\Omega$ corresponds to the volume in momentum space occupied by a single electronic state (where $\Omega$ denotes the total volume in real space). Using the linear dispersion of graphene, i.e. $E_F=\hbar v_F k_F$, we may write the electronic density in terms of the Fermi energy as follows: 
\begin{align*}
n=E_F^2/(\hbar^2 v_F^2\pi)\leftrightarrow E_F=\hbar v_F\sqrt{\pi n} \end{align*}
Therefore, writing the decay time in terms of the electronic density gives us: 
\begin{align*}\tau=\pi\hbar^2 n\mu/eE_F=\hbar \sqrt{\pi n}\mu/e v_F, \end{align*}
which is the result cited in \cite{jablan2009plasmonics}. We now use the values $n=3\times10^{13}/\text{cm}^2$ and $\mu=10^4 \text{ cm}^2/(\text{Volt}\times\text{s})$, $v_F=10^8 \text{cm}/\text{s}$, from which we obtain: 
\begin{align*}
&\tau=1.05\times 10^{-34}\text{Joules}\times\text{s}\sqrt{3\pi 10^{13}}\times10^4/(1.6\times 10^{-19}\text{Joules}\times 10^8)=\\&1.05\times 10^{-13}\text{Joules}\times\text{s}\sqrt{\pi 10}/(1.6\times\text{Joules})\approx 6.4\times 10^{-13} \text{s},
\end{align*}
as desired.

\section{Analytic forms for the generalized Mie coefficients (spheres)}
\label{section: sphere exact cpa}
In this section, we find the analytic form of the scattering coefficient, $s_l$, describing the amplitude/phase of an outgoing spherical wave when an incoming spherical wave is incident on a coated sphere. We note that the scattering of a plane wave/partial wave was already considered in \cite{christensen2015localized}, and our results will bear similarities to some of the equations reported in that paper. The only complication from the standard Mie scattering problem is the boundary condition set by the finite surface current density. This may be addressed by calculating the discontinuity in the normal component of the displacement field, $\mathbf{D}$, or by determining the discontinuity of the tangential component of the magnetic field, $\mathbf{H}$. We choose the latter method, giving us the following boundary condition (see, e.g. Jackson page 194 \cite{jackson1998classical} \footnote{We note that \cite{chen2011atomically} has a similar equation but is off by a minus sign (if the surface impedance is defined as $Z(\omega)=1/\sigma(\omega)$)}:
\begin{equation}
    e_\mathbf{r}\times(\mathbf{H}_>-\mathbf{H}_<)=\sigma(\omega)[\mathbf{E}-\mathbf{e}_r(\mathbf{E}\cdot\mathbf{e}_r)],
    \label{equation: conductivity boundary condition}
\end{equation}
where all fields are evaluated at the same angular coordinates, ($\theta, \phi$), and the subscripts denote the fields evaluated just outside ($>$) and just inside ($<$) of the sphere.  We note, for future reference in the derivation below, that the unit vectors in spherical coordinates have the following properties: $\mathbf{e}_r\times \mathbf{e}_\theta=\mathbf{e}_\phi$ and $\mathbf{e}_r\times \mathbf{e}_\phi=-\mathbf{e}_\theta$.

\subsection{$s_l$ for the TM channel}
We first consider TM polarized waves, for which the incoming magnetic field outside of the scatterer is given by $\mathbf{H}^{\text{incoming}}(\mathbf{r}, \omega)=h^{(2)}_l(r\omega/c)\mathbf{L}Y_{lm}(\theta, \phi)$ (note that the angular momentum operator, $\mathbf{L}$, does not operate on the radial variable), the internal field, which is required, of course, to be finite at the origin, is given by $\mathbf{H}^{\text{internal}}(\mathbf{r}, \omega)=a_lj_l(r\sqrt{\varepsilon}\omega/c)\mathbf{L}Y_{lm}(\theta, \phi)$, and the scattered field (outside of the scatterer) is given by $\mathbf{H}^{\text{outgoing}}(\mathbf{r}, \omega)=s_lh_l^{(1)}(r\omega/c)\mathbf{L}Y_{lm}(\theta, \phi)$. $a_l$ and $s_l$ are complex numbers, and, in the case of TM-polarized CPA, of angular momentum $l$, we require $s_l=0$. With these expressions for the magnetic field, the electric field may be readily calculated using Ampere's law:
\begin{align*}
\mathbf{E}(\mathbf{r}, \omega)=\frac{i}{\omega\varepsilon_0\varepsilon(\mathbf{r})}\nabla\times\mathbf{H}(\mathbf{r}, \omega)\rightarrow E_\phi(\mathbf{r}, \omega)=\frac{i}{\omega\varepsilon_0\varepsilon(\mathbf{r})r}\partial_r(rH_\theta(\mathbf{r}, \omega)), E_\theta(\mathbf{r}, \omega)=-\frac{i}{\omega\varepsilon_0\varepsilon(\mathbf{r})r}\partial_r(rH_\phi(\mathbf{r}, \omega)),
\end{align*}
where we used the fact that the magnetic field has no radial component (which also readily follows from the fact that $\mathbf{L}$ has no radial component). The electric field also has a radial component, but we need not solve for it for our derivation below. The relative permittivity, $\varepsilon(\mathbf{r})$, is a piecewise continuous function, given by:
\begin{equation}
\varepsilon(\mathbf{r})=\varepsilon + \Theta(|\mathbf{r}|-R)(1-\varepsilon),
\end{equation}
where $\Theta(x)$ is the Heaviside step function. Therefore, the boundary condition imposed by the finite surface current may be formulated as follows: 
\begin{align*}
h_l^{(2)}(k^+)+s_lh^{(1)}_l(k^+)-a_lj_l(k^-)=\frac{ia_l\sigma(\omega)}{\omega\varepsilon_0\varepsilon R}(k^-j_l(k^-))'=a_l(g(\omega)/\varepsilon) (k^-j_l(k^-))',
\end{align*}
where $g(\omega)$ is a dimensionless quantity defined in Equation 7 of \cite{christensen2015localized}. Additionally, we have to maintain continuity of the tangential component of the electric field across the boundary of the sphere, which gives us the following condition: 
\begin{align*}
a_l(k^-j_l(k^-))'=\varepsilon\Big[(k^+h^{(2)}_l(k^+))'+s_l(k^+h^{(1)}_l(k^+))'\Big]
\end{align*}
Therefore, putting both boundary conditions together allows us to eliminate the $a_l$ and gives us an equation in which the only unknown is the scattering coefficient:
\begin{align*}
(k^-j_l(k^-))'\frac{\Big[h_l^{(2)}(k^+)+s_l h_l^{(1)}(k^+) \Big]}{j_l(k^-)+(g(\omega)/\varepsilon)(k^-j_l(k^-))'}=\varepsilon\Big[(k^+h^{(2)}_l(k^+))'+s_l(k^+h^{(1)}_l(k^+))' \Big]
\end{align*}
By putting all terms proportional to the scattered field amplitude on one side of the equation, we readily obtain: 
\begin{align*}
&s_l\Big[(k^-j_l(k^-))'h_l^{(1)}(k^+)-\varepsilon j_l(k^-)(k^+h_l^{(1)}(k^+))'-g(\omega)(k^-j_l(k^-))'(k^+h_l^{(1)}(k^+))'\Big]= \\&\varepsilon j_l(k^-)(k^+h^{(2)}_l(k^+))'+g(\omega)(k^+h^{(2)}_l(k^+))'(k^-j_l(k^-))'-h^{(2)}_l(k^+)(k^-j_l(k^-))'
\end{align*}
We write this more explicitly as follows: 
\begin{equation}s_l= \frac{\varepsilon j_l(k^-)(k^+h^{(2)}_l(k^+))'+g(\omega)(k^+h^{(2)}_l(k^+))'(k^-j_l(k^-))'-h^{(2)}_l(k^+)(k^-j_l(k^-))'}{(k^-j_l(k^-))'h_l^{(1)}(k^+)-\varepsilon j_l(k^-)(k^+h_l^{(1)}(k^+))'-g(\omega)(k^-j_l(k^-))'(k^+h_l^{(1)}(k^+))'},
\label{EQUATION: TM SCATTERING FOR SPHERES}
\end{equation}
Note that this result is the same as Equation 6b of \cite{christensen2015localized} with the replacement $j_l(k^+)\rightarrow h_l^{(2)}(k^+)$. Crucially, we note that if we have no sphere (i.e. $g(\omega)=0$ and $\varepsilon=1$) then \cref{EQUATION: TM SCATTERING FOR SPHERES} naturally gives us $s_l=1$, as required. The scattering coefficient given by \cref{EQUATION: TM SCATTERING FOR SPHERES} is implemented in our code \cite{Coherent-Perfect-Absorption} by the function 
\texttt{e\_r\_cpa}. 

As in the main text, we denote by  $\sigma_{CPA}(k^+, l)$ the conductivity that makes the numerator of \cref{EQUATION: TM SCATTERING FOR SPHERES} vanish for a given choice of $k^+$ and $l$. In addition, since it will be useful below, we denote by $\sigma_{LSPR}(k^+, l)$ the conductivity at  which the denominator of \cref{EQUATION: TM SCATTERING FOR SPHERES} vanishes (corresponding to localized surface plasmon resonances). Note that from \cref{EQUATION: TM SCATTERING FOR SPHERES} we immediately have (suppressing $k^+, l$) $\Re(\sigma_{CPA})=-\Re(\sigma_{LSPR})$ and $\Im(\sigma_{CPA})=\Im(\sigma_{LSPR})$. This just means that a plasmonic laser will need precisely the same amount of gain as the perfect absorber needs loss. 

Lastly, we denote by $\sigma_{PS}(k^+, l)$ the passive sink conductivity, for which we have $s_l=-1$ for some $k^+, l$.  The passive sink conductivity follows readily from \cref{EQUATION: TM SCATTERING FOR SPHERES}:
\begin{equation}
\begin{split}
    &\varepsilon j_l(k^-)(k^+h_l^{(2)}(k^+))'-h_l^{(2)}(k^+)(k^-j_l(k^-))'+\frac{i\sigma_{PS}(k^+, l)}{k^+c\varepsilon_0}(k^-j_l(k^-))'(k^+h_l^{(2)}(k^+))' = \\ &\varepsilon j_l(k^-)(k^+h_l^{(1)}(k^+))'-h_l^{(1)}(k^+)(k^-j_l(k^-))'+\frac{i\sigma_{PS}(k^+, l)}{k^+c\varepsilon_0}(k^-j_l(k^-))'(k^+h_l^{(1)}(k^+))'
    \end{split}
\end{equation}
In this expression, all terms proportional to $j_l(k^+)$ or proportional to its derivatives cancel out, so we immediately have the following simple expression: 
\begin{equation}
    \sigma_{PS}(k^+, l)=ic\varepsilon_0 k^+\Bigg[\frac{\varepsilon j_l(k^-)}{(k^-j_l(k^-))'}- \frac{y_l(k^+)}{(k^+y_l(k^+)'} \Bigg],
\end{equation}
which is exactly Eq. (\equationpassivesink) \space of the main text. Note, of course, that since $\sigma_{PS}(k^+, l)$ corresponds to a scattering coefficient with \textit{unit modulus}, we naturally have $\Re \sigma_{PS}(k^+, l)=0$ for all $k^+$ and $l$ (for real permittivities, of course). 

Lastly, we note that since both the numerator and the denominator of \cref{EQUATION: TM SCATTERING FOR SPHERES} are linear in the conductivity, closed-form solutions of the conductivity for any desired value of $s_l$ may be found (this will be generically true for both polarizations and for both spheres and cylinders).  

\subsection{$s_l$ for the TE channel}
We now turn to TE polarized waves, which have no radial electric fields and thus are not associated with surface charge densities. The incoming electric field outside of the scatterer is given by $\mathbf{E}^{\text{incoming}}(\mathbf{r}, \omega)=h^{(2)}_l(r\omega/c)\mathbf{L}Y_{lm}(\theta, \phi)$, the internal field is given by $\mathbf{E}^{\text{internal}}(\mathbf{r}, \omega)=a_lj_l(r\sqrt{\varepsilon}\omega/c)\mathbf{L}Y_{lm}(\theta, \phi)$, and the outgoing field is given by $\mathbf{E}^{\text{outgoing}}(\mathbf{r}, \omega)=s_lh_l^{(1)}(r\omega/c)\mathbf{L}Y_{lm}(\theta, \phi)$. The matching conditions require $h^{(2)}_l(k^+)+s_lh^{(1)}_l(k^-)=a_lj_l(k^-)$. And we must also have: $e_\mathbf{r}\times(\nabla\times\mathbf{E}_>-\nabla\times\mathbf{E}_<)=i\omega\sigma(\omega)\mu_0\mathbf{E}$, which follows from Faraday's law. Writing the latter condition explicitly in terms of the spherical Bessel functions gives us:
\begin{equation}
-(k^+h^{(2)}_l(k^+))'-s_l(k^+h^{(1)}_l(k^+))'+a_l(k^-j_l(k^-))'=a_li\omega \sigma(\omega)\mu_0 R j_l(k^-)=a_lg(\omega)(k^+)^2j_l(k^-)
\end{equation}
Combining the two boundary conditions then gives us:
\begin{equation}
    j_l(k^-)(k^+h^{(2)}_l(k^+))'+s_lj_l(k^-)(k^+h^{(1)}_l(k^+))'=\Big[-g(\omega)(k^+)^2j_l(k^-)+(k^-j_l(k^-))'\Big]\Big[h^{(2)}_l(k^+)+s_lh_l^{(1)}(k^+) \Big]
\end{equation}
Finally, collecting terms gives us the TE scattering coefficient:
\begin{equation}
s_l=\frac{h^{(2)}_l(k^+)(k^-j_l(k^-))'-j_l(k^-)(k^+h^{(2)}_l(k^+))'-(k^+)^2g(\omega)j_l(k^-)h^{(2)}_l(k^+)}{j_l(k^-)(k^+h_l^{(1)}(k^+))'-h_l^{(1)}(k^+)(k^-j_l(k^-))'+(k^+)^2g(\omega)j_l(k^-)h_l^{(1)}(k^+)}
\label{EQUATION: TE SCATTERING FOR SPHERES}
\end{equation}
Note that this result is the same as Equation 6a of \cite{christensen2015localized} with the replacement $j_l(k^+)\rightarrow h_l^{(2)}(k^+)$. The scattering coefficient given by \cref{EQUATION: TE SCATTERING FOR SPHERES} is implemented in our code \cite{Coherent-Perfect-Absorption} by the function 
\texttt{h\_r\_cpa}. 

\section{Analytic forms for the generalized Mie coefficients (cylinders)}
\label{section: cylinder exact cpa}

In the following two subsections, we derive the Mie scattering coefficients, $s_m$, for coated cylinders for both polarizations. This is analogous to what we did in \cref{section: sphere exact cpa} but now for cylinders instead of spheres. The procedure is entirely the same as before (and arguably easier). We note, for future reference in the derivation below, that the unit vectors in cylindrical coordinates have the following properties: $\mathbf{e}_\mathbf{\rho}\times \mathbf{e}_\theta=\mathbf{e}_z$ and $\mathbf{e}_\mathbf{\rho}\times \mathbf{e}_z=-\mathbf{e}_\theta$. 

\subsection{$s_m$ for the TM channel}
We start with the transverse magnetic channel, for which we have $\mathbf{H}(\rho, z)=H_z(\rho, z)\mathbf{e}_z$, where the cylinder's axis is parallel to the $\mathbf{e}_z$ direction. Note that we are only considering fields that do not depend on the $z$ coordinate, which is why we may cleanly delineate our derivations by TE and TM polarizations(i.e. we do not have to worry about hybridization of the channels and cross-polarization coupling). In the TM case, the electric field has components along the $\mathbf{e}_\rho$ and $\mathbf{e}_\theta$ directions but no component along the cylinder's axis. We require continuity of the $\mathbf{e}_\theta$ component of the electric field as well as a change of the magnetic field across the cylinder's boundary given by 
\cref{equation: conductivity boundary condition}, which reads as follows in cylindrical coordinates:
\begin{align*}
H_{z, >}-H_{z, <}=-\sigma(\omega)E_\theta, H_{\theta, >}-H_{\theta, <}=\sigma(\omega)E_z \rightarrow 
H_m^{(2)}(k^+)+s_mH_m^{(1)}(k^+)-a_mJ_m(k^-)=-\sigma(\omega)E_\theta e^{-im\theta},
\end{align*}
where, in the second equality, we decomposed the magnetic field outside of the cylinder into an incoming wave (\textit{cylindrical} Hankel function of the second kind) with unit amplitude and an outgoing wave (cylindrical Hankel function of the first kind), with a corresponding scattering coefficient $s_m$. We also expressed the internal magnetic field as a Bessel function of the first kind, with coefficient, $a_m$. At any coordinate, we may express the electric field in terms of the magnetic field as follows:
\begin{align*}
\mathbf{E}=\frac{i}{\omega\varepsilon_0\varepsilon(\mathbf{r})}\Bigg[\frac{1}{\rho}\partial_\theta H_z \mathbf{e}_\rho-\partial_\rho H_z \mathbf{e}_\theta \Bigg]\end{align*}
Therefore, by using our expressions for the magnetic field, we may write the $\mathbf{e}_\theta$ component of the electric field as follows: 
\begin{align*}
E_\theta(\rho<R)=-\frac{ia_m}{c\varepsilon_0 \sqrt{\varepsilon}}J_m'(k^-), E_\theta(\rho>R)=-\frac{i}{c\varepsilon_0 }\Bigg[H_m^{(2)}(k^+)'+s_mH^{(1)}_m(k^+)' \Bigg]
\end{align*}
Therefore, our two boundary conditions are given as follows: 
\begin{align*}
a_m=\sqrt{\varepsilon}\Bigg[\frac{H_m^{(2)}(k^+)'+s_mH^{(1)}_m(k^+)'}{J_m'(k^-)}\Bigg],H_m^{(2)}(k^+)+s_mH_m^{(1)}(k^+)-a_mJ_m(k^-)=\frac{ia_m\sigma(\omega)}{c\varepsilon_0 \sqrt{\varepsilon}}J_m'(k^-) 
\end{align*}
Putting these boundary conditions together, we obtain:
\begin{equation}
H_m^{(2)}(k^+)+s_mH_m^{(1)}(k^+)=a_m\Bigg[J_m(k^-)+\frac{i\sigma(\omega)}{c\varepsilon_0\sqrt{\varepsilon}}J_m'(k^-)\Bigg]=\Bigg[\sqrt{\varepsilon}J_m(k^-)+\frac{i\sigma(\omega)}{c\varepsilon_0}J_m'(k^-)\Bigg]\Bigg[\frac{H_m^{(2)}(k^+)'+s_mH^{(1)}_m(k^+)'}{J_m'(k^-)}\Bigg]
\end{equation}
From which we obtain, for the scattering amplitude:
\begin{equation}
s_m=\frac{\sqrt{\varepsilon}J_m(k^-)H_m^{(2)}(k^+)'-H_m^{(2)}(k^+)J_m'(k^-)+\frac{i\sigma(\omega)}{c\varepsilon_0}J_m'(k^-)H_m^{(2)}(k^+)'}{H_m^{(1)}(k^+)J_m'(k^-)-\sqrt{\varepsilon}H^{(1)}_m(k^+)'J_m(k^-)-\frac{i\sigma(\omega)}{c\varepsilon_0}J_m'(k^-)H_m^{(1)}(k^+)'}, 
\label{EQUATION: TM SCATTERING FOR CYLINDERS}
\end{equation}
The scattering coefficient given by \cref{EQUATION: TM SCATTERING FOR CYLINDERS} is implemented in our code \cite{Coherent-Perfect-Absorption}
by the function \texttt{h\_z\_cpa}. 
\subsection{$s_m$ for the TE channel}
We now consider the TE channel, for which the electric (instead of magnetic) field is polarized along the cylinder's axis. As before, we require continuity of the electric field across the cylinder's boundary and a discontinuity of $H_\theta$ given by the conductivity of the coating. To start, we solve for the magnetic field: 
\begin{align*}\mathbf{H}=-\frac{i}{\omega\mu_0}\Bigg[\frac{1}{\rho}\partial_\theta E_z \mathbf{e}_\rho-\partial_\rho E_z \mathbf{e}_\theta \Bigg]\end{align*}
Therefore, using the notation of the previous subsection, we may write the boundary conditions explicitly as follows:
\begin{align*}
\frac{i}{c\mu_0}\Bigg[H_m^{(2)}(k^+)'+s_mH_m^{(1)}(k^+)'-a_m\sqrt{\varepsilon}J_m'(k^-) \Bigg]=a_m\sigma(\omega)J_m(k^-), H_m^{(2)}(k^+)+sH_m^{(1)}(k^+)=a_mJ_m(k^-)
\end{align*}
Putting these together, we obtain the following scattering coefficient for the TE channel: 
\begin{equation}
s_m = \frac{J_m(k^-)H_m^{(2)}(k^+)'-\sqrt{\varepsilon}J_m'(k^-)H_m^{(2)}(k^+)+\frac{i\sigma(\omega)}{c\varepsilon_0}J_m(k^+)H_m^{(2)}(k^+)}{\sqrt{\varepsilon}J_m'(k^-)H_m^{(1)}(k^+)-J_m(k^-)H_m^{(1)}(k^+)-\frac{i\sigma(\omega)}{c\varepsilon_0}J_m(k^-)H_m^{(1)}(k^+)}
\label{EQUATION: TE SCATTERING FOR CYLINDERS}
\end{equation}
The scattering coefficient given by \cref{EQUATION: TE SCATTERING FOR CYLINDERS} is implemented in our code \cite{Coherent-Perfect-Absorption}
by the function \texttt{e\_z\_cpa}. 

\section{Why is $\sigma'_{CPA}$ independent of $\varepsilon$ for lossless dielectric bulks?}
\label{Section: Real sigma independence}
In the main text, we made the assertion that if there is no loss in the bulk (i.e. if $\varepsilon\in \mathbb{R}$), then $\sigma'_{CPA}(k^+, m)$ for cylinders ($\sigma'_{CPA}(k^+, l)$ for spheres) will be independent of $\varepsilon$ (for all choices of $k^+$ and $l/m$). This neat property follows directly from the principle of conservation of energy. In the absence of outgoing power, the incoming power must be completely attenuated in the coating, immediately setting the value of $\sigma'_{CPA}$. In this section, we prove this result for cylinders (\cref{Subsection: energy conservation for cylinders}) and spheres (\cref{Subsection: energy conservation for spheres}), for both TM and TE polarizations. 

\subsection{$\Re(\sigma_{CPA}(k^+, m))$ from energy conservation for cylinders}
\label{Subsection: energy conservation for cylinders}
As a first example, we consider CPA in the TM channel by coated cylinders. Conservation of energy tells us that the total incoming radial flux must equal the power dissipated in the coating. Note that the Poynting vector will have both $\mathbf{e}_\theta$ and $\mathbf{e}_\rho$ components. However, only the component in the $\mathbf{e}_\rho$ component matters for this derivation.  Mathematically, conservation of energy tells us that the following equality must hold (we suppress the angular momentum dependence in the below until later when we specify the angular momentum):
\begin{equation}
    -\int_0^{2\pi} \Re\Big[\mathbf{E}(R, \theta)\times \mathbf{H}^*(R, \theta)\Big]\cdot\mathbf{e}_\rho R\text{d}\theta=\int_0^{2\pi}\int_0^\infty\sigma_{CPA}'(k^+)|\mathbf{E}(\rho,\theta)-\mathbf{e}_\rho (\mathbf{E}(\rho,\theta)\cdot\mathbf{e}_\rho)|^2\delta(\rho-R)\rho\text{d}\rho\text{d}\theta
    \label{Equation: Power conservation for real sigma cpa}
\end{equation}
The left side of \cref{Equation: Power conservation for real sigma cpa} is the power integrated (per unit length along the cylinder axis) over a cylindrical surface with the same radius as the scatterer, $R$. The right side is the Ohmic dissipation in the coating (of course, also per unit length). We are only concerned with CPA, so we write the magnetic field outside the cylinder as a purely incoming cylindrical wave, $\mathbf{H}(\rho, \theta)=H_m^{(2)}(\rho\omega/c)e^{im\theta}\mathbf{e}_z$. From this, we immediately get that \cref{Equation: Power conservation for real sigma cpa} may be written explicitly as follows (reintroducing, explicitly, the angular momentum dependence of the CPA conductivity):
\begin{equation}
   \frac{2\pi R}{c\varepsilon_0}\Re\Big[iH_m^{(1)}(k^+)H_m^{(2)}(k^+)'\Big] =\frac{2\pi R\sigma'_{CPA}(k^+, m)}{c^2\varepsilon^2_0}\Big[J'_m(k^+)^2+Y'_m(k^+)^2\Big],
\end{equation}
which, when simplified, gives us the following: 
\begin{equation}
    \sigma'_{CPA}(k^+, m)=c\varepsilon_0\Bigg[\frac{J_m(k^+)Y'_m(k^+)-Y_m(k^+)J'_m(k^+)}{J'_m(k^+)^2+Y'_m(k^+)^2} \Bigg]
\end{equation}
Note that this is precisely what one would obtain for $\sigma'_{CPA}(k^+, m)$ by setting the scattering coefficient given by \cref{EQUATION: TM SCATTERING FOR CYLINDERS} to zero. For small $k^+$, this may be approximated as follows: 
\begin{equation}
    \sigma'_{CPA}(k^+\ll 1, m)\approx \frac{c\varepsilon_0\Gamma(m)}{\pi\Gamma(m+1)}\frac{m(k^+)^m(k^+)^{-m-1}+m(k^+)^{m-1}(k^+)^{-m}}{[m^2\Gamma(m)^2/\pi^2](2^{2m})(k^+)^{-2m-2}}=\frac{c\varepsilon_0\pi(k^+)^{2m+1}}{\Gamma(m)\Gamma(m+1)m2^{2m-1}}
    \label{Equation: Real sigma small k TM cylinders}
\end{equation}
In the above, we neglected the $J_m'(k^+)^2$ term in the denominator since it is negligible compared to the $Y_m'(k^+)^2$ term. The inset of \cref{fig: supp figure 12}(b) shows the behavior of $\sigma'_{CPA}(k^+, m)$ compared to the approximation given by \cref{Equation: Real sigma small k TM cylinders}.

We now turn our attention to TE polarization, for which \cref{Equation: Power conservation for real sigma cpa} still applies but the field profile is given by $\mathbf{E}(\rho, \theta)=H^{(2)}(\rho\omega/c)e^{im\theta}\mathbf{e}_z$. In this case, \cref{Equation: Power conservation for real sigma cpa} gives us the following equality: 
\begin{equation}
\begin{split}
    &-\frac{1}{c\mu_0}\Re\Bigg[iH^{(2)}(k^+)H^{(1)}(k^+)' \Bigg]=\sigma'_{CPA}(k^+, m)\Bigg[J_m(k^+)^2+Y_m(k^+)^2 \Bigg]\rightarrow
    \\&\sigma'_{CPA}(k^+, m)=c\varepsilon_0\Bigg[\frac{J_m(k^+)Y'_m(k^+)-Y_m(k^+)J'_m(k^+)}{J_m(k^+)^2+Y_m(k^+)^2} \Bigg]
    \end{split}
\end{equation}
For small enough $k^+$, this may be approximated as follows: 
\begin{equation}
    \sigma'_{CPA}(k^+\ll 1, m)\approx \frac{c\varepsilon_0\Gamma(m)}{\pi\Gamma(m+1)}\frac{m(k^+)^m(k^+)^{-m-1}+m(k^+)^{m-1}(k^+)^{-m}}{[\Gamma(m)^2/\pi^2](2^{2m})(k^+)^{-2m}}=\frac{c\varepsilon_0m\pi(k^+)^{2m-1}}{\Gamma(m)\Gamma(m+1)2^{2m-1}}
\label{Equation: Real sigma small k TE cylinders}
\end{equation}
In this approximation, we neglected the contribution of $J_m(k^+)^2$ to the denominator (since it is small compared to the contribution from $Y_m(k^+)^2$). In the inset of \cref{fig: supp figure 13}(b) we show the behavior of $\sigma'_{CPA}(k^+, m)$ compared to the approximation given by \cref{Equation: Real sigma small k TE cylinders}.
\subsection{$\Re(\sigma_{CPA}(k^+, l))$ from energy conservation for spheres}
\label{Subsection: energy conservation for spheres}

The procedure in the previous subsection may be applied to the spherical case as well. For spheres, the analog of \cref{Equation: Power conservation for real sigma cpa} is given by the following (we suppress the angular momentum dependence in the below until later when we specify the angular momentum): 
\begin{equation}
    -\int_0^{2\pi}\int_0^\pi \Re\Big[\mathbf{E}(R, \theta)\times \mathbf{H}^*(R, \theta)\Big]\cdot\mathbf{e}_r R^2\sin(\theta)\text{d}\theta\text{d}\phi=\int_0^{2\pi}\int_0^\pi\sigma_{CPA}'(k^+)|\mathbf{E}(R,\theta)-\mathbf{e}_r(\mathbf{e}_r\cdot\mathbf{E}(R,\theta))|^2 R^2\sin(\theta )\text{d}\theta\text{d}\phi
    \label{Equation: Power conservation for real sigma cpa for spheres}
\end{equation}
The left side of \cref{Equation: Power conservation for real sigma cpa for spheres} is the power integrated over a spherical surface with the same radius as the scatterer. The right side is the Ohmic dissipation in the coating. We start with the TM modes, for which evaluating \cref{Equation: Power conservation for real sigma cpa for spheres} gives us the following equality:
\begin{equation}
\begin{split}
        &\frac{R}{\omega\varepsilon_0}\Re\Bigg[i(k^+h_l^{(2)}(k^+))h_l^{(1)}(k^+) \Bigg]\int_0^{2\pi}\int_0^\pi |\mathbf{L}Y_{lm}(\theta, \phi)|^2\sin(\theta)\text{d}\theta\text{d}\phi= \\&\frac{\sigma'_{CPA}(k^+)}{\omega^2\varepsilon_0^2}\Bigg[(k^+j_l(k^+))^2+ (k^+y_l(k^+))^2 \Bigg]\int_0^{2\pi}\int_0^\pi |\mathbf{L}Y_{lm}(\theta, \phi)|^2\sin(\theta)\text{d}\theta\text{d}\phi
        \end{split}
\end{equation}
From this, we obtain the following expression for $\sigma'_{CPA}(k^+, l)$:
\begin{equation}
    \sigma'_{CPA}(k^+, l)=c\varepsilon_0 k^+\Bigg[\frac{j_l(k^+)(k^+y_l(k^+))'-y_l(k^+)(k^+j_l(k^+))'}{(k^+j_l(k^+))'^2+(k^+y_l(k^+))'^2} \Bigg]
\end{equation}
This is exactly the real part of the expression for $\sigma_{CPA}(k^+, l)$ given by Equation \equationcpa \space in the main text. For small enough $k^+$, this expression is approximately given by:
\begin{equation}
    \sigma'_{CPA}(k^+\ll 1, l)\approx c\varepsilon_0 k^+\frac{(2l-1)!!}{(2l+1)!!}\Bigg[\frac{l(k^+)^l(k^+)^{-l-1}+(l+1)(k^+)^{-l-1}(k^+)^l}{[(2l-1)!!]^2l^2} \Bigg](k^+)^{2l+2}=\frac{c\varepsilon_0 (k^+)^{2l+2}}{[(2l-1)!!]^2l^2}
    \label{Equation: Subwavelength real sigma CPA for spheres}
\end{equation}

We close this section with $\sigma'_{CPA}(k^+, l)$ for  CPA in the TE channel for spheres. In this case, \cref{Equation: Power conservation for real sigma cpa for spheres} gives us the following equality: 
\begin{equation}
\begin{split}
       &-\frac{1}{R\omega\mu_0}\Re\Bigg[ih_l^{(2)}(k^+)(k^+h_l^{(1)}(k^+))' \Bigg]\int_0^{2\pi} \int_0^{\pi} |\mathbf{L}Y_{lm}(\theta, \phi)|^2\sin(\theta)\text{d}\theta\text{d}\phi= \\&\sigma'_{CPA}(k^+, l)\Big[j_l(k^+)^2+y_l(k^+)^2 \Big]\int_0^{2\pi}\int_0^\pi |\mathbf{L}Y_{lm}(\theta, \phi)|^2  \sin(\theta)\text{d}\theta\text{d}\phi
       \end{split}
\end{equation}
From this, we obtain the following for $\sigma'_{CPA}(k^+, l)$:
\begin{equation}
    \sigma'_{CPA}(k^+, l)=\frac{c\varepsilon_0}{k^+}\Bigg[\frac{j_l(k^+)(k^+y_l(k^+))'-y_l(k^+)(k^+j_l(k^+))'}{j_l(k^+)^2+y_l(k^+)^2} \Bigg]
\end{equation}
For small $k^+$, we may neglect the contribution of $(k^+j_l(k^+))^2$ in the denominator, allowing us to write $\sigma'_{CPA}(k^+, l)$ in approximate form as follows: 
\begin{equation}
    \sigma'_{CPA}(k^+\ll 1, l)\approx \frac{c\varepsilon_0}{k^+}\frac{(2l-1)!!}{(2l+1)!!}\Bigg[\frac{l(k^+)^l(k^+)^{-l-1}+(l+1)(k^+)^{-l-1}(k^+)^l}{(2l-1)!!(2l-1)!!} \Bigg](k^+)^{2l+2}=\frac{c\varepsilon_0 (k^+)^{2l}}{(2l-1)!!(2l-1)!!}
    \label{Equation: Real sigma small k TE spheres}
\end{equation}
The inset of \cref{fig: supp figure 16}(b) shows the exact behavior of $\sigma'_{CPA}(k^+, l)$ for $0.05<k^+<0.35$ compared to the approximation given by \cref{Equation: Real sigma small k TE spheres}. 

\section{Quasistatic limit of the fully retarded CPA conductivity for spheres}
\label{section: quasistatic 1}

We use expansions for the spherical Bessel functions from Equations 5.1 and 5.2 of the book by Bohren and Huffman \cite{bohren2008absorption}. Each term in the denominator of the TM scattering coefficient (\cref{EQUATION: TM SCATTERING FOR SPHERES}) involves a product of $j_l(k^-)$ or $h_l^{(1)}(k^+)$ (or their derivatives) (similarly for the numerator, with $h_l^{(1)}(k^+)\rightarrow h_l^{(2)}(k^+)$). Since we are concerned with the small $k^+, k^-$ limit, we need only consider the $y_l(k^+)$ term in the Hankel functions. Both the numerator and the denominator of the TM scattering coefficients read (considering only the most divergent components and neglecting overall constants of proportionality since we wish to see where the denominator/numerator go to zero): 
\begin{equation}
    y_l(k^+)[k^-j_l(k^-)]'-\varepsilon j_l(k^-)[k^+y_l(k^+)]'-g(\omega)[k^-j_l(k^-)]'[k^+y_l(k^+)]'
\end{equation}
Our approximations for the various spherical Bessel functions are given by: 
\begin{equation}
j_l(x)\approx\frac{x^l}{(2l+1)!!}, [xj_l(x)]'\approx\frac{(l+1)x^l}{(2l+1)!!}, y_l(x)\approx-\frac{(2l-1)!!}{x^{l+1}}, [xy_l(x)]'\approx\frac{l(2l-1)!!}{x^{l+1}}
\label{Equation: spherical bessel function approximations}
\end{equation}
Therefore, we may write the numerator/denominator of the TM scattering coefficient as (up to a constant of proportionality): 
\begin{equation}
\begin{split}
    &(l+1)(k^-)^l(k^+)^{-l+1}+l(k^-)^{l+2}(k^+)^{-l-1}+g(\omega)(k^+)^2l(l+1)(k^-)^l(k^+)^{-l-1}=\\&(k^-)^l(k^+)^{-l-1}\Big[(l+1)(k^+)^2+l(k^-)^2+g(\omega)(k^+)^2l(l+1) \Big]
    \end{split}
\end{equation}
From which we obtain the following for $\sigma_{CPA}''(k^+, l)$ in the quasistatic limit:
\begin{equation}(l+1)+l\varepsilon+g(\omega)l(l+1)=0\rightarrow \frac{\sigma_{CPA}''(k^+, l)}{c\varepsilon_0}\approx k^+\Bigg[\frac{(l+1)+l\varepsilon}{l(l+1)}\Bigg]
\label{Equation: Qausistatic conductivity for TM modes in spheres}
\end{equation}
This is the quasistatic approximation we plotted in Fig. (\figurecpaconductivity) of the main text. As we showed there, \cref{Equation: Qausistatic conductivity for TM modes in spheres} agrees quite well with its fully retarded counterpart in the subwavelength regime. 

\subsection{Analogous result for the TE channel}
\begin{figure}[htb!]
\centering
\includegraphics[scale=1]{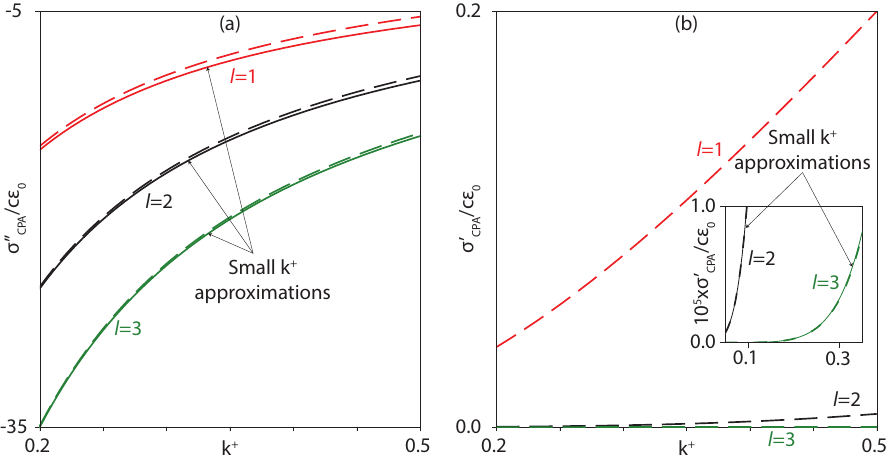}
  \caption{\textbf{$\sigma_{CPA}(k^+, l)$ for the TE channel of a coated sphere for small (subwavelength) $k^+$:} (a) $\sigma''_{CPA}(k^+, l)$ for the TE channel for $l=1, 2, 3$. Solid lines are the small $k^+$ approximations to the imaginary part of the CPA conductivity (which scale inversely with $k^+$) given by \cref{Equation: small k te im sigma for spheres}. (b) $\sigma'_{CPA}(k^+, l)$ for the TE channel for $l=1, 2, 3$. Inset shows a zoomed in version of $\sigma'_{CPA}(k^+, l)$ for $l=2$ and $l=3$ (dashed lines) with the small-$k^+$ approximations given by \cref{Equation: Real sigma small k TE spheres}. The bulk permittivity is taken to be $\varepsilon=2.1$. }
  \label{fig: supp figure 16}
\end{figure}

In this case, the most divergent parts of the numerator/denominator of the TE scattering coefficient, \cref{EQUATION: TE SCATTERING FOR SPHERES},  must be zero at CPA, giving us the following condition: 
\begin{align*}
j_l(k^-)[k^+y_l(k^+)]'-y_l(k^+)[k^-j_l(k^-)]'+\frac{ik^+\sigma_{CPA}(k^+, l)}{c\varepsilon_0}j_l(k^-)y_l(k^+)=0
\end{align*}
Using the approximations given by \cref{Equation: spherical bessel function approximations}, we obtain the following condition:
\begin{align*}
l(k^-)^l(k^+)^{-l-1}+(l+1)(k^+)^{-l-1}(k^-)^l-\frac{ik^+\sigma_{CPA}(k^+, l)}{c\varepsilon_0}(k^-)^l(k^+)^{-l-1}=0
\end{align*}
Therefore, the transverse electric CPA conductivity will be given by, at small $k^+$,
\begin{equation}
 \sigma_{CPA}(k^+, l)\approx-\frac{i(2l+1)c\varepsilon_0}{k^+} 
 \label{Equation: small k te im sigma for spheres}
\end{equation}
This result is plotted against its exact counterpart in \cref{fig: supp figure 16}(a), showing good agreement in the subwavelength regime. Furthermore, we draw the reader's attention to the neat fact that \cref{Equation: small k te im sigma for spheres} does not depend on the permittivity, $\varepsilon$. This means that the CPA conductivity in the TE channel (both real and imaginary parts) has no dependence on the bulk permittivity for small enough scatterers. This is in stark contrast to the CPA conductivity in the TM channel, for which only $\sigma'_{CPA}$ is independent of permittivity (for all size regimes). 

\section{Quasistatic result for the CPA conductivity from scalar electromagnetism (spheres)}
\label{section: quasistatic 2}
In this section, we derive the result of \cref{section: quasistatic 1} through scalar electromagnetism. Surface plasmon resonances are typically calculated from the less complicated scalar theory, and we have already established (see \cref{section: sphere exact cpa}) that $\sigma''_{CPA}(k^+, l)=\sigma_{LSPR}''(k^+, l)$. Thus, we may obtain an approximation for the imaginary part of the CPA conductivity (in the plasmonic, $k^+\ll 1$, regime) from simple scalar electromagnetism. We take the electric field to be the gradient of a scalar potential, $\mathbf{E}(\mathbf{r}, \omega)=-\nabla \phi(\mathbf{r}, \omega)$. Since we require $\nabla^2\phi(\mathbf{r})=0$ everywhere (except at the boundary of the sphere), we know that we may take the following expansion for the scalar potential: 
\begin{align*}
\phi(r, \theta, \phi)=\sum_l \Bigg[\frac{a_l}{r^{l+1}}+b_l r^l\Bigg]P_l(\cos(\theta)),
\end{align*}
where we have taken the variation in the azimuthal direction to be zero (since the modes of the system and thus the polarizabilities are degenerate with respect to the azimuthal quantum number). In the \textit{absence of a coating}, our two boundary conditions on the scalar potential and the requirement that it may not diverge at the origin, necessarily gives us the following two stipulations (with $<, >$ indicating quantities inside and outside the sphere, respectively): 
\begin{align*}b_{l, <}R^l=\frac{a_{l,>}}{R^{l+1}}+b_{l, >}R^l \text{ and } \varepsilon lb_{l, <}R^{l-1}= -\frac{(l+1)a_{l, >}}{R^{l+2}}+ lb_{l, >}R^{l-1},\end{align*}
where the first boundary condition corresponds to the continuity of the tangential component of the electric field (equivalently, continuity of the potential), and the second boundary condition corresponds to the continuity of the normal component of the displacement field. Without loss of generality, we set $b_{l, >}$ to unity (corresponding to an overall rescaling of all fields). Our goal is to find $a_{l, >}$:
\begin{align*}
\varepsilon l\Bigg[\frac{a_{l, >}}{R^{2l+1}}+1\Bigg]= -\frac{(l+1)a_{l, >}}{R^{2l+1}}+ l\rightarrow a_{l, >}\Bigg[\varepsilon l+(l+1)\Bigg]=R^{2l+1}l\Big[1-\varepsilon \Big]\rightarrow a_{l, >}=R^{2l+1}\frac{l(1-\varepsilon)}{(l+1)+\varepsilon l}
\end{align*}
The polarizability is defined as the ratio of the induced multipole moment to the applied field \cite{fuchs1987multipolar}. The moment is given by $4\pi \varepsilon_0 a_{l, >}$ and the applied field is given by $-b_l$ \footnote{We note that the particular prefactors here do not really matter, as the reader may verify by reading through to the end of the derivation.}. Therefore, we get that the multipolar polarizability, $\alpha_l$, is given by the following expression: 
\begin{equation}
\alpha_l = 4\pi\varepsilon_0R^{2l+1}\frac{l(\varepsilon-1)}{(l+1)+\varepsilon l}
\label{Equation: Polarizability of sphere without coating}
\end{equation}
Since the plasmon resonances correspond to the poles of the polarizability (as plasmon resonances exist in the absence of an external field), we see that $\varepsilon=-(l+1)/l$ for a plasmon. For the dipolar, $l=1$, mode, this gives us the familiar condition that $\varepsilon=-2$. Now we consider the effect of non-vanishing surface conductivity, i.e. a coating. In this case, the displacement field is not continuous across the boundary. This is because the finite surface conductivity $\sigma(\omega)$ yields a non-zero charge density given implicitly by the continuity equation:
\begin{align*}
-i\omega\rho(\mathbf{r}, \omega)+\nabla\cdot \mathbf{J}(\mathbf{r}, \omega)=0\rightarrow \rho(\mathbf{r}, \omega)=-\frac{i}{\omega}\nabla\cdot\mathbf{J}(\mathbf{r}, \omega), 
\end{align*}
where the current density is given by: 
\begin{align*}
\mathbf{J}(r, \theta, \omega)=\sigma(\omega)E_\theta(r, \theta, \omega)\delta(r-a) \mathbf{e}_\theta=-\sigma(\omega)\Big[\mathbf{e}_\theta\cdot\nabla\phi(r, \theta, \omega)\Big]\delta(r-a)\mathbf{e}_\theta
\end{align*}
Note that the divergence in spherical coordinates may be written as: 
\begin{align*}
-\frac{i}{\omega}\nabla\cdot\mathbf{J}(\mathbf{r}, \omega)=\frac{ib_{l, <}R^l\sigma(\omega)}{\omega R^2\sin(\theta)}\partial_\theta (\sin(\theta)\partial_\theta P_l(\cos(\theta))) \delta(r-a)\end{align*}
Furthermore, as suggested by their name,  Legendre polynomials satisfy Legendre's equation, meaning:
\begin{align*}\Big[(1-x^2)P'_l(x) \Big]'=-l(l+1)P_l(x) \rightarrow \frac{1}{\sin(\theta)}\partial_\theta\Big[\sin(\theta)\partial_\theta P_l(\cos(\theta)) \Big]=-l(l+1)P_l(\cos(\theta)), 
\end{align*}
where we used $\partial_x=-(\sin(\theta))^{-1}\partial_\theta$ for $x=\cos(\theta)$.
This gives us the following expression for the induced charge density (per unit area):
\begin{align*}\rho(\mathbf{r}, \omega)=-\frac{i\sigma(\omega)l(l+1)R^lb_{l, <}}{\omega R^2}P_l(\cos(\theta)) \end{align*}
Therefore, our generalized boundary conditions (in the presence of a coating) are given by: 
\begin{equation}
\begin{split}
&b_{l, <}R^l=\frac{a_{l,>}}{R^{l+1}}+R^l \text{ (continuity of } E_\theta \text{)}\\ &\Bigg[\frac{(l+1)a_{l, >}}{R^{l+2}}-lR^{l-1}\Bigg]+\varepsilon lb_{l, <}R^{l-1}=-\frac{i\sigma(\omega)l(l+1)}{\varepsilon_0\omega R^2}R^lb_{l, <}\text{ (discontinuity in } D_r \text{)} 
\end{split}
\end{equation}
We combine these two boundary conditions to give us:
\begin{align*}
&-\frac{i\sigma(\omega)l(l+1)}{\varepsilon_0\omega R^2}\Bigg[ \frac{a_{l,>}}{R^{l+1}}+R^l\Bigg]=\Bigg[\frac{(l+1)a_{l, >}}{R^{l+2}}-lR^{l-1}\Bigg]+\varepsilon l\Bigg[\frac{a_{l, >}}{R^{l+2}}+R^{l-1} \Bigg]\rightarrow \\ &a_{l, >}\Big[-l\varepsilon-(l+1)- g(\omega)l(l+1)\Big]=R^{2l+1}\Big[l\varepsilon-l+g(\omega)l(l+1) \Big]
\end{align*}
Thus, we get the non-retarded polarizability in the presence of a coating (generalization of \cref{Equation: Polarizability of sphere without coating}; see also Equation 9 of \cite{christensen2015localized}): 
\begin{equation}
\alpha_l=\frac{4\pi\varepsilon_0 l\Big[\varepsilon-1+(l+1)g(\omega) \Big]}{l\varepsilon+(l+1)+g(\omega)l(l+1)} 
\label{Equation: Polarizability of sphere with coating}
\end{equation}
Note that the poles of \cref{Equation: Polarizability of sphere without coating} are entirely consistent with the approximate form for $\sigma''_{CPA}(k^+, l)$ we derived in \cref{Equation: Qausistatic conductivity for TM modes in spheres}.

\begin{figure}[htb!]
\centering
\includegraphics[scale=1]{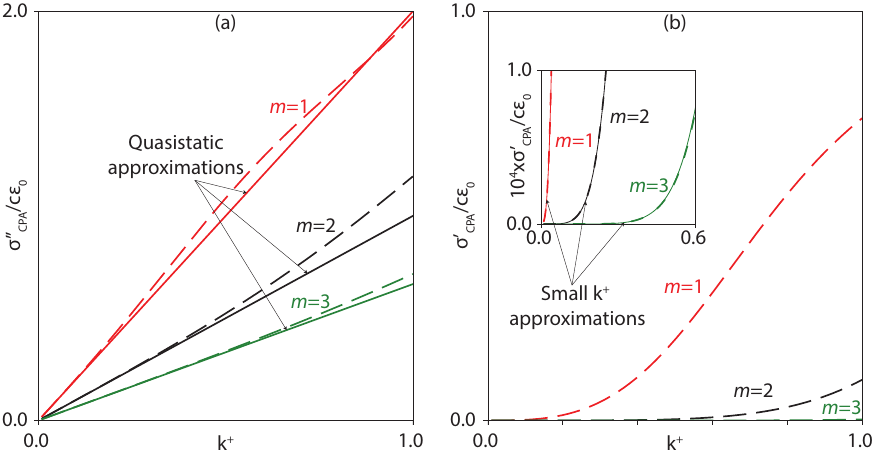}
  \caption{\textbf{$\sigma_{CPA}(k^+, m)$ for the TM channel of a coated cylinder for small (subwavelength) $k^+$:} (a) Imaginary part of CPA conductivity, $\sigma''_{CPA}(k^+, m)$, in units of $c\varepsilon_0$ for different values of angular momentum number, $m$. Solid lines correspond to the quasistatic (linear in $k^+$) approximation given by \cref{Equation: TM cylinder CPA for small k}. (b) Real part of CPA conductivity, $\sigma'_{CPA}(k^+, m)$. Inset shows the small $k^+$ behavior for $m=1, 2, 3$. The small $k^+$ approximation is given by \cref{Equation: Real sigma small k TM cylinders}. The bulk permittivity is set to $\varepsilon=1$. }
  \label{fig: supp figure 12}
\end{figure}

\begin{figure}[htb!]
\centering
\includegraphics[scale=1]{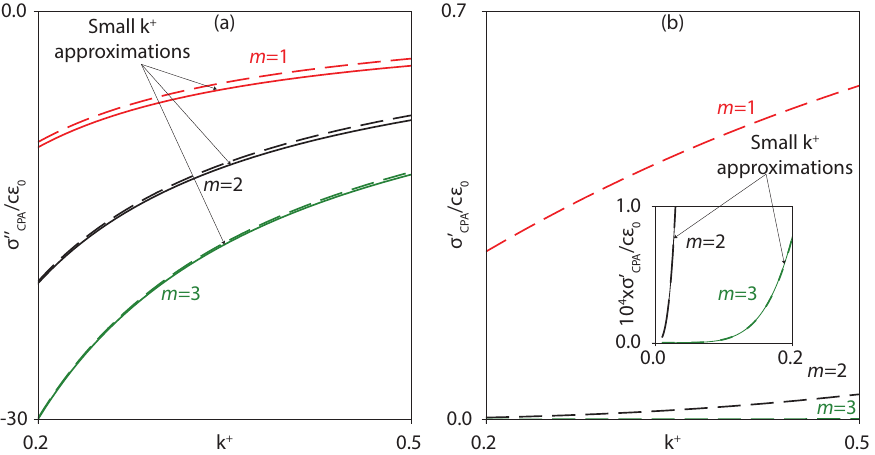}
  \caption{\textbf{$\sigma_{CPA}(k^+, m)$ for the TE channel of a coated cylinder for small (subwavelength) $k^+$:} (a) Imaginary part of CPA conductivity, $\sigma''_{CPA}(k^+, m)$, in units of $c\varepsilon_0$ for different values of the angular momentum number, $m$. Solid lines correspond to the small $k^+$ approximation (which scales inversely with $k^+$) given by \cref{Equation: TE cylinder CPA for small k}. (b) Real part of CPA conductivity, $\sigma'_{CPA}$. The bulk permittivity is set to $\varepsilon=1$. Inset shows the exact small $k^+$ behavior of $\sigma'_{CPA}(k^+)$ compared with the approximation given by \cref{Equation: Real sigma small k TE cylinders}. }
  \label{fig: supp figure 13}
\end{figure}

\section{Quasistatic result for the CPA conductivity for cylinders}
\label{section: quasistatic 3}
\begin{figure}[htb!]
\centering
\includegraphics[scale=1]{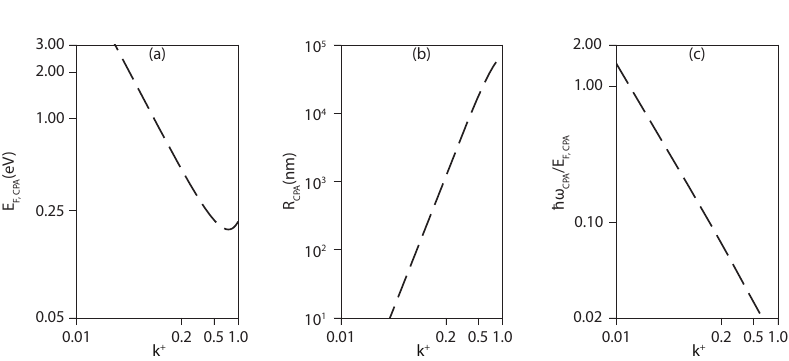}
  \caption{\textbf{$E_{F, CPA}$, $R_{CPA}$, and $\omega_{CPA}$ for CPA in the $m=1$ (dipolar) TM channel by a coated cylinder:} All subplots correspond to a coated cylinder with bulk permittivity $\varepsilon=1$, in accordance with the values for $\sigma_{CPA}$ shown in \cref{fig: supp figure 12}.  (a) $E_{F, CPA}$ as a function of $k^+$. (b) $R_{CPA}$ as a function of $k^+$. (c) $\hbar\omega_{CPA}$ normalized to $E_{F, CPA}$ as a function of $k^+$. Note that, for $k^+$ more than a few tenths of an eV, $\hbar\omega_{CPA}$ is much smaller than $E_{F, CPA}$, meaning we can safely neglect the contribution of interband electronic transitions to the conductivity.}
  \label{fig: supp figure 15}
\end{figure}

\begin{figure}[htb!]
\centering
\includegraphics[scale=1]{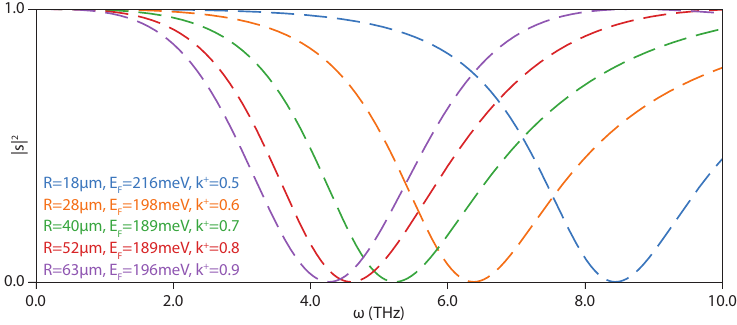}
  \caption{\textbf{Bandwidth of TM polarized, $m=1$, CPA for graphene-coated cylinders:} For five values of $k^+$, we fix the coating's Fermi energy to $E_{F, CPA}$ as calculated from Equation \equationefomega \space of the main text and $\sigma_{CPA}(k^+)$ from \cref{fig: supp figure 12}. We then sweep the frequency to show the behavior of $|s|^2$ around the CPA frequency, $\omega_{CPA}$. We show, in color-coded text, the values of $k^+$, $R_{CPA}$ and $E_{F, CPA}$ corresponding to each of the plotted lines. }
  \label{fig: supp figure 14}
\end{figure}

Just as we approximated (see \cref{section: quasistatic 1} and \cref{section: quasistatic 2}) the imaginary part of the CPA conductivity for spheres (in the subwavelength regime) by calculating $\sigma''_{LSPR}(k^+, l)$, the conductivity associated with localized surface plasmon resonances \footnote{Note that we only did this for the TM modes, which are associated with non-zero surface charge densities.}, in the quasistatic regime, we repeat, in this section, the same procedure for cylinders. Instead of calculating the multipolar polarizability and then finding its poles, we take a shortcut and require, from the outset, that there be no externally applied field (giving us the plasmonic conductivity directly). Laplace's equation for the scalar potential in a cylindrical system is given by the following: 
\begin{equation}
\frac{1}{\rho}\partial_\rho(\rho \partial_\rho\Psi(\rho, \theta, z))+\frac{1}{\rho^2}\partial_\theta^2\Psi(\rho, \theta, z)+\partial_z^2\Psi(\rho, \theta, z)=0
\end{equation}
We take the following ansatz for the scalar potential: $\Psi(\mathbf{r})=f(\rho)e^{im\theta}e^{ikz}$, with $m$ an integer. Furthermore, we only consider solutions with no variation along the cylinder axis; i.e. we set $k=0$. This then gives us the following differential equation for $f(\rho)$: 
\begin{equation}
\rho\partial_\rho(\rho\partial_\rho f(\rho))=m^2 f(\rho)\rightarrow \Psi(\rho, \theta, z)=\Bigg[\frac{a_m}{\rho^m}+b_m\rho^m \Bigg]e^{im\theta}
\end{equation}
As we require the potential to be regular everywhere in space, we take, for the solution inside the cylinder ($\rho<R$), $f(\rho<R)=b_m\rho^m$. Outside the cylinder, we take the solution to be $f(\rho>R)=a_m\rho^{-m}$. We require the potential to match at the boundary (equivalently, we require the tangential component of the electric field to be continuous across the boundary), which gives us the following relation between the internal and external fields: 
\begin{align*}
a_mR^{-m}=b_m R^m\rightarrow a_m=b_mR^{2m}
\end{align*}
\textit{If we have no surface current}, the second boundary condition will be given by continuity of the normal component of the displacement field: 
\begin{equation}
-a_m a^{-m-1}=\varepsilon b_ma^{m-1}\rightarrow \varepsilon=-1,
\label{Equation: Quasistatic TM Cylinder no coating}
\end{equation}
which is the familiar surface plasmon resonance condition for a bulk cylinder. Note that unlike the surface plasmon condition for a sphere, \cref{Equation: Quasistatic TM Cylinder no coating} is independent of the angular momentum number. The same result may be obtained by considering the zeros of the  denominator of the  retarded scattering function (\cref{EQUATION: TM SCATTERING FOR CYLINDERS} with $\sigma(\omega)=0$),
\begin{align*}
J'_m(k^-)H_m^{(1)}(k^+)-\sqrt{\varepsilon} J_m(k^-)H_m^{(1)}(k^+)'=0
\end{align*}
In the quasistatic limit, only the Bessel function of the second kind contributes in the Hankel function (since the Bessel function of the second kind will diverge at the origin). Furthermore, $J_m\approx z^m$ and $Y_m\approx z^{-m}$ (constants of proportionality do not matter since we have zero on the right hand side of the equation and each term on the left hand side has exactly one $J_m$ and one $Y_m$), which means that we have: 
\begin{equation}
(k^-)^{m-1}(k^+)^{-m}+\sqrt{\varepsilon} (k^-)^{m}(k^+)^{-m-1}=0\rightarrow \frac{1}{\sqrt{\varepsilon}}+\sqrt{\varepsilon}=0\rightarrow1+\varepsilon=0,
\end{equation}
which is precisely the quasistatic relation given by \cref{Equation: Quasistatic TM Cylinder no coating}. We now consider the situation \textit{with a finite surface current}. In this case, the continuity of the potential (equivalently, the continuity of the tangential part of the electric field) still gives us: 
\begin{equation}
a_mR^{-m}=b_m R^m\rightarrow a_m=b_mR^{2m}
\end{equation}
If we have a finite surface current, the second boundary condition must, however, be modified and expressed as follows:
\begin{equation}
a_m R^{-m-1}+\varepsilon b_mR^{m-1}=\rho_s/m\varepsilon_0\rightarrow (1+\varepsilon)b_m R^{m-1}=\rho_s/m\varepsilon_0,
\end{equation}
where $\rho_s e^{im\theta}$ is the surface charge density defined by $\tilde{\rho}(\rho, \theta, z)=\rho_s e^{im\theta}\delta(\rho-R)$, and $\tilde{\rho}$ is the volume charge density, that is determined from the conductivity, $\sigma(\omega)$, via the continuity equation (briefly reintroducing the frequency dependence for clarity): 
\begin{equation}
\tilde{\rho}(\mathbf{r}, \omega)=-\frac{i}{\omega}\nabla\cdot\mathbf{J}(\mathbf{r}, \omega)=\frac{i}{\omega R^2}\partial_\theta(\sigma(\omega)\partial_\theta(b_mR^me^{im\theta}\delta(\rho-R)))=\frac{i\sigma(\omega)}{\omega R^2}\partial^2_\theta \Big(b_m R^{m}e^{im\theta}\Big)\delta(\rho-R)
\end{equation}
Therefore, we get the following dispersion relation: 
\begin{equation}
\Big[1+\varepsilon\Big]R^{m-1}=-\frac{im^2\sigma(\omega)}{m\omega R^2\varepsilon_0}R^m\rightarrow \varepsilon+1=-\frac{im\sigma(\omega)}{\omega R\varepsilon_0}, 
\end{equation}
from which we obtain the following approximation for $\sigma_{CPA}(k^+, m)$ for the transverse magnetic channel of a cylinder (in the subwavelength regime):
\begin{equation}
    \sigma_{CPA}''(k^+, m) =\sigma''_{LSPR}(k^+, m)\approx\frac{c\varepsilon_0 k^+(\varepsilon+1)}{m}
    \label{Equation: TM cylinder CPA for small k}
\end{equation}
In \cref{fig: supp figure 12}, we compare the quasistatic result given by \cref{Equation: TM cylinder CPA for small k} to the exact CPA conductivity. Using the values of $\sigma_{CPA}(k^+)$, we may find the required $E_{F, CPA}, R_{CPA}$ and $\omega_{CPA}$ for subwavelength coherent perfect absorption for graphene-coated cylinders. In \cref{fig: supp figure 14}, we show the bandwidth of CPA for five values of $k^+$ ranging from $k^+=0.5$ to $k^+=0.9$ (see \cref{fig: supp figure 15} to see how the CPA parameters scale with $k^+$). The dispersive conductivity of graphene was implemented via the function \texttt{graphene\_conductivity} in our code \cite{Coherent-Perfect-Absorption}.
Of course, we may obtain the same result from the zeros of the denominator \footnote{Or the zeros of the numerator, since we are only concerned with the imaginary part of the CPA conductivity} of the scattering amplitude, \cref{EQUATION: TM SCATTERING FOR CYLINDERS}.
\begin{equation}
J_m'(k^-)H_m^{(1)}(k^+)-\sqrt{\varepsilon} J_m(k^-)H_m^{(1)}(k^+)'-\frac{i\sigma(\omega)}{c\varepsilon_0}J_m'(k^-)H_m^{(1)}(k^+)'=0
\end{equation}
By again only considering the contribution of the Bessel function of the second kind in the Hankel function terms, we obtain the following:
\begin{equation}
m(k^-)^{m-1}(k^+)^{-m}+m\sqrt{\varepsilon}(k^-)^{m}(k^+)^{-m-1}+\frac{im^2\sigma(\omega)}{c\varepsilon_0}(k^-)^{m-1}(k^+)^{-m-1}=0
\end{equation}
Which we readily simplify to give us:
\begin{equation}
(k^-)^{-1}+\sqrt{\varepsilon}(k^+)^{-1}+\frac{im\sigma(\omega)}{c\varepsilon_0}(k^-)^{-1}(k^+)^{-1}=0\rightarrow 1+\varepsilon+\frac{im\sigma(\omega)}{\varepsilon_0\omega R}=0,
\end{equation}
which is the non-retarded result derived before. Thus, as in the case for spheres, both scalar electromagnetism and the small $k^+$ expansion of the fully retarded scattering amplitudes give us the same result for $\sigma_{CPA}(k^+, m)$. 

\subsection{Analogous result for the TE channel}
For the TE channel, we cannot use results from scalar electromagnetism, but we can still examine the zeros of the numerator/denominator of the scattering coefficient in the small $k^+, k^-$ limit. This gives us the  following condition:
\begin{equation} 
J_m(k^-)H_m^{(1)}(k^+)'-\sqrt{\varepsilon}J_m'(k^-)H^{(1)}_m(k^+)+\frac{i\sigma_{CPA}(k^+, m)}{c\varepsilon_0}J_m(k^-)H_m^{(1)}(k^+)=0 
\end{equation}
In the $k^+\ll1$ limit, we may Taylor expand, as we did in the previous subsection, to give the simpler expression:
\begin{equation}
-n(k^-)^m(k^+)^{-m-1}-n\sqrt{\varepsilon}(k^-)^{m-1}(k^+)^{-m}+\frac{i\sigma_{CPA}(k^+, m)}{c\varepsilon_0}(k^-)^m(k^+)^{-m}=0
\end{equation}
From this, we obtain the following approximation for the imaginary part of the CPA conductivity for TE modes at small $k^+$: 
\begin{equation}
\sigma_{CPA}(k^+, m)\approx-\frac{2imc\varepsilon_0}{k^+}
\label{Equation: TE cylinder CPA for small k}
\end{equation}
Note that unlike \cref{Equation: TM cylinder CPA for small k}, $\sigma_{CPA}(k^+, m)$ for the TE modes does not, at lowest order, depend on the permittivity of the bulk. In \cref{fig: supp figure 13}(a), we compare the approximate CPA conductivity, given by \cref{Equation: TE cylinder CPA for small k} with the exact CPA conductivity, showing good agreement for small $k^+$, as expected. 
\FloatBarrier
\newpage
\section{Deriving \cref{EQUATION: TM SCATTERING FOR SPHERES} and \cref{EQUATION: TE SCATTERING FOR SPHERES} from a transfer matrix formalism}
\label{section: transfer matrix formalism for thin film spheres}
\begin{figure}[htb!]
\centering
\includegraphics[scale=1]{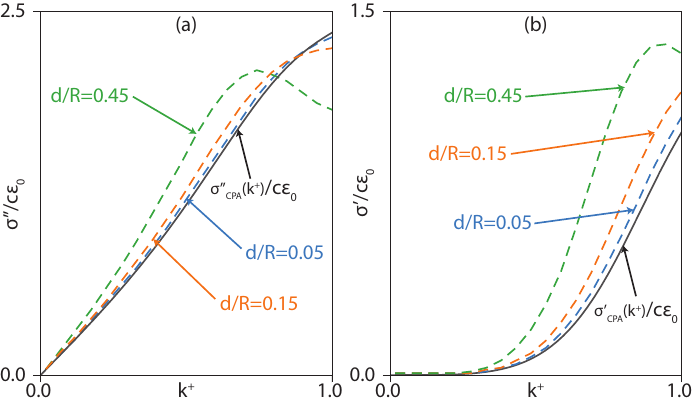}
  \caption{\textbf{ $\sigma_{CPA}(k^+)\equiv \sigma_{CPA}(k^+, l=1)$ for the TM channel as a thin film limit in spheres:} (a) For three values of thin film thickness, we find the conductivity corresponding to perfect absorption by modeling the thin film according to a thickness dependent permittivity (see discussion in \cref{Subsection: Transfer matrix TM spheres}). The permittivity of the bulk is taken to be $\varepsilon=2.1$. Solid black lines corresponds to the CPA conductivity (for an infinitely thin coating) for the dipolar TM channel of a sphere. (a) ((b)) corresponds to the imaginary (real) parts of the conductivities.}
  \label{fig: supp figure 10}
\end{figure}

\begin{figure}[htb!]
\centering
\includegraphics[scale=1]{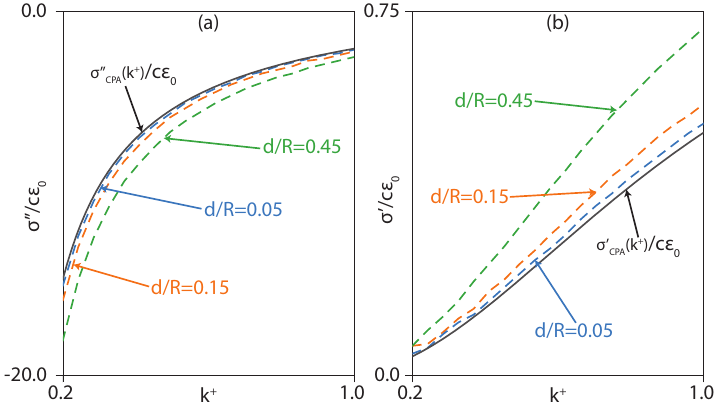}
  \caption{\textbf{ $\sigma_{CPA}(k^+)\equiv \sigma_{CPA}(k^+, l=1)$ for the TE channel as a thin film limit in spheres:} For three values of thin film thickness, we find the conductivity corresponding to perfect absorption by modeling the thin film according to a thickness dependent permittivity (see discussion in \cref{Subsection: Transfer matrix TM spheres}). The permittivity of the bulk is taken to be $\varepsilon=2.1$. Solid black lines corresponds to the CPA conductivity (for an infinitely thin coating) for the dipolar TM channel of a sphere. (a) ((b)) corresponds to the imaginary (real) parts of the conductivities.}
  \label{fig: supp figure 11}
\end{figure}

As stated in the main text, we may obtain the scattering coefficients corresponding to a coated sphere by considering the coating to be an infinitely thin film with a permittivity that depends on the thickness of the film (corresponding to a volume polarization density that depends on the film's thickness).
\subsection{Scattering coefficient for the TM ($H_r=0$) channel (spheres)}
\label{Subsection: Transfer matrix TM spheres}
 Although we are ultimately interested in the three-layer case, we set up the general transfer matrix formalism for completeness. We are concerned with the scattering of a system composed of concentric shells (layers), each with its own permittivity. We index the layers from $n=0, 1, 2..., N$ (where each layer covers the region $r_{n}<r<r_{n+1}$ (with $r_0\equiv 0$)), and the $n$'th layer has a corresponding refractive index, $n_n$. for TM modes, we have for the magnetic field in each region: 
\begin{align}
    \mathbf{H}_n(r, \theta, \phi)=[a_nh_l^{(1)}(n_nkr)+b_nh_l^{(2)}(n_nkr)]\mathbf{L}Y_{lm}(\theta, \phi),
\end{align}
where $k\equiv \omega/c$. We have the following boundary condition (continuity of the tangential part of the magnetic field since there is no surface current):
\begin{align*}
a_nh_l^{(1)}(n_nkr_{n+1})+b_nh_l^{(2)}(n_nkr_{n+1})=a_{n+1}h_l^{(1)}(n_{n+1}kr_{n+1})+b_{n+1}h_l^{(2)}(n_{n+1}kr_{n+1})
\end{align*}
We define $x_n\equiv n_n kr_n$ and $y_n\equiv n_n kr_{n+1}$ in order to rewrite this boundary condition as follows: 
\begin{equation}
a_nh_l^{(1)}(y_n)+b_nh_l^{(2)}(y_n)=a_{n+1}h_l^{(1)}(x_{n+1})+b_{n+1}h_l^{(2)}(x_{n+1})
\label{Equation: BC 1 sphere transfer}
\end{equation}
Similarly, our second boundary condition (continuity of the tangential component of the electric field) gives us the following: 
\begin{equation}
\varepsilon_{n+1}\Big[a_n(y_nh_l^{(1)}(y_n))'+b_n(y_nh_l^{(2)}(y_n))'\Big]=\varepsilon_{n}\Big[a_{n+1}(x_{n+1}h_l^{(1)}(x_{n+1}))'+b_{n+1}(x_{n+1}h_l^{(2)}(x_{n+1}))'\Big]
\label{Equation: BC 2 sphere transfer}
\end{equation}
We may combine the boundary conditions given by \cref{Equation: BC 1 sphere transfer} and \cref{Equation: BC 2 sphere transfer} to form a $2\times 2$ transfer matrix given by the following:
\begin{align*}
\begin{bmatrix}h_l^{(1)}(y_n) && h_l^{(2)}(y_n)\\ \varepsilon_{n+1}(y_nh_l^{(1)}(y_n))'  && \varepsilon_{n+1}(y_n h_l^{(2)}(y_n))' \end{bmatrix}\begin{bmatrix}a_n \\ b_n \end{bmatrix}=\begin{bmatrix}h_l^{(1)}(x_{n+1}) && h_l^{(2)}(x_{n+1})\\ \varepsilon_n(x_{n+1}h_l^{(1)}(x_{n+1}))'  && \varepsilon_n(x_{n+1} h_l^{(2)}(x_{n+1}))' \end{bmatrix}\begin{bmatrix}a_{n+1} \\ b_{n+1} \end{bmatrix}
\end{align*}
By inverting this matrix relation, we obtain the following relation between the amplitudes in the $n$'th and $n+1$'th layers: 
\begin{align*}
\begin{bmatrix}a_{n+1} \\ b_{n+1} \end{bmatrix} =\frac{1}{\varepsilon_n\Big[h_l^{(1)}(x_{n+1})(x_{n+1}h_l^{(2)}(x_{n+1}))'-h_l^{(2)}(x_{n+1})(x_{n+1}h_l^{(1)}(x_{n+1}))' \Big]}\times\begin{bmatrix} \varepsilon_n(x_{n+1} h_l^{(2)}(x_{n+1}))'&& -h_l^{(2)}(x_{n+1})\\ -\varepsilon_n(x_{n+1}h_l^{(1)}(x_{n+1}))'  &&  h_l^{(1)}(x_{n+1})\end{bmatrix} \times\\\begin{bmatrix}h_l^{(1)}(y_n) && h_l^{(2)}(y_n)\\ \varepsilon_{n+1}(y_nh_l^{(1)}(y_n))'  && \varepsilon_{n+1}(y_n h_l^{(2)}(y_n))' \end{bmatrix}\begin{bmatrix}a_n \\ b_n 
\end{bmatrix}
\end{align*}
Note that since we are ultimately concerned with the scattering coefficient, we only care about the ratio of $a_N/b_N$ (the ratio of outgoing to incoming wave amplitudes in the environment in which the entire scatterer is placed), so we may rid ourselves of the annoying denominator, giving:  
\begin{equation}
\begin{bmatrix}a_{n+1} \\ b_{n+1} \end{bmatrix}\propto\begin{bmatrix} \varepsilon_n(x_{n+1} h_l^{(2)}(x_{n+1}))'&& -h_l^{(2)}(x_{n+1})\\ -\varepsilon_n(x_{n+1}h_l^{(1)}(x_{n+1}))'  &&  h_l^{(1)}(x_{n+1})\end{bmatrix}\begin{bmatrix}h_l^{(1)}(y_n) && h_l^{(2)}(y_n)\\ \varepsilon_{n+1}(y_nh_l^{(1)}(y_n))'  && \varepsilon_{n+1}(y_n h_l^{(2)}(y_n))' \end{bmatrix}\begin{bmatrix}a_n \\ b_n
\end{bmatrix}
\label{Equation: Transfer matrix spheres}
\end{equation}
With the transfer matrix formalism, given by \cref{Equation: Transfer matrix spheres}, now set up, we consider an $N=2$ layer case, with a two dimensional coating, where the inner core has dielectric constant $\varepsilon_0$, the environment has a dielectric constant $\varepsilon_2$ and the two dimensional coating has a thickness dependent dielectric constant $\varepsilon_1=1+i\sigma(\omega)/(d\omega\tilde{\varepsilon}_0)$ (and we take the limit $d\rightarrow 0$ at the end of the calculation). Note that we introduce the somewhat unconventional notation $\tilde{\varepsilon}_0$ as the vacuum permittivity. We do this to avoid confusion with the previously introduced relative permittivity of the bulk, $\varepsilon_0$. 

With the definitions of $x_i$ and $y_i$ from the preceding sections, we have, therefore: $y_0=\sqrt{\varepsilon_0}(\omega/c)R$, $x_1=\sqrt{\varepsilon_1}(\omega/c)R$, $y_1=\sqrt{\varepsilon_1}(\omega/c)(R+d)$ and $x_2=\sqrt{\varepsilon_2}(\omega/c)(R+d)$, where we have introduced the bulk sphere radius, $R$. With these definitions in mind, our expression for the amplitudes in region two (the environment) is given by: 
\begin{align*}
\small\begin{bmatrix}a_2\\ b_2\end{bmatrix}=\begin{bmatrix} \varepsilon_1(x_{2} h_l^{(2)}(x_{2}))'&& -h_l^{(2)}(x_{2})\\ -\varepsilon_1(x_{2}h_l^{(1)}(x_{2}))'  &&  h_l^{(1)}(x_{2})\end{bmatrix}\times\begin{bmatrix}h_l^{(1)}(y_1) && h_l^{(2)}(y_1)\\ \varepsilon_{2}(y_1h_l^{(1)}(y_1))'  && \varepsilon_{2}(y_1 h_l^{(2)}(y_1))' \end{bmatrix}\times \begin{bmatrix} \varepsilon_0(x_1 h_l^{(2)}(x_1))'&& -h_l^{(2)}(x_1)\\ -\varepsilon_0(x_1h_l^{(1)}(x_1))'  &&  h_l^{(1)}(x_1)\end{bmatrix}\begin{bmatrix} j_l(y_0)\\\varepsilon_{1}(y_0 j_l(y_0))' \end{bmatrix}
\end{align*}
We multiply the penultimate matrix by the last matrix to give us: 
\begin{align*}\small\begin{bmatrix}a_2\\ b_2\end{bmatrix}=\begin{bmatrix} \varepsilon_1(x_{2} h_l^{(2)}(x_{2}))'&& -h_l^{(2)}(x_{2})\\ -\varepsilon_1(x_{2}h_l^{(1)}(x_{2}))'  &&  h_l^{(1)}(x_{2})\end{bmatrix}\times\begin{bmatrix}h_l^{(1)}(y_1) && h_l^{(2)}(y_1)\\ \varepsilon_{2}(y_1h_l^{(1)}(y_1))'  && \varepsilon_{2}(y_1 h_l^{(2)}(y_1))' \end{bmatrix}\times \begin{bmatrix} \varepsilon_0(x_1 h_l^{(2)}(x_1))'j_l(y_0) -\varepsilon_{1}h_l^{(2)}(x_1)(y_0 j_l(y_0))' \\ -\varepsilon_0(x_1h_l^{(1)}(x_1))'j_l(y_0)  +  \varepsilon_{1}h_l^{(1)}(x_1)(y_0 j_l(y_0))' \end{bmatrix},\end{align*}

Note that the last element of this expression is a $2\times 1$ vector not a $2\times 2$ matrix. We next multiply the first two matrices to give us: 
\begin{align*}
\begin{bmatrix}a_2\\ b_2\end{bmatrix}=\begin{bmatrix} \varepsilon_1 (x_2h_l^{(2)}(x_2))'h_l^{(1)}(y_1)-\varepsilon_2 (y_1h_l^{(1)}(y_1))'h_l^{(2)}(x_2) && \varepsilon_1(x_2h_l^{(2)}(x_2))'h_l^{(2)}(y_1)-\varepsilon_2 h_l^{(2)}(x_2)(y_1h_l^{(2)}(y_1))'\\\varepsilon_2(y_1h_l^{(1)}(y_1))'h_l^{(1)}(x_2)-\varepsilon_1(x_2h_l^{(1)}(x_2))' h_l^{(1)}(y_1) && \varepsilon_2 (y_1 h_l^{(2)}(y_1))'h_l^{(1)}(x_2)-\varepsilon_1 h_l^{(2)}(y_1)(x_2h_l^{(1)}(x_2))'\end{bmatrix}\\\times\begin{bmatrix} \varepsilon_0(x_1 h_l^{(2)}(x_1))'j_l(y_0) -\varepsilon_{1}h_l^{(2)}(x_1)(y_0 j_l(y_0))' \\ -\varepsilon_0(x_1h_l^{(1)}(x_1))'j_l(y_0)  +  \varepsilon_{1}h_l^{(1)}(x_1)(y_0 j_l(y_0))' \end{bmatrix}
\end{align*}
We now expand each of $a_2$ and $b_2$ into terms proportional to $\varepsilon_1^2, \varepsilon_1\varepsilon_{0, 2}$ and $\varepsilon_2\varepsilon_0$. We denote these by superscripts, i.e. $a_2=a_{2}^{11}+a_2^{10}+a_2^{12}+a_2^{02}$. Let's first restrict ourselves to $a_2$ and consider the terms proportional to $\varepsilon_2\varepsilon_0$:
\begin{equation}
\begin{split}
&a_2^{02}=\varepsilon_0\varepsilon_2j_l(y_0)h_l^{(2)}(x_2)\Big[(x_1h_l^{(1)}(x_1))'(y_1h_l^{(2)}(y_1))'-(x_1h_l^{(2)}(x_1))'(y_1h_l^{(1)}(y_1))' \Big]\approx
\\&\varepsilon_0\varepsilon_2 j_l(y_0) h_l^{(2)}(x_2)\Big[e^{i(x_1-y_1)}-e^{-i(x_1-y_1)} \Big]\approx -2i\varepsilon_0\varepsilon_2 j_l(y_0)h_l^{(2)}(x_2)\sqrt{\varepsilon_1}(\omega/c)d\rightarrow 0,
\end{split}
\label{Equation: atwozero}
\end{equation}
where we used the fact that $|x_1|, |y_1|\gg 1$ (in the limit $d\rightarrow 0$) in order to use the asymptotic expansions of the spherical Hankel functions (given by \cref{Equation: Asymptotic hankel and spherical bessel}). Note that although $x_1$ and $y_1$ individually diverge in the zero thickness limit, their difference, $y_1-x_1=\sqrt{\varepsilon_1}(d\omega/c)\rightarrow 0$ as $d\rightarrow 0$, which is why we were justified in Taylor expanding the exponentials in \cref{Equation: atwozero} . Furthermore, we note that, $a_{2}^{02}$ as given by \cref{Equation: atwozero} scales as $d^{1/2}$ as $d\rightarrow 0$. We now move on to $a_{2}^{11}$, which reads as follows: 
\begin{equation}
\begin{split}
&a_{2}^{11}=\varepsilon_1^2(y_0j_l(y_0))'(x_2h_l^{(2)}(x_2))'[h_l^{(1)}(x_1)h_l^{(2)}(y_1)-h_l^{(2)}(x_1)h_l^{(1)}(y_1)] \approx
\\&\varepsilon_1^2(y_0j_l(y_0))'(x_2h_l^{(2)}(x_2))'\frac{1}{x_1y_1}\Bigg[e^{i(x_1-y_1)}-e^{-i(x_1-y_1)} \Bigg]=
-i\varepsilon_1(y_0j_l(y_0))'(x_2h_l^{(2)}(x_2))'\frac{2\sqrt{\varepsilon_1}(\omega/c)d}{(\omega/c)^2[R^2+Rd]}
\\&\approx -\frac{2i\sqrt{\varepsilon_1}}{(\omega/c)R}(y_0j_l(y_0))'(x_2 h_l^{(2)}(x_2))'(i\sigma(\omega)/(R\omega\tilde{\varepsilon}_0))
\end{split}
\label{Equation: aoneone}
\end{equation}
Note that $a_2^{11}$, as given by \cref{Equation: aoneone}, scales as $d^{-1/2}$ as $d\rightarrow 0$. This means that we are definitely justified in ignoring the contribution of $a_{2}^{02}$ in the thin film limit. Now we investigate the remaining terms, i.e. $a_{2}^{01}$ and $a_{2}^{12}$:
\begin{align*}
a_{2}^{10}=\varepsilon_0\varepsilon_1 j_l(y_0)(x_2h_l^{(2)}(x_2))'\Big[h^{(1)}_l(y_1)(x_1h_l^{(2)}(x_1))'-h_l^{2}(y_1)(x_1h_l^{(1)}(x_1)' \Big] \approx-2i\varepsilon_0\sqrt{\varepsilon_1} j_l(y_0)(x_2h_l^{(2)}(x_2))'\frac{1}{(\omega/c)R} \\a_2^{12}=\varepsilon_2\varepsilon_1 h_l^{(2)}(x_2)(y_0j_l(y_0))'[h_l^{(2)}(x_1)(y_1 h_l^{(1)}(y_1))'-h_l^{(1)}(x_1)(y_1h_l^{(2)}(y_1))']  \approx  2i\varepsilon_2\sqrt{\varepsilon_1}h_l^{(2)}(x_2)(y_0j_l(y_0))'\frac{1}{(\omega/c)R}
\end{align*}
Note that both of these quantities also scale as $d^{-1/2}$. Therefore, we may combine all three terms that actually contribute in the thin film limit to give us the following: 
\begin{align*}
a_2^{11}+a_{2}^{10}+a_{2}^{12}\approx\frac{2i\sqrt{\varepsilon_1}}{R\omega/c}\Bigg[-\varepsilon_0 j_l(y_0)(x_2h_l^{(2)}(x_1))'-(i\sigma(\omega)/\omega\tilde{\varepsilon}_0)(y_0 j_l(y_0))'(x_2h_l^{(2)}(x_2))'+\varepsilon_2 h_l^{(2)}(x_2)(y_0j_l(y_0))'\Bigg]
\end{align*}
Note that we do not have to independently calculate $b_{2}$. We can simply note that the matrix elements involved in $b_2$ are the same as those for $a_2$ with the simple replacement $h_l^{(2)}(x_2)\rightarrow -h_l^{(1)}(x_2)$. Therefore, we get:
\begin{equation}
\frac{a_2}{b_2}=\frac{-\varepsilon_0 j_l(y_0)(x_2h_l^{(2)}(x_1))'-(i\sigma(\omega)/R\omega\tilde{\varepsilon}_0)(y_0 j_l(y_0))'(x_2h_l^{(2)}(x_2))'+\varepsilon_2 h_l^{(2)}(x_2)(y_0j_l(y_0))'}{\varepsilon_0 j_l(y_0)(x_2h_l^{(1)}(x_1))'+(i\sigma(\omega)/R\omega\tilde{\varepsilon}_0)(y_0 j_l(y_0))'(x_2h_l^{(1)}(x_2))'-\varepsilon_2 h_l^{(1)}(x_2)(y_0j_l(y_0))'},
\label{Equation: slfromtransfer}
\end{equation}
which is exactly \cref{EQUATION: TM SCATTERING FOR SPHERES} (well, \cref{Equation: slfromtransfer} is more general than \cref{EQUATION: TM SCATTERING FOR SPHERES} since it accounts for the possibility of an environment with a relative permittivity not equal to unity). We show in \cref{fig: supp figure 10} how the CPA conductivity for a thin film asymptotically approaches $\sigma_{CPA}(k^+, l)$ \footnote{$\sigma_{CPA}(k^+, l)$ being the TM CPA conductivity for coated spheres in this case.} as the thickness of the film approaches zero. 

\subsection{Scattering coefficient for the TE ($E_r=0$) channel (spheres)}
Since the derivation is almost identical to the one presented in \cref{Subsection: Transfer matrix TM spheres}, we omit the derivation of \cref{EQUATION: TE SCATTERING FOR SPHERES} as the thin film limit of a transfer matrix calculation. However, we include numerical evidence that the CPA conductivity of a very thin film approaches $\sigma_{CPA}(k^+, l)$ \footnote{$\sigma_{CPA}(k^+, l)$ being the TE CPA conductivity for coated spheres in this case.} for the TE channel as well (see \cref{fig: supp figure 11}). 

\section{Deriving equation \cref{EQUATION: TM SCATTERING FOR CYLINDERS} and \cref{EQUATION: TE SCATTERING FOR CYLINDERS} from a transfer matrix formalism}
\label{section: transfer matrix formalism for thin film cylinders}
\begin{figure}[h]
\centering
\includegraphics[scale=1]{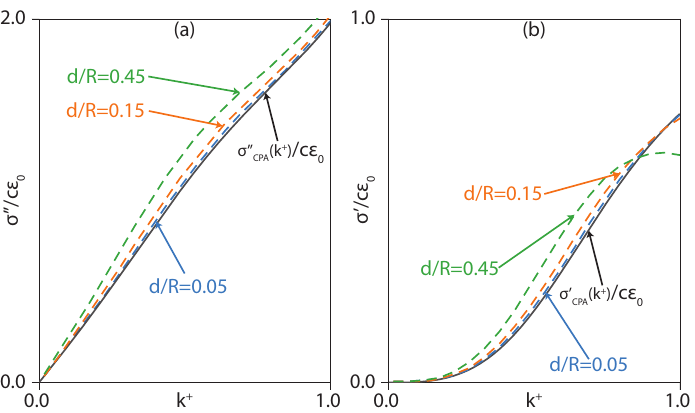}
  \caption{\textbf{ $\sigma_{CPA}(k^+)\equiv \sigma_{CPA}(k^+, m=1)$ for the TM channel as a thin film limit in cylinders:} For three values of thin film thickness, we find the conductivity corresponding to perfect absorption by modeling the thin film according to a thickness dependent permittivity (see \cref{Equation: Thickness dependent permittivity}). The permittivity of the bulk is taken to be that of vacuum, i.e. $\varepsilon=1$. Solid black line shows $\sigma_{CPA}(k^+, m=1)$ (the CPA conductivity for an infinitely thin coating) for the TM channel. (a) ((b)) corresponds to imaginary (real) parts of the conductivities.}
  \label{fig: supp figure 8}
\end{figure}
As we did for spheres in the previous section, we show that the CPA conductivity for cylinders may be found through a transfer matrix formalism for a thin film. We again take the layers to be indexed by $n=0, 1, 2,...,N$, with each layer corresponding to the region $r_n<r<r_{n+1}$ and we define $x_n\equiv n_nkr_n$, $y_n\equiv n_nkr_{n+1}$, with $k\equiv \omega/c$. We consider the TM channel, taking $\mathbf{H}_{n}(r, \theta)=e^{im\theta}\Bigg[a_nH^{(1)}_{m}(n_nkr)+b_nH_m^{(2)}(n_nkr) \Bigg]\mathbf{e}_z$. As before, we require $a_0=b_0$ so that the field does not diverge at the origin (i.e. so we have no nonzero coefficient of the Bessel function of the second kind at the origin). In addition, our boundary conditions give us the following relations: 
\begin{equation}
\begin{split}
    &a_{n}H_m^{(1)}(y_n)+b_n H_m^{(2)}(y_n)=a_{n+1}H_m^{(1)}(x_{n+1})+b_{n+1} H_m^{(2)}(x_{n+1}) 
\\& n_{n+1}\Bigg[a_{n}H_m^{(1)}(y_n)'+b_n H_m^{(2)}(y_n)'\Bigg]=n_n\Bigg[a_{n+1}H_m^{(1)}(x_{n+1})'+b_{n+1} H_m^{(2)}(x_{n+1})'\Bigg] 
\end{split}
\label{Equation: BC1 and BC2 Transfer matrix cylinders}
\end{equation}
As in the case of the sphere, we combine the boundary conditions given by \cref{Equation: BC1 and BC2 Transfer matrix cylinders}, giving us the following $2\times 2$ transfer matrix:
\begin{align*}
\begin{bmatrix}H_m^{(1)}(y_n) && H_m^{(2)}(y_n)\\ n_{n+1}H_m^{(1)}(y_n)'  && n_{n+1} H_m^{(2)}(y_n)'\end{bmatrix}\begin{bmatrix}a_n\\ b_n \end{bmatrix}=\begin{bmatrix}H_m^{(1)}(x_{n+1}) && H_m^{(2)}(x_{n+1})\\ n_{n}H_m^{(1)}(x_{n+1})'  && n_{n} H_m^{(2)}(x_{n+1})'\end{bmatrix}\begin{bmatrix}a_{n+1}\\ b_{n+1} \end{bmatrix}
\end{align*}
We invert this relationship (note that, as before, the determinant is irrelevant if we only care about the ratio of the fields). Therefore, we obtain the following: 
\begin{equation}
\begin{bmatrix}a_{n+1}\\ b_{n+1} \end{bmatrix} \propto\begin{bmatrix} n_n H_m^{(2)}(x_{n+1})' && -H_m^{(2)}(x_{n+1})\\ -n_nH_m^{(1)}(x_{n+1})' && H_m^{(1)}(x_{n+1})\end{bmatrix}\begin{bmatrix}H_m^{(1)}(y_n) && H_m^{(2)}(y_n)\\ n_{n+1}H_m^{(1)}(y_n)'  && n_{n+1} H_m^{(2)}(y_n)'\end{bmatrix}\begin{bmatrix}a_{n}\\ b_{n} \end{bmatrix}
\label{Equation: Transfer matrix cylinder}
\end{equation}
\cref{Equation: Transfer matrix cylinder} is a general transfer matrix, which may be applied recursively to obtain the scattering properties of a system composed of an arbitrary number of concentric cylindrical shells. We are, of course, only concerned with the three-layer case, in which the normalized radial coordinates and the thickness-dependent permittivity of the middle layer are given by:
\begin{equation}
n_1^2=1+\frac{i\sigma(\omega)}{\tilde{\varepsilon}_0\omega d}, y_0=n_0(\omega/c)R, x_1=n_1(\omega/c)R, y_1=n_1(\omega/c)(R+d), x_2=n_2(\omega/c)(R+d),
\label{Equation: Thickness dependent permittivity}
\end{equation}
where we are concerned with the $d\rightarrow 0$ limit (with $\sigma(\omega)$ remaining fixed). Solving for $a_2$ and $b_2$ then gives us the following:
\begin{align*}
\begin{bmatrix}a_2 \\ b_2 \end{bmatrix}=\begin{bmatrix} n_1 H_m^{(2)}(x_{2})' && -H_m^{(2)}(x_{2})\\ -n_1H_m^{(1)}(x_{2})' && H_m^{(1)}(x_{2})\end{bmatrix}\begin{bmatrix}H_m^{(1)}(y_1) && H_m^{(2)}(y_1)\\ n_{2}H_m^{(1)}(y_1)'  && n_{2} H_m^{(2)}(y_1)'\end{bmatrix}\begin{bmatrix} n_0H_m^{(2)}(x_1)'J_m(y_0)-n_1J_m'(y_0)H_m^{(2)}(x_1)\\ n_1J_m'(y_0)H_m^{(1)}(x_1)-n_0J_m(y_0)H_m^{(1)}(x_1)'\end{bmatrix},
\end{align*}
where the last element is a $2\times 1$ vector, \textit{not} a $2\times2$ matrix. We multiply the first two matrices to give us: 
\begin{align*}
\begin{bmatrix}a_2 \\ b_2 \end{bmatrix}=\begin{bmatrix}n_1H_m^{(2)}(x_2)'H_m^{(1)}(y_1)-n_2H_m^{(1)}(y_1)'H_m^{(2)}(x_2) && n_1H_m^{(2)}(x_2)'H_m^{(2)}(y_1)-n_2H_m^{(2)}(y_1)'H_m^{(2)}(x_2) \\ n_2H_m^{(1)}(y_1)'H_m^{(1)}(x_2)-n_1H_m^{(1)}(x_2)'H_m^{(1)}(y_1) && n_2H_m^{(2)}(y_1)'H_m^{(1)}(x_2)-n_1H_m^{(1)}(x_2)'H_m^{(2)}(y_1)\end{bmatrix}\times\\ \begin{bmatrix} n_0H_m^{(2)}(x_1)'J_m(y_0)-n_1J_m'(y_0)H_m^{(2)}(x_1)\\ n_1J_m'(y_0)H_m^{(1)}(x_1)-n_0J_m(y_0)H_m^{(1)}(x_1)'\end{bmatrix}
\end{align*}
As in the case of the sphere, we do not need to solve for $b_2$ if we solve for $a_2$. We can simply make the replacement $H_m^{(2)}(x_2)\rightarrow -H_m^{(1)}(x_2)$. Therefore, we just solve for $a_2$. In the following, we separate $a_2$ into terms proportional to $n_0n_1$, $n_0n_2$, $n_1^2$, and $n_1n_2$. We start with the term proportional to $n_1n_2$:
\begin{equation}
n_1n_2H_m^{(2)}(x_2)J_m'(y_0)\Bigg[H_m^{(1)}(y_1)'H_m^{(2)}(x_1)-H_m^{(2)}(y_1)'H_m^{(1)}(x_1) \Bigg]\approx \frac{2in_1n_2H_m^{(2)}(x_2)J_m'(y_0)}{n_1(\omega/c)R}\frac{2}{\pi},
\label{Equation: aonetwocylinder}
\end{equation}
where we used the asymptotic expansions of the Hankel functions given by \cref{Equation: Asymptotic cylindrical Hankel functions}. The term proportional to $n_0n_1$ reads:
\begin{equation}
n_0n_1H_m^{(2)}(x_2)'J_m(y_0)\Bigg[H_m^{(2)}(x_1)'H_m^{(1)}(y_1)-H_m^{(1)}(x_1)'H_m^{(2)}(y_1) \Bigg]\approx-\frac{2in_0n_1H_m^{(2)}(x_2)'J_m(y_0)}{n_1(\omega/c)R}\frac{2}{\pi}
\label{Equation: aonezerocylinder}
\end{equation}

\begin{figure}[htb!]
\centering
\includegraphics[scale=1]{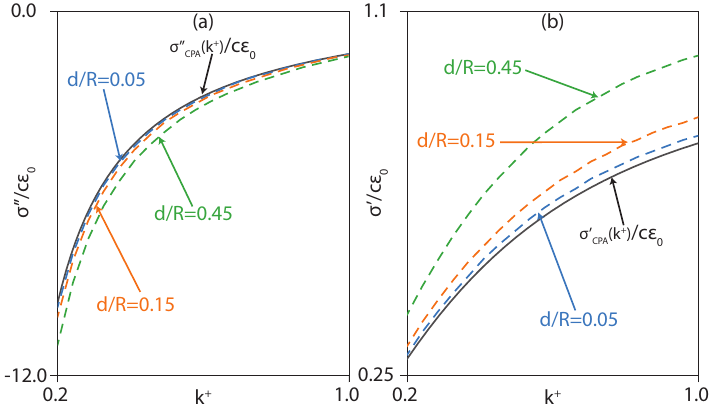}
  \caption{\textbf{ $\sigma_{CPA}(k^+)\equiv \sigma_{CPA}(k^+, m=1)$ for the TE channel as a thin film limit in cylinders:} For three values of thin film thickness, we find the conductivity corresponding to perfect absorption by modeling the thin film according to a thickness dependent permittivity (see \cref{Equation: Thickness dependent permittivity}). The permittivity of the bulk is taken to be $\varepsilon=1$. Solid black line shows $\sigma_{CPA}(k^+, m=1)$ (the CPA conductivity for an infinitely thin coating) for the TE channel. (a) shows the imaginary parts of the conductivities (note that in the TE channel these are negative in the subwavelength regime) and (b) shows the real parts of the conductivities (which are always positive since we require a coating that absorbs radiation).}
  \label{fig: supp figure 9}
\end{figure}

The term proportional to $n_1^2$ reads: 
\begin{equation} 
n_1^2J_m'(y_0)H_m^{(2)}(x_2)'\Bigg[H_m^{(1)}(x_1)H_m^{(2)}(y_1)-H_m^{(2)}(x_1)H_m^{(1)}(y_1) \Bigg]\approx -\frac{2in^3_1(\omega/c)d J_m'(y_0)H_m^{(2)}(x_2)'}{n_1(\omega/c)R}\frac{2}{\pi}
\label{Equation: aoneonecylinder}
\end{equation}
We stress that \cref{Equation: aonetwocylinder}, \cref{Equation: aonezerocylinder} and \cref{Equation: aoneonecylinder} all scale in the same way with the thickness of the thin film. The remaining term, proportional to $n_1^2$, will not have the same scaling relation, and, in particular, it will vanish in the $d\rightarrow 0$ limit (as we now show). The term proportional to $n_0n_2$ reads: 
\begin{equation}
n_0n_2J_m(y_0)H_m^{(2)}(x_2)\Bigg[H_m^{(1)}(x_1)'H_m^{(2)}(y_1)'-H_m^{(1)}(y_1)'H_m^{(2)}(x_1)' \Bigg]\approx 0 
\label{Equation: atwozerocylinder}
\end{equation}
Therefore, neglect the contribution of \cref{Equation: atwozerocylinder} to $a_2$ and we obtain (getting rid of constants of proportionality that will cancel out in the scattering coefficient):
\begin{equation}
\begin{split}
    &a_2\approx n_2H_m^{(2)}(x_2)J_m'(y_0)-n_0H_m^{(2)}(x_2)'J_m(y_0)-\frac{i\sigma(\omega)}{\tilde{\varepsilon}_0\omega d}\frac{\omega}{c}dJ_m'(y_0)H_m^{(2)}(x_2)'=\\&n_2H_m^{(2)}(x_2)J_m'(y_0)-n_0H_m^{(2)}(x_2)'J_m(y_0)-\frac{i\sigma(\omega)}{c\tilde{\varepsilon}_0}J_m'(y_0)H_m^{(2)}(x_2)' 
    \end{split}
\end{equation}
So the scattering coefficient is given by: 
\begin{align*}\frac{a_2}{b_2}=\frac{n_2H_m^{(2)}(x_2)J_m'(y_0)-n_0H_m^{(2)}(x_2)'J_m(y_0)-\frac{i\sigma(\omega)}{ c\tilde{\varepsilon}_0}J_m'(y_0)H_m^{(2)}(x_2)'}{-n_2H_m^{(1)}(x_2)J_m'(y_0)+n_0H_m^{(1)}(x_2)'J_m(y_0)+\frac{i\sigma(\omega)}{c\tilde{\varepsilon}_0}J_m'(y_0)H_m^{(1)}(x_2)'}, \end{align*}
which is the same as the TM scattering coefficient for coated cylinders given by \cref{EQUATION: TM SCATTERING FOR CYLINDERS}. We note that a similar procedure holds for the TE scattering coefficient (and we omit the derivation here so as to not be repetitive). In \cref{fig: supp figure 8} and \cref{fig: supp figure 9} we show how the CPA conductivity for a thin film of finite thickness asymptotically approaches $\sigma_{CPA}(k^+, m)$ as $d\rightarrow 0$ for TM and TE modes, respectively. 

\FloatBarrier
\section{CPA in the large $k^+$ limit for spheres}
\label{section: large k cpa spheres}

\begin{figure}[htb!]
\centering
\includegraphics[scale=1]{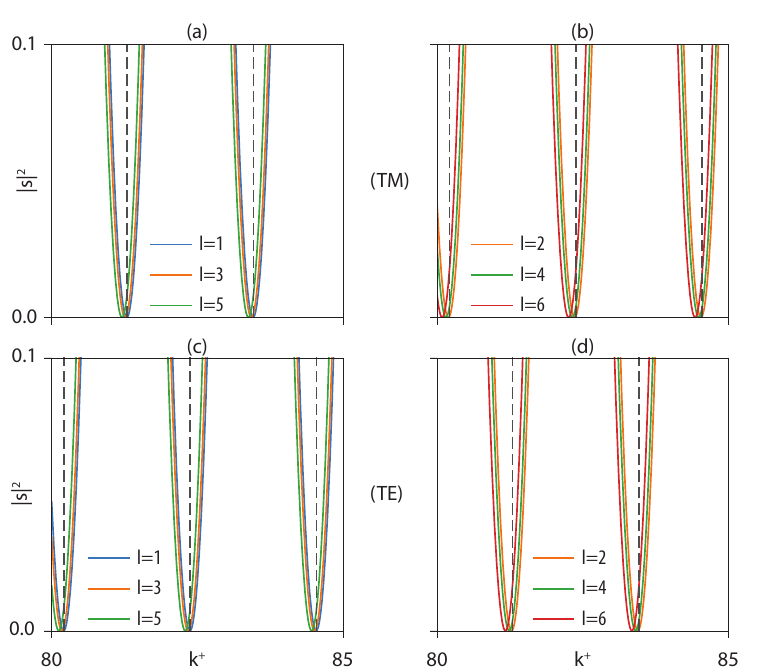}
  \caption{\textbf{Large $k^+$ CPA for spheres:} Scattering coefficient $|s_l|^2$ for a coated sphere with bulk permittivity $\varepsilon=2.1$ and surface conductivity $\sigma=c\varepsilon_0$. (a) and (c) correspond to odd values of the angular momentum quantum number and (b) and (c) correspond to even values. Vertical dashed lines are at coordinates $k^-=\pi/2+p\pi$ (for subplots (a) and (d)) and $k^-=p\pi$ (for subplots (b) and (c)), where $p$ is any integer. \cref{Equation: Large k sphere TM} and \cref{Equation: Large k sphere TE} imply the total absorption that we see at the locations of the vertical dashed lines.}
  \label{fig: supp figure 4}
\end{figure}

In this section, we show that in the large $k^+$ limit, CPA of a coated sphere (with a dispersion-free conductivity) for both polarizations occurs periodically in frequency, with periodicity, $\Delta\omega$, given by the following: 
\begin{equation}
    \Delta\omega=\frac{\pi c}{R\sqrt{\epsilon}}
\end{equation}
We start with the TM modes. To find the condition for CPA, we require the numerator of the TM scattering coefficient to vanish. That is, we require the following (see also \cite{christensen2015localized}): 
\begin{align*}-h^{(2)}_l(k^+)\Big[k^-j_l(k^-) \Big]'+\varepsilon j_l(k^-)\Big[k^+ h^{(2)}_l(k^+) \Big]'+g(\omega)\Big[k^-j_l(k^-) \Big]'\Big[k^+ h^{(2)}_l(k^+) \Big]'=0,\end{align*}
where $g(\omega)=\frac{i\sigma(\omega)}{R\omega \varepsilon_0}$, where $\sigma(\omega)$ is the conductivity of the surface layer. We now examine the asymptotic limit of this expression, using the following expansions for the spherical Bessel and Hankel functions:
\begin{equation}
j_l(x)\approx \frac{\sin(x-l\pi/2)}{x}, y_l(x)\approx -\frac{\cos(x-l\pi/2)}{x}, h^{(1)}_l(x)\approx -i\frac{\exp(i(x-l\pi/2))}{x}, h^{(2)}_l(x)\approx i\frac{\exp(-i(x-l\pi/2))}{x}
\label{Equation: Asymptotic hankel and spherical bessel}
\end{equation}
Using these asymptotic expansions, we may readily rewrite our CPA condition as follows: 
\begin{align*}
&-\frac{i}{k^+}\cos(k^--l\pi/2)+\frac{\varepsilon}{k^-}\sin(k^--l\pi/2)+g(\omega)\cos(k^--l\pi/2)=0 \rightarrow  \\&i\Bigg[\frac{\sigma'(\omega)}{c\varepsilon_0}-1\Bigg]\cos(k^--l\pi/2)+\Bigg[\sqrt{\varepsilon}\sin(k^--l\pi/2)-\frac{\sigma''(\omega)}{c\varepsilon_0}\cos(k^--l\pi/2)\Bigg]=0
\end{align*}
For this to vanish, both its real and imaginary parts have to vanish, which gives us the following values for the real and imaginary parts of the CPA conductivity:
\begin{equation}
\sigma'_{CPA}(k^+, l)=c\varepsilon_0, \sigma''_{CPA}(k^+, l)=c\varepsilon_0\sqrt{\varepsilon} \tan(k^--l\pi/2)
\label{Equation: Large k sphere TM}
\end{equation}
For TE modes, a similar derivation holds. In the TE case, we require the following:
\begin{align*}
-j_l(k^-)\Big[k^+h_l^{(2)}(k^+) \Big]'+h_l^{(2)}(k^+)\Big[k^-j_l(k^-) \Big]'-g(\omega)(k^+)^2j_l(k^-)h^{(2)}_l(k^+)=0  \end{align*}
Using the asymptotic expansions above, we obtain the following condition for $k^+\gg 1$
\begin{align*}-\sin(k^--l\pi/2)+i\sqrt{\varepsilon}\cos(k^--l\pi/2)+\frac{\sigma(\omega)}{c\varepsilon_0}\sin(k^--l\pi/2)=0 \end{align*}
From which we may read off the real and imaginary parts of the CPA conductivity: 
\begin{equation}
    \sigma'_{CPA}(k^+, l)=c\varepsilon_0, \sigma''_{CPA}(k^+, l)=-c\varepsilon_0\sqrt{\varepsilon}\cot(k^--l\pi/2)
\label{Equation: Large k sphere TE}
\end{equation}
To further determine the large $k^+$ behavior of CPA, in \cref{fig: supp figure 4}, we calculate the scattering coefficient for a coated sphere with a nondispersive surface conductivity, $\sigma=c\varepsilon_0$. From \cref{Equation: Large k sphere TM} and \cref{Equation: Large k sphere TE}, we expect CPA to occur at $k^-=\pi/2+p\pi$ (where $p$ is any integer) for  odd (even) $l$ modes for TM (TE) polarization. Furthermore, we expect  CPA to occur at $k^-=p\pi$ for even (odd) $l$ modes for TM (TE) polarization. These expectations are clearly consistent with the numerical results shown in \cref{fig: supp figure 4}. 

\section{CPA in the large $k^+$ limit for cylinders}
\label{section: large k cpa cylinders}
We start with the TM modes. In order to find $\sigma_{CPA}(k^+, m)$ at large $k^+$, we use the following expansions for the cylindrical Hankel functions (appropriate for large arguments):
\begin{equation}
H_m^{(1)}(x)\approx\sqrt{\frac{2}{\pi x}}e^{ix} (-i)^m e^{-i\pi/4}, H_m^{(2)}(x)\approx\sqrt{\frac{2}{\pi x}}e^{-ix} (i)^m e^{i\pi/4},
\label{Equation: Asymptotic cylindrical Hankel functions}
\end{equation}
which can be found on page 198 of Bohren and Huffman \cite{bohren2008absorption}. From these approximations, we may also readily read off the asymptotic expansions of the Bessel functions of the first and second kind for large arguments: 
\begin{align*}
&J_m(x)\equiv \frac{H_m^{(1)}(x)+H_m^{(2)}(x)}{2}\approx\sqrt{\frac{1}{2\pi x}}\Bigg[(-i)^m e^{-i\pi/4}e^{ix}+i^m e^{i\pi/4}e^{-ix} \Bigg],
\\&Y_m(x)\equiv \frac{H_m^{(1)}(x)-H_m^{(2)}(x)}{2i}\approx (-i)\sqrt{\frac{1}{2\pi x}}\Bigg[(-i)^m e^{-i\pi/4}e^{ix}-i^m e^{i\pi/4}e^{-ix} \Bigg]
\end{align*}
For CPA, we need the numerator of the scattering coefficient, \cref{EQUATION: TM SCATTERING FOR CYLINDERS}, to vanish. That is, we require the following: 
\begin{align*}
\sqrt{\varepsilon}J_m(k^-)H_m^{(2)}(k^+)'-H_m^{(2)}(k^+)J_m'(k^-)+\frac{i\sigma_{CPA}(k^+, m)}{c\varepsilon_0}J_m'(k^-)H_m^{(2)}(k^+)' = 0 
\end{align*}
Note that we are concerned, in this section, with the large $k^+$ limit, so we may disregard terms in the derivatives of the Bessel/Hankel functions proportional to $(k^{+/-})^{-3/2}$ because these will be small compared to terms proportional to $(k^{+/-})^{-1/2}$. We therefore have the following approximations:
\begin{align*}
&J_m'(k^-)\approx\sqrt{\frac{1}{2\pi k^-}}\Bigg[(-i)^m e^{i\pi/4}e^{ik^-}+i^me^{-i\pi/4}e^{-ik^-} \Bigg], \\&H_m^{(2)}(k^+)'\approx i^m\sqrt{\frac{2}{\pi k^+}}e^{-ik^+}e^{-i\pi/4}, 
\\&\sqrt{\varepsilon}J_m(k^-)H_m^{(2)}(k^+)'\approx \frac{\varepsilon^{1/4}}{\pi k^+}\Big[-ie^{i(k^--k^+)}+(-1)^m e^{-i(k^-+k^+)} \Big], 
\\&J'_m(k^-)H_m^{(2)}(k^+)\approx \frac{1}{\pi k^+\varepsilon^{1/4}}\Big[ie^{i(k^--k^+)}+(-1)^m e^{-i(k^-+k^+)} \Big],
\\ & J'_m(k^-)H_m^{(2)}(k^+)'\approx \frac{1}{\pi k^+\varepsilon^{1/4}}\Big[e^{i(k^--k^+)}+(-i)(-1)^m e^{-i(k^-+k^+)} \Big]
\end{align*}
Therefore, in order to have perfect absorption, for a given $k^+$ and $m$, we must have the following equality hold: 
\begin{align*}
(-i)e^{ik^-}\Bigg[\sqrt{\varepsilon}+1-\frac{\sigma_{CPA}(k^+, m)}{c\varepsilon_0} \Bigg]+(-1)^me^{-ik^-}\Bigg[\sqrt{\varepsilon}-1+\frac{\sigma_{CPA}(k^+, m)}{c\varepsilon_0} \Bigg]=0
\end{align*}
From this, we readily obtain the asymptotic limit of the CPA conductivity:
\begin{equation}
\begin{split}
    &\frac{\sigma_{CPA}(k^+, m)}{c\varepsilon_0}=\frac{(-1)^m(1-\sqrt{\varepsilon})e^{-ik^-}+i(\sqrt{\varepsilon}+1)e^{ik^-}}{(-1)^me^{-ik^-}+ie^{ik^-}}=1-\sqrt{\varepsilon}\Bigg[\frac{e^{i m\pi/2}e^{-ik^-}e^{-i\pi/4}-e^{i\pi/4}e^{-im\pi/2}e^{ik^-}}{e^{im\pi/2}e^{-ik^-}e^{-i\pi/4}+e^{i\pi/4}e^{-im\pi/2}e^{ik^-}} \Bigg]= 
    \\&1+i\sqrt{\varepsilon}\tan\Big[k^--(m-1/2)\pi/2\Big]
    \end{split}
    \label{Equation: Large k cylinder TM}
\end{equation}
In the second equality, we judiciously multiplied both the numerator and the denominator of the ratio in the second expression by a factor of $e^{-im\pi/2}e^{-i\pi/4}$. As in the spherical case, $\Re(\sigma_{CPA}(k^+, m))=c\varepsilon_0$ asymptotically and $\Im(\sigma_{CPA}(k^+, m))$ oscillates with period $\Delta k^+=\pi/\sqrt{\varepsilon}\leftrightarrow \Delta k^-=\pi$. 

\begin{figure}[htb!]
\centering
\includegraphics[scale=1]{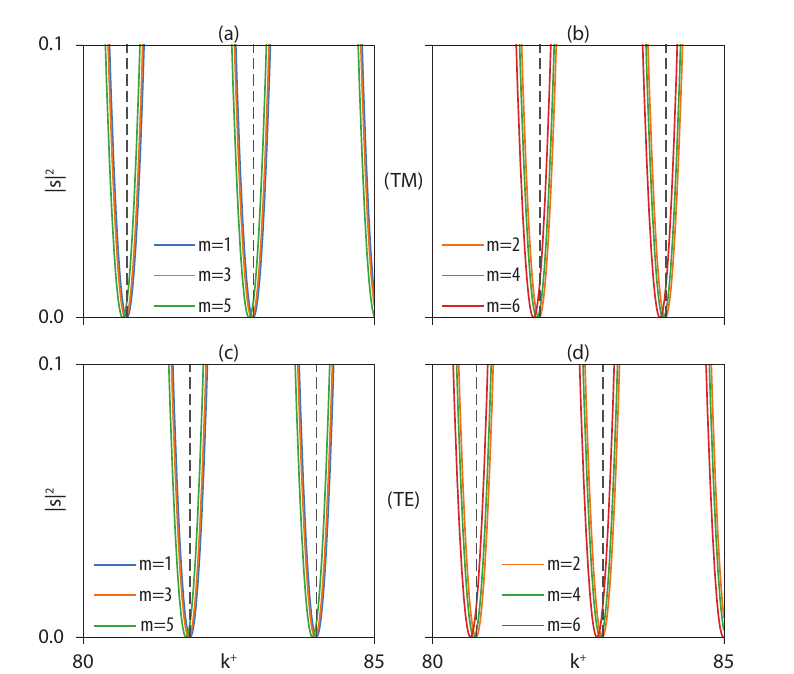}
  \caption{\textbf{Large $k^+$ CPA for cylinders:} Scattering coefficient $|s_m|^2$ for a coated cylinder with bulk permittivity $\varepsilon =2.1$ and surface conductivity $\sigma=c\varepsilon_0$. (a) and (c) correspond to odd values of the angular momentum quantum number, $m$, and (b) and (d) correspond to even values. Vertical dashed lines are delineate the coordinates $k^-=\frac{\pi}{4}+p\pi$ (for subplots (a) and (d)) and $k^-=-\frac{\pi}{4}+p\pi$ (for subplots (b) and (c)), where $p$ is any integer. See \cref{Equation: Large k cylinder TM} and \cref{Equation: Large k cylinder TE} to understand why the system is perfectly absorbing at the locations of the vertical dashed lines. }
  \label{fig: supp figure 7}
\end{figure}

For the TE modes, the derivation is similar. For CPA, we require the following to hold:
\begin{align*}
J_m(k^-)H_m^{(2)}(k^+)'-\sqrt{\varepsilon}H^{(2)}_m(k^+)J_m'(k^-)+\frac{i\sigma_{CPA}(k^+, m)}{c\varepsilon_0}J_m(k^-)H_m^{(2)}(k^+)=0
\end{align*}
Using our previously derived asymptotic expansions, we obtain the following for $\sigma_{CPA}(k^+, m)/c\varepsilon_0$:
\begin{equation}
\frac{(-1)^{(m+1)}(1-\sqrt{\varepsilon})e^{-ik^-}+i(\sqrt{\varepsilon}+1)e^{ik^-}}{(-1)^{m+1}e^{-ik^-}+ie^{ik^-}}=1+i\sqrt{\varepsilon}\tan\Big[k^--(m+1/2)\pi/2 \Big]
\label{Equation: Large k cylinder TE}
\end{equation}
As in the case of the TM channel, $\Re(\sigma_{CPA}(k^+, m))=c\varepsilon_0$ asymptotically and we have the same oscillation period for $\Im\sigma_{CPA}(k^+, m)$, $\Delta k^{-}=\pi$. To illustrate the scattering behavior of cylinders for large $k^+$, we calculate $|s_m|^2$ for a cylinder with $\sigma=c\varepsilon_0$ (for all frequencies; i.e. no dispersion) in \cref{fig: supp figure 7}. As predicted by \cref{Equation: Large k cylinder TM} (for TM modes), we observe periodic CPA for all odd angular momentum modes for  $k^-$ that satisfy $k^-=\frac{\pi}{4}+p\pi$, with $p$ being any integer. Similarly, we observe CPA for $k^-=-\frac{\pi}{4}+p\pi$ for the even angular momentum modes, in agreement with \cref{Equation: Large k cylinder TM}.

\section{CPA in the large $m$ limit for cylinders}
\label{section: Whispering gallery modes}
We now consider the regime of high angular momentum, $m\gg 1$. In this regime, we may make the following approximation to the Bessel functions of the first and second kinds \cite{nockel1997resonances, gradshteyn2014table}: 
\begin{equation}
\begin{split}
    & J_m(x<m)\approx \frac{\exp(g(m, \alpha))}{\sqrt{2m\pi\tanh(\alpha)}} 
    , Y_m(x<m)\approx \frac{-2\exp(-g(m, \alpha))}{\sqrt{2m\pi\tanh(\alpha)}} 
    \\ & J_{m}(x>m)\approx \sqrt{\frac{2}{m\pi\tan(\beta)}}\cos(f(m, \beta)), Y_m(x>m)\approx \sqrt{\frac{2}{m\pi\tan(\beta)}}\sin(f(m, \beta))
    \end{split},
\end{equation}
where we defined $f(m, \beta)\equiv m\tan(\beta)-m\beta-\pi/4$ and $g(m, \alpha)\equiv m(\tanh(\alpha)-\alpha)$, with $\cosh(\alpha)\equiv m/x, \cos(\beta)\equiv m/x$. Similarly, one may make reasonably accurate approximations to the derivatives of the Bessel functions, given as follows: 
\begin{equation}
    \begin{split}
    &J'_m(x<m)= \frac{\exp(g(m, \alpha))}{\sqrt{2\pi m\tanh(\alpha)}}\sinh(\alpha)
  , J'_m(x>m)= -\sqrt{\frac{2}{m\pi\tan(\beta)}}\sin(f(m, \beta))\sin(\beta), 
    \\& Y'_m(x<m)= \frac{2\exp(-g(m, \alpha))}{\sqrt{2\pi m\tanh(\alpha)}}\sinh(\alpha),
     Y'_m(x>m)= \sqrt{\frac{2}{m\pi\tan(\beta)}}\cos(f(m, \beta))\sin(\beta)
    \end{split}
\end{equation}
We are concerned, ultimately, with the whispering gallery regime, for which $k^+<m<k^-$. In this regime, the real part of the CPA conductivity is negligible compared to the imaginary part, and we may write an implicit equation for the CPA conductivity as follows:
\begin{equation}
\begin{split}
    &J_m'(k^-)Y_m(k^+)-\sqrt{\varepsilon} J_m(k^-)Y_m(k^+)'+\frac{\sigma''_{CPA}(k^+, m)}{c\varepsilon_0}J_m'(k^-)Y_m(k^+)'\approx 0 \rightarrow  \\ 
&\sigma_{CPA}''(k^+, m)\approx c\varepsilon_0\Bigg[\sqrt{\varepsilon}\frac{J_m(k^-)}{J_m'(k^-)}-\frac{Y_m(k^+)}{Y_m'(k^+)} \Bigg]\approx c\varepsilon_0\Bigg[-\frac{\sqrt{\varepsilon}\cot(f(m, \beta_{-}))}{\sin(\beta_-)}\Bigg],
\end{split}
\end{equation}
where $\beta_-$ is defined by $\cos(\beta_-)=m/k^-$. We have checked that this is a good approximation for $m=190$ for $k^-\approx m$ (corresponding to the whispering gallery regime \cite{novotny2012principles}). Now, we note that:
\begin{equation}
    \frac{m+1}{k^-+1}\approx\frac{m}{k^-}\Big[(1+1/m)(1-1/k^-) \Big]\approx\frac{m}{k^-},
\end{equation}
where, in the last equality we used the fact that $k^-\approx m$. This means that $\sigma''_{CPA}(k^+, m)$ has the following property: 
\begin{equation}
    \sigma''_{CPA}(k^+, m)\approx \sigma''_{CPA}\Bigg(k^++\frac{1}{\sqrt{\varepsilon}}, m+1\Bigg)
\end{equation}
This differs significantly from the behavior in the large $k^+$, small $m$ regime (i.e. \cref{Equation: Large k cylinder TM}). In that case, the symmetry property was: 
\begin{equation}
    \sigma''_{CPA}(k^+, m)\approx \sigma''_{CPA}\Bigg(k^++\frac{\pi}{2\sqrt{\varepsilon}}, m+1\Bigg)
\end{equation}
This is the essence of the difference between the Fabry-Perot regime. Note that $(\pi/2)=2\pi R/(4R)$, which is exactly the ratio of the round trip distance in the whispering gallery regime to the round trip distance in the Fabry-Perot regime.

\section{Divergence of the imaginary part of the CPA conductivity}
\label{section: first divergence}
Unlike the quasistatic approximations for the CPA conductivities, the fully retarded values of $\sigma''_{CPA}(k^+, l)$ (for spheres) and $\sigma''_{CPA}(k^+, m)$ (for cylinders) diverge at discrete values of $k^+$. For cylinders, in the TM channel, these values are given by the roots of the derivative of the Bessel function inside the cylinder (i.e. $J_m'(k^-)=0$), and for spheres (also in the TM channel), they are given by the zeros of the derivative of the Ricatti-Bessel function inside the sphere (i.e. $(k^-j_l(k^-))'=0$). For the dipolar, $l/m=1$, TM channel, the first divergence in the cylindrical (spherical) case thus occurs at $k^+\approx 1.8412/\sqrt{\varepsilon}$ ($k^+\approx 2.744/\sqrt{\varepsilon}$). 

For the TE channel, the divergences of $\sigma_{CPA}''(k^+, l)$ for spheres occur at the zeros of the spherical Bessel function inside the sphere; i.e. $j_l(k^-)=0$. Thus, the first divergence for the dipolar channel occurs at $k^+\approx 4.493/\sqrt{\varepsilon}$. Similarly, for cylinders, the divergences for $\sigma''_{CPA}(k^+, m)$ for the TE channel occur at the zeros of the cylindrical Bessel function inside the cylinder; i.e. $J_m(k^-)=0$, which yields $k^+\approx 3.8317/\sqrt{\varepsilon}$ as the first divergence for the dipolar channel.

\section{Bandwidth of subwavelength CPA in the TM channel for spherical systems}
\label{section: cpa bandwidth}

\begin{figure}[htb!]
\centering
\includegraphics[scale=1]{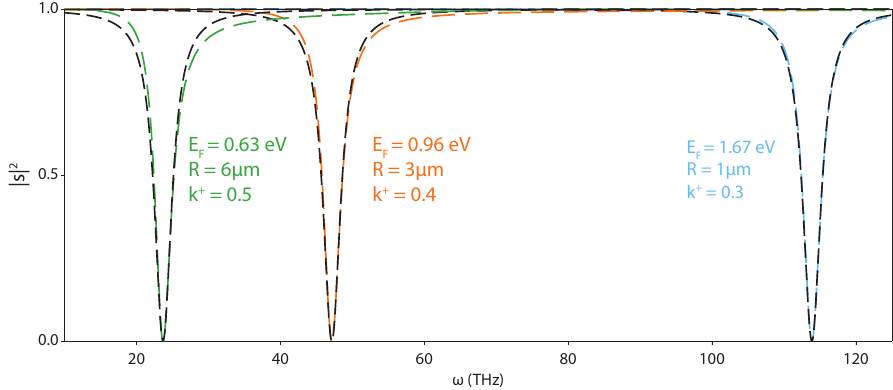}
  \caption{\textbf{Bandwidth of CPA :} We calculate $|s|^2\equiv |s_1|^2$ for graphene coated spheres tuned to perfect absorption for $k^+=0.3, 0.4$ and $0.5$. Colored lines correspond to values obtained through \cref{EQUATION: TM SCATTERING FOR SPHERES} while the black lines correspond to the approximate formula, \cref{Equation: approximate s squared}. For each value of $k^+$, we also indicate the radius of the scatterer, $R$, and the corresponding Fermi energy, $E_F$.}
  \label{fig: supp figure 1}
\end{figure}

In this section, we find the CPA bandwidth (i.e., the half width at half maximum of the absorption $1-|s|^2$). We focus on subwavelength TM waves in the dipolar (i.e. $s\equiv s_1$) channel for coated spherical systems. For small spheres, $k^+\ll 1$, we may approximate the scattering coefficient as follows: 
\begin{equation}s(\omega)\approx -\frac{2\Bigg[(1-\varepsilon)-2g(\omega)\Bigg](k^+)^3+3i\Bigg[(2+\varepsilon)+2g(\omega)\Bigg]}{2\Bigg[(1-\varepsilon)-2g(\omega)\Bigg](k^+)^3-3i\Bigg[(2+\varepsilon)+2g(\omega)\Bigg]},
\label{Equation: bandwidth s}
\end{equation}
which follows from  using the approximations of the spherical Bessel and Hankel functions given by \cref{Equation: spherical bessel function approximations}. Note that, for both the numerator and the denominator,  we have maintained the lowest as well as the next to next to lowest order terms, in contrast to \cref{Equation: Qausistatic conductivity for TM modes in spheres}. The next to next to lowest order terms (proportional to $(k^+)^3$) are required to include the effects of radiative loss. $g(\omega)\equiv i\sigma(\omega)/(k^+c\varepsilon_0)$ is the dimensionless quantity previously introduced in \cref{EQUATION: TM SCATTERING FOR SPHERES}.
The CPA condition at a given frequency, $\omega_{CPA}$, will be satisfied if we have 
\begin{align*}
g(\omega_{CPA})=\frac{-(2+\varepsilon)/2+(i/3)(1-\varepsilon)(k^+)^3}{1+(2i/3)(k^+)^3}\approx-(2+\varepsilon)/2+i(k^+)^3
\end{align*}
In order to find the half width at half maximum, we calculate $s$ at a neighboring frequency, $\omega_{CPA}+\delta$:
\begin{equation}
\begin{split}
        &s(\omega_{CPA}+\delta)\approx-\frac{3(k^+)^2(\delta R_{CPA}/c) +i\Re [g'(\omega_{CPA})]\delta}{2(k^+)^3-i\Re [g'(\omega_{CPA})]\delta}\approx \frac{\Re [g'(\omega)]\delta}{\Re [g'(\omega_{CPA})]\delta + 2i(k^+)^3} \rightarrow\\ &|s(\omega_{CPA}+\delta)|^2\approx 1-\frac{4(k^+)^6}{4(k^+)^6+\Re [g'(\omega_{CPA})]^2\delta^2}
        \end{split}
        \label{Equation: approximate s squared}
\end{equation}
In the expression above, we made several (valid) approximations which we will now justify. First, we ignored the contribution of $\Im g'(\omega_{CPA})$, since this is much smaller in magnitude compared to $\Re g'(\omega_{CPA})$ for $k^+<1$ and, correspondingly, $\omega\tau\gg1$. In addition, since $\Re g'(\omega_{CPA})\approx -2\Re g(\omega_{CPA})/\omega_{CPA}\approx (2+\varepsilon)/\omega_{CPA}$, we know that terms of the form $\Re g'(\omega_{CPA})$ may be considered large compared to $(k^+)^2(R_{CPA}/c)$. From the above, we may read off the half width at half maximum, $\delta_{hwhm}$, as:
\begin{align*}
\delta_{hwhm}\approx \frac{2(k^+)^3}{|\Re g'(\omega_{CPA})|}
\approx\frac{2(k^+)^3\omega_{CPA}}{2+\varepsilon} \approx \frac{1}{\tau},
\end{align*}
where we used $\omega_{CPA} \tau=\Im\sigma_{CPA}(k^+, l=1)/\Re\sigma_{CPA}(k^+, l=1)$ and used the appropriate subwavelength approximations for the CPA conductivities given by \cref{Equation: Qausistatic conductivity for TM modes in spheres} and \cref{Equation: Subwavelength real sigma CPA for spheres}. In order to validate the approximate model used above, we calculate $|s|^2$ from both \cref{EQUATION: TM SCATTERING FOR SPHERES} and \cref{Equation: approximate s squared} in \cref{fig: supp figure 1}. Clearly, the approximate expression for $|s|^2$ is in good agreement with the exact expression for small values of $k^+$.

\section{Tuning permittivity and complex frequency to bring the system back to CPA}
\label{Section: complex frequencies}

\begin{figure}[htb!]
\centering
\includegraphics[scale=1]{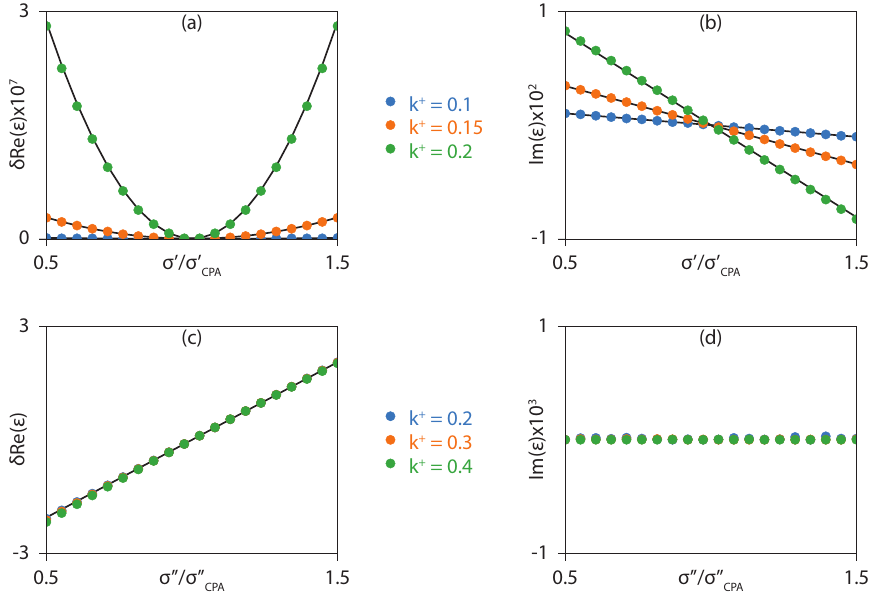}
  \caption{\textbf{Tuning permittivity to achieve CPA when $\sigma\neq \sigma_{CPA}(k^+)$:} (a, b) For three values of $k^+$, we sweep the real part of the conductivity (fixing the imaginary part to $\sigma''_{CPA}$) and find the change in permittivity required to achieve CPA. (c, d) Same as (a, b) except we sweep the imaginary part of the conductivity and fix the real part to $\sigma'_{CPA}$. Solid lines in (b) and (c) correspond to the lowest order linear approximations given by \cref{Equation: change real sigma change im epsilon} and \cref{Equation: change im sigma change real epsilon}, respectively. Solid line in (a) corresponds to the quadratic approximation given by \cref{Equation: change real sigma change real epsilon}. For all subplots, the system in consideration is a sphere, the polarization of the field is TM and the permittivity before tuning is $\varepsilon=2.1$. Note that we only consider the dipolar mode (and suppress angular momentum labels in this caption and in the figure as a result). }
  \label{fig: supp figure 2}
\end{figure}

Up to this point, we have considered a scatterer with fixed permittivity and fixed normalized radius, $k^+$. For a given value of these two parameters, we have shown that CPA can be achieved if the coating has the appropriate conductivity, $\sigma_{CPA}(k^+, l/m)$ (for spheres/cylinders, respectively). In this section, we answer three related questions:
\begin{enumerate}
    \item If we change the real part of the conductivity, $\sigma'\rightarrow \sigma'_{CPA}+\delta\sigma'_{CPA}$, how must we change the permittivity to preserve the CPA condition? \footnote{In this case, for simplicity, we keep the imaginary part of the conductivity fixed at its CPA value.}
    \label{enum: track cpa zeros 1}
    \item If we change the imaginary part of the conductivity, $\sigma''\rightarrow \sigma''_{CPA}+\delta\sigma''_{CPA}$, how must we change the permittivity to preserve the CPA condition?
    \footnote{In this case, for simplicity, we keep the real part of the conductivity fixed at its CPA value.}
    \label{enum: track cpa zeros 2}
    \item If we change the radius of the scatterer, $R\rightarrow R_{CPA}+\delta_R$, where will the CPA zeros exist in the complex frequency plane?
    \label{enum: track cpa zeros 3}
\end{enumerate}
At first glance, one would perhaps be inclined to think that the answers to \cref{enum: track cpa zeros 1} and \cref{enum: track cpa zeros 2} should be similar. However, in the derivations below, and in \cref{fig: supp figure 2}, we show that their respective answers actually differ substantially. Unlike the questions posed by \cref{enum: track cpa zeros 1} and \cref{enum: track cpa zeros 2}, the answer to \cref{enum: track cpa zeros 3} clearly requires some model for the frequency dependence of the coating's conductivity. In keeping with the rest of this paper, we assume a Drude conductivity (e.g. graphene in the intraband approximation). Lastly, we note that the system we consider in this section is a sphere in the dipolar TM channel (and we thus suppress angular momentum indices for brevity). We start this section by answering question (1) and (2). From the numerator of \cref{Equation: bandwidth s} (which must be zero at CPA), 
\begin{equation}
    2\Bigg[(1-\varepsilon)-2g(\omega)\Bigg](k^+)^3+3i\Bigg[(2+\varepsilon)+2g(\omega)\Bigg],
    \label{Equation: s numerator}
\end{equation}
we see that if the conductivity is moved off its CPA value, this can be counteracted by judiciously tuning the permittivity. Concretely, if we have $\varepsilon, g(\omega)$ for which \cref{Equation: s numerator} is zero, and we tune both $g(\omega)$ and $\varepsilon$ such that $g\rightarrow g+\delta_g$, $\varepsilon\rightarrow \varepsilon+\delta_\varepsilon$, then, to maintain CPA, one must require:
\begin{equation}
    \Big[3i-2(k^+)^3\Big]\delta_\varepsilon +\Big[6i-4(k^+)^3\Big]\delta_g=0\rightarrow \delta_\varepsilon=-\frac{6i-4(k^+)^3}{3i-2(k^+)^3}\delta_g= -2i\delta_\sigma/(c\varepsilon_0k^+),
\end{equation}
where we wrote the change in $g$ in terms of the change in conductivity, $\delta_\sigma$ \footnote{Note that we are keeping $k^+$ fixed.}. Note that we previously showed (see \cref{Equation: Qausistatic conductivity for TM modes in spheres}) that $\sigma''_{CPA}(k^+)\approx c\varepsilon_0 k^+(2+\varepsilon)/2$. Therefore, we may equivalently express the change in the real part of the permittivity as follows:
\begin{equation}
\Re \delta_{\varepsilon} \approx (2+\varepsilon)\Im(\delta_\sigma)/\sigma''_{CPA}(k^+)
\label{Equation: change im sigma change real epsilon}
\end{equation}
Similarly, since $\sigma_{CPA}'(k^+)\approx c\varepsilon_0 (k^+)^4$, we may write the change in the imaginary part of the permittivity as follows:
\begin{equation}
    \Im \delta_\varepsilon\approx -2\Re(\delta_\sigma)(k^+)^3/\sigma'_{CPA}(k^+)
    \label{Equation: change real sigma change im epsilon}
\end{equation}
These relations hold well for small values of $k^+$ as may be seen in \cref{fig: supp figure 2}(b, c). Interestingly, at this order of the calculation, we see that $\Re\delta_\varepsilon$ has no dependence on $\Re\delta_\sigma$ and $\Im \delta_\varepsilon$ has no dependence on $\Im\delta_\sigma$. However, as we show below, $\Re\delta_\varepsilon$ actually has a weak  quadratic dependence on $\Re\delta_\sigma$. In contrast, however, $\Im\delta_\varepsilon$ has no dependence on $\Im\delta_\sigma$. To demonstrate this, we need to include additional terms in the expansion of the scattering coefficient's numerator.
We expand the numerator (up to a constant of proportionality) as follows: 
\begin{equation}-i\Bigg[\varepsilon+2+2g(\omega)\Bigg]+\Bigg[\frac{2\varepsilon}{3}-\frac{2}{3}+\frac{4g(\omega)}{3} \Bigg](k^+)^3+i\Bigg[\varepsilon\Bigg[\frac{\varepsilon}{10}+\frac{1}{2} \Bigg]+\Bigg[\frac{2\varepsilon}{5}-1 \Bigg]+g(\omega)\Bigg[\frac{2\varepsilon}{5}+1\Bigg] \Bigg](k^+)^2
\end{equation}
Requiring CPA upon a change $\varepsilon\rightarrow\varepsilon+\delta_\varepsilon, g\rightarrow g+\delta_g$ gives us the following condition:
\begin{equation}
\Bigg[\frac{2}{3}(k^+)^3 -i\Bigg]\Bigg[\delta_\varepsilon+2\delta_g \Bigg]+i\Bigg[\frac{\varepsilon}{5}\delta_\varepsilon+\frac{1}{2}\delta_\varepsilon+\frac{2}{5}\delta_\varepsilon+\frac{2}{5}g\delta_\varepsilon+\frac{2}{5}\varepsilon\delta_g+\delta_g +\frac{2}{5}\delta_g\delta_\varepsilon+\frac{\delta_\varepsilon^2}{10} \Bigg](k^+)^2=0 
\label{Equation: higher order s num}
\end{equation}
Note that, crucially, we included a term that is quadratic in the change in permittivity. For convenience, we introduce a parameter, $\delta$, which we define by the relation $\delta_\varepsilon\equiv\delta-2\delta_g$ \footnote{We do this so that we may isolate the higher order contribution to the change in permittivity}, allowing us to write \cref{Equation: higher order s num} as follows: 
\begin{equation}
\begin{split}
    &\Bigg[\frac{2}{3}(k^+)^3 -i\Bigg]\delta+i\Bigg[(\varepsilon/5+1/2)\delta-\frac{4}{5}(1+g)\delta_g +\frac{2}{5}(1+g)\delta+\frac{(\delta-2\delta_g)^2}{10}-\frac{4}{5}\delta_g^2+\frac{2}{5}\delta_g\delta\Bigg](k^+)^2=\\ &\Bigg[\frac{2}{3}(k^+)^3 -i\Bigg]\delta+i\Bigg[(\varepsilon/5+1/2)\delta-\frac{4}{5}(1+g)\delta_g +\frac{2}{5}(1+g)\delta+\frac{\delta^2-4\delta_g^2}{10}\Bigg](k^+)^2 \approx \\ & -\delta+\Bigg[\delta/2+(2\varepsilon/5)\delta_g-\frac{2}{5}\delta_g^2 \Bigg](k^+)^2=0\rightarrow \delta\approx-\frac{2}{5}(k^+)^2(\delta_g^2-\varepsilon\delta_g)
    \end{split}
    \label{Equation: change real sigma change real epsilon}
\end{equation}
Note that $\delta$ is always real if we tune \textit{only} the imaginary part of the conductivity or \textit{only} the real part of the conductivity. This is why tuning the real part of the conductivity yields a quadratic change in the real part of the permittivity. However, tuning the imaginary part of the conductivity does not yield an analogous quadratic change in the imaginary part of the permittivity. The quadratic relation shown above holds well in the small as shown in \cref{fig: supp figure 2}(a). 

Interestingly, $\Im\delta_\varepsilon$ actually has no dependence on $\Im \delta_\sigma$. This actually follows from \cref{section: first divergence} and the fact that our expression for $\sigma'_{CPA}(k^+)$ has no dependence on the permittivity as long as the permittivity is real. These two facts combined mean that if $\sigma_{CPA}=\sigma'_{CPA}+i\sigma''_{CPA}$ for a given value of $\varepsilon$, then $\tilde{\sigma}_{CPA}=\sigma'_{CPA}+i(\sigma''_{CPA}+\delta_{\sigma''})$ \footnote{Where $\delta_{\sigma''}$ is real valued} is also a solution just with a different (real valued) permittivity (at the same $k^+$). 
\subsection{Zeros in the complex frequency plane}
\begin{figure}[h!]
\centering
\includegraphics[scale=1]{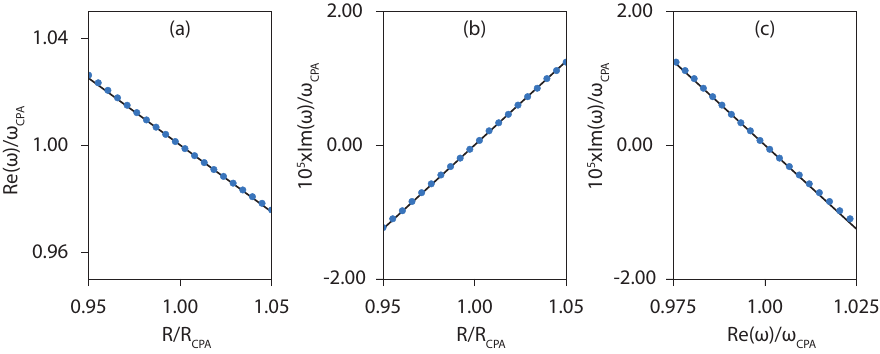}
  \caption{\textbf{CPA frequencies in the complex plane:} For a system that achieves CPA at $k^+=0.1$, we sweep the radius and find complex frequencies at which the CPA condition is achieved. (a, b) show the real and imaginary parts of the frequency as the radius is changed, respectively. (c) shows the same data as (a) and (b) but in the complex frequency plane. Solid lines in (a) and (b) correspond to analytic approximations given by \cref{Equation: change radius change re omega} and \cref{Equation: Change radius change im omega}, respectively. Solid line in (c) corresponds to using \cref{Equation: change radius change re omega} for the real part of the frequency and \cref{Equation: Change radius change im omega} for the imaginary part of the frequency. For all subplots, the system under consideration is a sphere with $\varepsilon=2.1$, and we only consider the dipolar TM channel.  }
  \label{fig: supp figure 3}
\end{figure}
We now turn to \cref{enum: track cpa zeros 3}. 
By the same procedure as above, we may find the complex frequency satisfying CPA if the scatterer's radius is altered slightly from its CPA value, $R_{CPA}$. To do so, we consider tuning the radius, $R\rightarrow R_{CPA}+\delta_R$ as well as the frequency, $\omega\rightarrow \omega_{CPA}+\delta_\omega$. Furthermore, we note that $g(\omega)$ is both a function of $R$ and of $\omega$, so we use the notation $g(R, \omega)$ in the remainder of this subsection. Lastly, we note that for a Drude material at CPA, we have the following convenient expressions for the first derivatives of $g(R, \omega)$: 
\begin{equation}
    \partial_\omega g(R, \omega)|_{(R_{CPA}, \omega_{CPA})}\approx \frac{2+\varepsilon}{\omega_{CPA}}-\frac{3i(k^+)^3}{\omega_{CPA}}, \partial_R g(R, \omega)|_{(R_{CPA}, \omega_{CPA})}\approx \frac{2+\varepsilon}{2R_{CPA}}-\frac{i(k^+)^3}{R_{CPA}}
\end{equation}
The lowest order change in the frequency is real valued  and follows immediately from expanding \cref{Equation: s numerator} at the shifted radius and complex frequency (ignoring terms of order $(k^+)^3$): 
\begin{equation}
    \delta_R\partial_R g(R, \omega)|_{(R_{CPA}, \omega_{CPA})}+\delta_\omega\partial_\omega g(R, \omega)|_{(R_{CPA}, \omega_{CPA})}\approx0\rightarrow (\delta_\omega/\omega_{CPA})\approx- (\delta_R/2R_{CPA}),
    \label{Equation: change radius change re omega}
\end{equation}
which may be easily verified as seen in \cref{fig: supp figure 3}(a). The shift in the imaginary part of the frequency is slightly more involved. We expand \cref{Equation: s numerator} and maintain terms of order $(k^+)^3$. Furthermore, we write the shift in frequency as $\delta_\omega=-(\delta_R/2R_{CPA})\omega_{CPA}+i\delta$ in order to isolate the imaginary part of the frequency shift. 
\begin{equation}
\begin{split}
        &6\Bigg[(1-\varepsilon)-2g\Bigg](k^+)^3\Bigg[\frac{\delta_R}{R_{CPA}}+\frac{\delta_\omega}{\omega_{CPA}}\Bigg]-4(k^+)^3\Bigg[\partial_R g\delta_R +\partial_\omega g\delta_\omega \Bigg]+6i\Bigg[\partial_Rg\delta_R +\partial_\omega g\delta_\omega\Bigg] \approx\\  &9(k^+)^3\Bigg[\frac{\delta_R}{R_{CPA}}\Bigg]-3(k^+)^3\Bigg[\frac{\delta_R}{R_{CPA}} \Bigg]-6\frac{2+\varepsilon}{\omega_{CPA}}\delta\approx 0 \rightarrow \frac{\delta}{\omega_{CPA}}\approx \frac{(k^+)^3}{2+\varepsilon}\frac{\delta_R}{R_{CPA}},
    \label{Equation: Change radius change im omega}
    \end{split}
\end{equation}
As shown in \cref{fig: supp figure 3}, this is a good fit to the imaginary part of the CPA zero. 

\section{Effect of nonlocality on the CPA conductivity for coated spheres}
\label{section: nonlocal cpa conductivity for spheres}
As mentioned in the main text, our closed form solutions for the CPA conductivity may be readily extended to cases in which the response function of the coating is spatially nonlocal. Spatial nonlocality may be introduced by a modification of Ohm's law as follows \cite{ christensen2015localized}:
\begin{equation}
    \frac{\beta^2}{\omega^2}\nabla(\nabla\cdot\mathbf{J}(\mathbf{r},\omega))+\mathbf{J}(\mathbf{r},\omega)=\sigma(\omega)\mathbf{E}(\mathbf{r},\omega),
    \label{Equation: Nonlocal Ohm's Law}
\end{equation}
where $\beta$ is a quantity that has units of velocity and is on the order of the Fermi velocity, $v_F$. The exact value of $\beta$ depends on the underlying electronic structure of the coating, so we keep it generic in the following derivation. We note, for convenience, that linear and quadratic dispersions in two dimensions yield the same  value for $\beta$ given by $\beta^2=3v_F^2/4$. In the following, we show that inclusion of spatial nonlocality introduces a (closed-form) modification to $\sigma_{CPA}(k^+, l)$ for TM polarized waves (given by \cref{Equation: Nonlocal sigma cpa spheres}) but does not modify $\sigma_{CPA}(k^+, l)$ for TE polarized waves. 

\subsection{Effect of nonlocality for TM polarized waves}
For TM waves, the surface current in the coating, $\mathbf{J}(\mathbf{r}, \omega)$, must be expressible in the following form: 
\begin{equation}
\begin{split}
    \mathbf{J}(\mathbf{r}, \omega)=\delta(r-R)\Bigg[\nabla\times\mathbf{r}\times\nabla f(\mathbf{r}, \omega)-\mathbf{e}_r\cdot\Big[\nabla\times\mathbf{r}\times\nabla f(\mathbf{r}, \omega)\Big]\mathbf{e}_r\Bigg],
   \end{split}
\end{equation}
where the Dirac delta function ensures that we only have a non-zero current density at the radial coordinate of the coating, and we have, of course, ensured that the current is purely tangential to the sphere by explicitly subtracting its radial component. In the above, the function $f(\mathbf{r}, \omega)\equiv jY_{lm}(\theta, \phi)$, where $j$ is a constant to be determined later. For future convenience, we write the current density explicitly in spherical coordinates as follows: 
\begin{equation}
    \mathbf{J}(\mathbf{r}, \omega)=-\frac{\delta(r-R)}{r}\Bigg[\partial_r(r\partial_\theta f(\mathbf{r}, \omega))\mathbf{e}_\theta+\frac{1}{\sin(\theta)}\partial_r(r\partial_\phi f(\mathbf{r}, \omega)) \mathbf{e}_\phi\Bigg]
\end{equation}
In order to use \cref{Equation: Nonlocal Ohm's Law}, we calculate the divergence of this surface current, which gives us:
\begin{equation}
\begin{split}    
    &\nabla\cdot\mathbf{J}(\mathbf{r}, \omega)=\Bigg[\partial_r\delta(r-R)\Bigg]\mathbf{e}_r\cdot\Bigg[\nabla\times\mathbf{r}\times\nabla f(\mathbf{r}, \omega)\Bigg]-\frac{1}{r^2}\partial_r\Bigg[r^2\delta(r-R)\mathbf{e}_r\cdot\Big[\nabla\times\mathbf{r}\times\nabla f(\mathbf{r}, \omega )\Big]\Bigg]= \\&-\frac{\delta(r-R)}{r^2}\partial_r\Bigg[r^2\mathbf{e}_r\cdot\Big[\nabla\times\mathbf{r}\times\nabla f(\mathbf{r}, \omega)\Big] \Bigg]
    \end{split}
\end{equation}
Note that, quite conveniently, the annoying radial derivatives on the Dirac delta functions cancel out exactly. Therefore, we may readily evaluate the divergence of the current density, giving us the following expression:
\begin{equation}
\begin{split}    
    &\nabla\cdot\mathbf{J}(\mathbf{r}, \omega)=-\frac{\delta(r-R)}{r^2}\partial_r\Bigg[r^2\mathbf{e}_r\cdot\Bigg[\nabla\times\Big[\partial_\theta f(\mathbf{r}, \omega) \mathbf{e}_\phi-\frac{1}{\sin(\theta)}\partial_\phi f(\mathbf{r}, \omega) \mathbf{e}_\theta\Big]\Bigg]\Bigg]=\\&-\frac{\delta(r-R)}{r^2}\partial_r\Bigg[\frac{r}{\sin(\theta)}\Bigg[\partial_\theta\Big(\sin(\theta)\partial_\theta f(\mathbf{r}, \omega)\Big)+\frac{1}{\sin(\theta)}\partial_\phi^2 f(\mathbf{r}, \omega)\Bigg]\Bigg] =\frac{\delta(r-R)l(l+1)}{r^2}\partial_r\Big[rf(\mathbf{r}, \omega) \Big],
\end{split}
\end{equation}  
where we used the fact that the spherical harmonics, $Y_{lm}(\theta, \phi)$, are eigenfunctions of the square of the angular momentum operator. To complete our derivation, we now calculate the gradient of the divergence obtained above:
\begin{equation}
\begin{split}  
    \nabla(\nabla\cdot \mathbf{J}(\mathbf{r}, \omega))=\frac{\delta(r-R)l(l+1)}{r^3}\Bigg[\partial_r\Big[r\partial_\theta f(\mathbf{r}, \omega)\Big]\mathbf{e}_\theta+\frac{1}{\sin(\theta)}\partial_r\Big[r\partial_\phi f(\mathbf{r}, \omega)\Big]\mathbf{e}_\phi \Bigg]
\end{split}
\end{equation}
Therefore, using \cref{Equation: Nonlocal Ohm's Law}, we obtain the following constitutive equation connecting the current density to the total field: 
\begin{equation}
    \frac{\beta^2}{\omega^2}\nabla\cdot(\nabla\cdot\mathbf{J}(\mathbf{r}, \omega))+\mathbf{J}(\mathbf{r}, \omega)=\Bigg[1-\frac{\beta^2l(l+1)}{\omega^2R^2} \Bigg]\mathbf{J}(\mathbf{r}, \omega)\rightarrow \mathbf{J}(\mathbf{r}, \omega)=\frac{\sigma(\omega)}{1-\beta^2l(l+1)/(R^2\omega^2)}\mathbf{E}(\mathbf{r}, \omega)
\end{equation}
Thus, we immediately have the following nonlocal correction to $\sigma_{CPA}$:
\begin{equation}
    \sigma_{CPA}^{NL}(k^+, l)=\Bigg[1-\frac{(\beta/c)^2l(l+1)}{(k^+)^2} \Bigg]\sigma_{CPA}(k^+, l)
    \label{Equation: Nonlocal sigma cpa spheres}
\end{equation}

\subsection{Effect of nonlocality for TE polarized waves}

For TE polarized waves, there is no correction due to nonlocality. This follows simply from the fact that, in the TE polarized case, the surface current admits the following form: 
\begin{equation}
    \mathbf{J}(\mathbf{r}, \omega)=\delta(r-R)\nabla\times (\mathbf{r}f(\mathbf{r}, \omega)),
\end{equation}
which has no divergence. Note that in the above, we did not have to subtract any radial component since the angular momentum operator has no radial component. 

\section{Effect of nonlocality on the CPA conductivity for coated cylinders}
\label{section: nonlocal cpa conductivity for cylinders}
We now turn our attention to cylindrical scatterers. We focus only on the TM polarized channel, as there is no correction to $\sigma_{CPA}(k^+, m)$ for the TE polarized channel. In this case, the current density is given by (with $f(\mathbf{r}, \omega)\equiv j\rho$, with $j$ being a constant--note that $f$ requires a dependence on $\rho$ due to the derivatives with respect to $\rho$ below):
\begin{equation}\mathbf{J}(\mathbf{r}, \omega)=\delta(r-\rho)\Big[\nabla\times(f(\mathbf{r}, \omega)e^{im\theta}\mathbf{e}_z)-\mathbf{e}_\rho\cdot\Big[\nabla\times(f(\mathbf{r}, \omega)e^{im\theta}\mathbf{e}_z) \Big]\Big],
\end{equation}
where, just as in the last section, we explicitly ensured a null normal component of the surface current density. We write the current density explicitly in cylindrical coordinates as follows: 
\begin{align*}\mathbf{J}(\mathbf{r}, \omega)=-\delta(\rho-R)e^{im\theta}\partial_\rho f(\mathbf{r}, \omega)\mathbf{e}_\phi \end{align*}
We now turn to the relevant derivatives of the current (so that we may use \cref{Equation: Nonlocal Ohm's Law}): 
\begin{align*}\nabla\cdot\mathbf{J}(\mathbf{r}, \omega)=-\frac{im}{R}\delta(\rho-R)e^{im\theta}\partial_\rho f(\mathbf{r}), \nabla(\nabla\cdot\mathbf{J}(\mathbf{r}, \omega))=\frac{m^2}{R^2}\delta(\rho-R)e^{im\theta}\partial_\rho f(\mathbf{r}, \omega) \end{align*}
Therefore, using \cref{Equation: Nonlocal Ohm's Law}, we obtain the following: 
\begin{align*}\frac{\beta^2}{\omega^2}\nabla(\nabla\cdot\mathbf{J}(\mathbf{r}, \omega))+\mathbf{J}(\mathbf{r}, \omega)=\Bigg[1-\frac{\beta^2m^2}{\omega^2R^2}\Bigg]\mathbf{J}(\mathbf{r}, \omega)=\sigma(\omega)\mathbf{E}(\mathbf{r}, \omega) \end{align*}
Thus, we obtain the following nonlocal correction to the CPA conductivity:
\begin{equation}
    \sigma_{CPA}^{NL}(k^+, m)=\Bigg[1-\frac{(\beta/c)^2m^2}{(k^+)^2} \Bigg]\sigma_{CPA}(k^+, m)
    \label{Equation: nonlocal sigma cpa for cylinders}
\end{equation}
Note that the ratio, $\beta/c$, is quite small (on the order of $\approx 0.01$), so for $k^+>0.01$, nonlocality can be neglected. 

\section{Lattice of coupled coated cylinders}
\label{section: Dipole lattice model}

In this section, we model a lattice of coated cylinders coupled via their electric polarizabilities. The calculation rests on three quantities: 
\begin{enumerate}
    \item The Dyadic Green's function in two dimensions (\cref{subsection: Green's function}).
    \item The lattice sum of the Dyadic Green's function in two dimensions (\cref{subsection: lattice sum}).
    \item The polarizability of an isolated coated cylinder (\cref{subsection: polarizability of a coated cylinder}).
\end{enumerate}
Using the lattice sum of the Dyadic Green's function, we find the polarizability required to give reflection/transmission coefficients corresponding to CPA (\cref{subsection: lattice cpa condition}).  
\subsection{Dyadic Green's function in two dimensions}
\label{subsection: Green's function}
We start with the Dyadic Green's function in two dimensions which gives the electric field at a point $\mathbf{r}$ due to a dipole at $\mathbf{r}'$. Of course, since we are in two dimensions, the dipole is actually a dipole moment per unit length. The electric field produced by such a line dipole source is given by the following:
\begin{equation}
\mathbf{E}_{\text{dipole}}(\mathbf{r})=\mu_0\omega^2 \Bigg[1+\frac{c^2}{\varepsilon\omega^2}\nabla\nabla \Bigg]\text{G}(\mathbf{r}, \mathbf{r}')\mathbf{p}(\mathbf{r}')=\frac{1}{\varepsilon_0}\Bigg[\frac{\omega^2}{c^2}+\nabla\nabla \Bigg]\text{G}(\mathbf{r}, \mathbf{r}')\mathbf{p}(\mathbf{r}').
\label{Equation: Electric field of dipole}
\end{equation}
where the scalar Green's function in two dimensions is given by an outgoing Hankel function \footnote{Note that in papers (e.g. \cite{rahmanzadeh2021analytical}) that use the opposite time convention, i.e. $e^{i\omega t}$, the outgoing Hankel function will actually be the Hankel function of the second kind. In our work, we consistently stick to the physics time convention of $e^{-i\omega t}$.} \cite{morse1946methods}: 
\begin{equation}
\text{G}(\mathbf{r}, \mathbf{r}')=\frac{i}{4}H_0^{(1)}(|\mathbf{r}-\mathbf{r}'|\omega/c) = \frac{1}{4\pi^2}\int\int \frac{e^{i\mathbf{k}\cdot(\mathbf{r}-\mathbf{r}')}}{|\mathbf{k}|^2-(\omega/c)^2}\text{d}k_x\text{d}k_y,
\label{Equation: Green's function}
\end{equation}
where we also introduced a Fourier representation, which will be useful later when we calculate the dipole lattice sum in \cref{subsection: lattice sum}. In \cref{Equation: Green's function}, $\mathbf{k}=k_x\mathbf{e}_x+k_y\mathbf{e}_y$ and the cylinder is along the $\mathbf{e}_z$ axis. We are ultimately interested in, not just isolated scatterers, but also with periodic systems of scatterers. For periodic systems, we must define two additional Green's functions:
\begin{equation}
\begin{split}
        &G_{\text{lattice}}\equiv \partial_y^2\sum_{m\neq n} G(\mathbf{r}_n, \mathbf{r}_m)\\
        &G_{\text{far-field}}\equiv \partial_y^2\sum_{m\neq n} G(\mathbf{r}, \mathbf{r}_m)
\end{split}
\label{Equation: Greens function sums}
\end{equation}
As we will see shortly, $G_{\text{lattice}}$ governs the interaction of a given cylinder with all other cylinders (for p-polarization), and $G_{\text{far-field}}$ dictates the reflection and transmission coefficients of the periodic system.
\subsection{Far-field Green's function lattice sum}
\label{subsection: lattice sum}
\begin{figure}[htb!]
\centering
\includegraphics[scale=1]{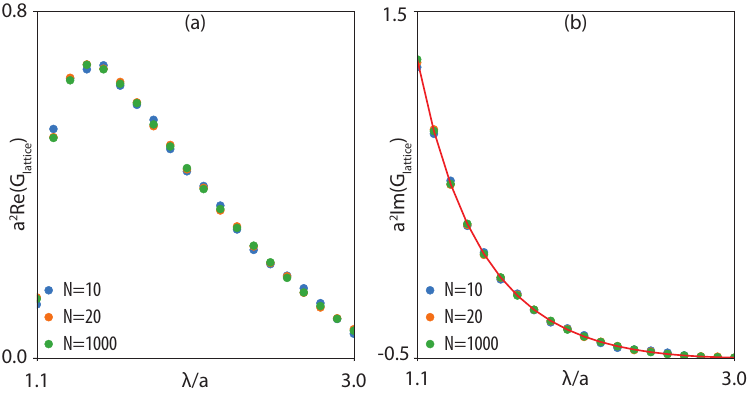}
  \caption{\textbf{Lattice Green's function sum for normal incidence:} We plot the real (a) and imaginary (b) parts of the lattice Green's function sum given by \cref{Equation: Greens function sums} as a function of the ratio of wavelength, $\lambda$, to unit cell length, $a$. In both (a) and (b), $N$ is the number of lattice sites included in the sum. In (b), we additionally overlay the analytic form of the imaginary part of the lattice Green's function sum, as given by \cref{Equation: lattice greens function sum imaginary}. Note that we restrict $\lambda$ to the region $\lambda>a$ in which there is only one propagating channel.}
  \label{fig: supp figure 6}
\end{figure}

In this subsection, we calculate the field radiated from a one-dimensional lattice of dipole cylinders (cylinder axis along the $\mathbf{e}_z$ direction and periodicity direction along the $\mathbf{e}_x$ axis), with each cylinder having the same dipole moment. In particular, we are concerned with the far-field ($|y|>>\lambda$, with $\lambda=2\pi c/\omega$ denoting the wavelength of light under consideration). The quantity we wish to calculate is given by (see \cref{Equation: Green's function}): 
\begin{align*}\sum_n \text{G}(\mathbf{r}, \mathbf{r}_n)=\frac{1}{4\pi^2}\sum_n\int\int \frac{e^{i\mathbf{k}\cdot(\mathbf{r}-\mathbf{r}_n)}}{|\mathbf{k}|^2-(\omega/c)^2}\text{d}k_x\text{d}k_y, \end{align*}
where $\mathbf{r}-\mathbf{r}_n=(x-na, y)$, with $\mathbf{r}_n$ labeling a lattice site at coordinate $(x=na, y=0)$ and $\mathbf{k}=(k_x, k_y)$.
We first perform the sum over the real space lattice sites (of which there are $N$ total), 
\begin{align*}
\sum_{n=0}^{N-1} e^{i(x-na)k_x}=N\sum_{m=-\infty}^\infty e^{ik_x x} \delta_{k_x, 2\pi m/a}=\frac{2\pi N}{L}\sum_{m=-\infty}^\infty e^{ik_x x}\delta\Big(k_x-\frac{2\pi m}{a}\Big),
\end{align*}
where $a$ denotes the lattice constant (i.e. distance between adjacent cylinders) and we used the following relation between the Kronecker delta and the delta function in reciprocal space: 
$\delta_{k,k'}=(2\pi/L)\delta(k-k')$.
Therefore, our lattice-summed scalar Green's function is given by:
\begin{align*}
\sum_n \text{G}(\mathbf{r}, \mathbf{r}_n)=\frac{1}{2\pi a}\int_{-\infty}^{\infty} e^{ik_yy}\sum_{m=-\infty}^{\infty} e^{i2\pi m x/a}\frac{1}{k_y^2+(2\pi m/a)^2-(\omega/c)^2}\text{d}k_y
\end{align*}
By expanding the denominator into a sum of simple poles, our lattice-summed scalar Green's function may be written as follows: 
\begin{equation}
\frac{1}{4\pi a}\sum_{m=-\infty}^{\infty} \frac{e^{i2\pi m x/a}}{\sqrt{(\omega/c)^2-(2\pi m/a)^2}}\int_{-\infty}^{\infty} \Bigg[\frac{e^{ik_yy}}{k_y-\sqrt{(\omega/c)^2-(2\pi m/a)^2}} -\frac{e^{ik_yy}}{k_y+\sqrt{(\omega/c)^2-(2\pi m/a)^2}} \Bigg]\text{d}k_y
\end{equation}
We focus on the term corresponding to lowest order diffraction, i.e. $m=0$. In order to evaluate the contour integral, we need to add a small imaginary part to the angular frequency, $\omega\rightarrow\omega+i\delta$. This corresponds to a Green's function having no poles in the upper half of the complex plane (i.e. being causal). For $y>0$, we must close the contour in the upper half of the complex plane, which results in picking up a residue at $k_y=(\omega+i\delta)/c$. For $y<0$, we close the contour in the lower half of the complex plane and pick up a residue at $k_z=-(\omega+i\delta)/c$. Our lattice sum then gives: 
\begin{equation}
\sum_n \text{G}(\mathbf{r}, \mathbf{r}_n)=\frac{ie^{i|y|\omega/c}}{2 (a\omega/c)} +\text{terms corresponding to } m\neq 0
\label{Equation: lattice sum result}
\end{equation}
Note that for $y>0$, this goes as $e^{i2\pi y/\lambda}$ (plane wave propagating in the positive $y$ direction). And for $y<0$, it corresponds to a plane wave propagating in the minus $y$ direction. Now we consider the terms with $m\neq 0$. If $2\pi/a >\omega/c\leftrightarrow \lambda > a$, we have necessarily that all the poles will lie on the imaginary axis. In particular, for each such term, we will obtain a contribution to the lattice sum given by: 
\begin{equation}
\frac{ie^{i2\pi m x/a}}{2a\sqrt{(\omega/c)^2-(2\pi m/a)^2}}e^{i\sqrt{(\omega/c)^2-(2\pi m/a)^2}|y|} 
\end{equation}

However, note that all of these terms are evanescent and do not contribute in the far-field or carry power (and we therefore do not have to consider these contributions to the reflection/transmission coefficients). In particular, if each cylinder has a dipole moment $p\mathbf{e}_x$, the electric field will be given by (using \cref{Equation: lattice sum result} in conjunction with \cref{Equation: Electric field of dipole}): 
\begin{equation}
    \mathbf{E}(\mathbf{r}, \omega)=-G_{\text{far-field}}(\mathbf{r})p\mathbf{e}_x=\frac{\omega^2}{\varepsilon_0c^2}\frac{ie^{i|y|\omega/c}}{2(a\omega/c)}p\mathbf{e}_x
    \label{Equation: lattice sum result 2}
\end{equation}
\subsection{Lattice Green's function sum and its analytic evaluation}
In this section, we wish to find an analytic expression for the imaginary part of the lattice Green's function, $G_\text{lattice}$ (given by \cref{Equation: Green's function})  corresponding to p-polarization. Concretely, we wish to calculate the sum of the quantity $\partial_y^2\Im \text{G}(\mathbf{r}_n, \mathbf{r}_m)$ over all $m\neq n$. 
We first calculate the imaginary part of the Green's function corresponding to the self-energy:
\begin{equation}
\begin{split}
&\lim_{\mathbf{r}\rightarrow\mathbf{r}_0}\partial_y^2\Im \text{G}(\mathbf{r}, \mathbf{r}_0)=\lim_{\mathbf{r}\rightarrow \mathbf{r}_0}\frac{1}{4}\partial_y^2J_0(\omega|\mathbf{r}-\mathbf{r}_0|/c)=\\&\lim_{r\rightarrow 0}\frac{1}{4}\Bigg[\frac{(\omega/c)J_0'(r\omega /c)}{r}(1-(y/r)^2)+(\omega/c)^2(y/r)^2J_0''(r\omega/c)\Bigg]\approx -\frac{1}{8}(\omega/c)^2,
\end{split}
\end{equation}
where we used $J_0'(z)=-J_1(z)\approx -z/2+O(z^3)$, which allowed us to write \begin{equation}
    \frac{\omega}{c}\Bigg[(\omega/c)J_0''(r\omega/c )-J'_0(r\omega/c)/r\Bigg](y/r)^2\approx (\omega/c)^2O((r\omega/c)^2)(y/r)^2,
\end{equation}
which vanishes in the limit $r\rightarrow 0$ (regardless of how we take the limit). Therefore, the lattice sum's imaginary part will be: 
\begin{equation}-\lim_{\mathbf{r}\rightarrow\mathbf{r}_0}\partial_y^2\Im \text{G}(\mathbf{r}, \mathbf{r}_0)+\sum_{m}\partial_y^2 \Im \text{G}(\mathbf{r}_n, \mathbf{r}_m)=-(\omega/c)/(2a)+(\omega/c)^2/8
\label{Equation: lattice greens function sum imaginary}
\end{equation}
The first term in the above, $-k/2a$ corresponds to the zeroth-order diffraction term in the lattice Green's function (which we already derived in \cref{Equation: lattice sum result}). In \cref{fig: supp figure 6} we show the real and imaginary parts of the lattice Green's function sum. From \cref{fig: supp figure 6}(b), we clearly see that numerical evaluation over lattice sites matches well with the analytic expression derived in \cref{Equation: lattice greens function sum imaginary}.

\subsection{Retarded polarizability of a coated cylinder (perpendicular to the cylinder's axis)}
\label{subsection: polarizability of a coated cylinder}

\begin{figure}[htb!]
\centering
\includegraphics[scale=1]{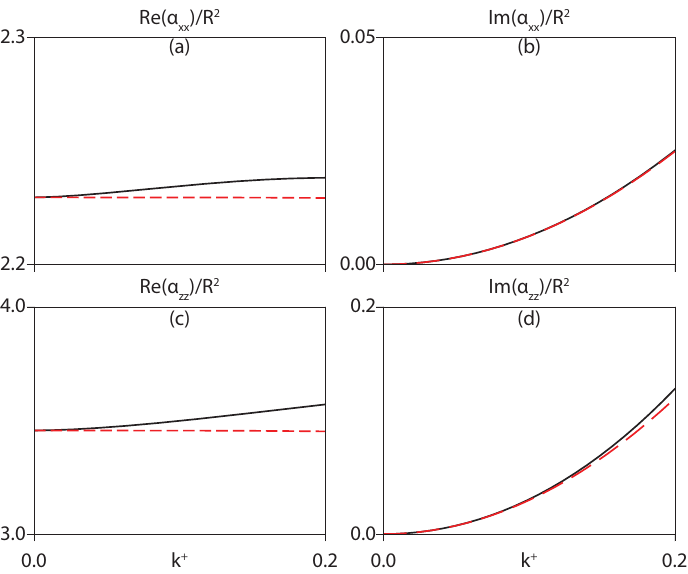}
  \caption{\textbf{Retarded polarizability of isolated cylindrical scatterers:} Polarizability perpendicular to (a, b) and parallel to (c, d) the cylinder axis. Dashed red lines correspond to the small cylinder limit, given by \cref{Equation: alphaxx quasistatic} for (a, b) and \cref{Equation: alphazz quasistatic} for (c, d). Bulk permittivity is set to $\varepsilon=2.1$ for all subplots. We put explicit subscripts (e.g. ``xx") unlike in the text to disambiguate between the parallel and perpendicular polarizabilities.}
  \label{fig: supp figure 5}
\end{figure}
In this section, we calculate the retarded polarizability of a cylinder (perpendicular to the cylinder axis) in terms of the cylindrical Mie coefficients. We do this as follows: First, we calculate the scattered field of a dipole, with polarizability $\alpha$, due to an incident magnetic field polarized along the cylinder's axis, $\mathbf{H}=H_{\text{inc, z}}\mathbf{e}_z$. We then match this to the scattered field that is predicted by the Mie coefficients. 
Concretely, we consider an electric dipole, $\mathbf{p}=\varepsilon_0\alpha_{xx}E_{inc, x}\mathbf{e}_x+\varepsilon_0\alpha_{yy}E_{inc, y}\mathbf{e}_y$, where $\alpha_{xx}=\alpha_{yy}\equiv \alpha$ and $\mathbf{E}_{\text{inc}}=(i/\omega\varepsilon_0)(\partial_y H_{\text{inc}, z}, -\partial_xH_{\text{inc}, z} )$ (and the derivatives are understood to be evaluated at the dipole's origin). The current density of such a dipole is given by $\mathbf{J}=-i\omega\mathbf{p}\delta(\mathbf{r}-\mathbf{r}_0)=\delta(\mathbf{r}-\mathbf{r}_0)\alpha(\partial_y H_{\text{inc}, z}, -\partial_xH_{\text{inc}, z} )$, where $\mathbf{r}_0$ denotes the location of the dipole. At this point, we may find the scattered magnetic field, which is also polarized along the cylinder's axis: $H_{\text{scat}, z}=\partial_x G(\mathbf{r}, \mathbf{r}_0) J_y(\mathbf{r}_0)-\partial_y G(\mathbf{r}, \mathbf{r}_0) J_x(\mathbf{r}_0)$ (where it's understood that the derivatives on the Green's function are with respect to the observation point, $\mathbf{r}$). Thus, we may write the scattered field in terms of the incident field as follows:  
\begin{align*}
&\mathbf{H}_{\text{scat}}(\mathbf{r})=-\alpha\Big[\partial_x \text{G}(\mathbf{r}, \mathbf{r}_0)\partial_x H_{\text{inc}, z}(\mathbf{r}_0)+\partial_y\text{G}(\mathbf{r}, \mathbf{r}_0)\partial_yH_{\text{inc}, z}(\mathbf{r}_0)  \Big]\mathbf{e}_z=\\&\frac{i\alpha \omega}{4c}H_1^{(1)}(\omega|\mathbf{r}-\mathbf{r}_0|/c)\Bigg[\cos(\theta)\partial_x H_{\text{inc}, z}(\mathbf{r}_0)+\sin(\theta)\partial_yH_{\text{inc}, z}(\mathbf{r}_0) \Bigg]\mathbf{e}_z\end{align*}
In order to avail ourselves of the Mie coefficients, we take  $H_{\text{inc}, z}(\mathbf{r})=2J_1(k|\mathbf{r}-\mathbf{r}_0|)e^{i\theta}$ as the incident magnetic field. In addition, in the above we introduced the angle, $\theta$, which is given by $\tan(\theta)=(y-y_0)/(x-x_0)$. In order to calculate the derivatives, we use the chain rule to write  $\partial_x = \cos(\theta)\partial_r-\frac{\sin(\theta)}{r}\partial_\theta$ and $\partial_y =\sin(\theta)\partial_r +\frac{\cos(\theta)}{r}\partial_\theta$, from which we immediately obtain: 
\begin{align*}
s-1= \frac{i\alpha k^2}{2}\Bigg[ J'_1(0)\Bigg]\rightarrow \frac{\sqrt{\varepsilon} J_1(k^-)J'_l(k^+)-J'_1(k^-)J_1(k^+)}{\sqrt{\varepsilon} J_1(k^-)H_1^{(1)}(k^+)'-J'_1(k^-)H_1^{(1)}(k^+)}=-\frac{i\alpha \omega^2}{4c^2}\Bigg[ J'_1(0)\Bigg],
\end{align*}
where we have introduced the Mie scattering coefficient, $s$. In order to sanity check our results, we examine the small cylinder, quasistatic, limit. In this limit, we may set $H^{(1)}_1\approx iY_1, H_1^{(2)}\approx -iY_1$, giving us the following expression: 
\begin{align*}\frac{\sqrt{\varepsilon}J_1(k^-)J'_1(k^+)-J'_1(k^-)J_1(k^+)}{\sqrt{\varepsilon} J_1(k^-)Y'_1(k^+)-J'_1(k^-)Y_1(k^+)}\approx \frac{\alpha \omega^2}{4c^2}J_1'(0) \end{align*}
We use the expansions, $J_1(z)\approx z/2, Y_1(z)\approx -2/(\pi z)$, from which we obtain: 
\begin{equation}
\frac{\varepsilon k^+/4-k^+/4}{\varepsilon/(\pi k^+)+1/(\pi k^+)}=\frac{\alpha \omega^2}{8c^2}\rightarrow \alpha = (2\pi R^2)\Bigg[\frac{\varepsilon-1}{\varepsilon+1}\Bigg],
\label{Equation: Static polarizability of uncoated cylinder perpendicular to axis without radiative loss}
\end{equation}
where we used $k^+\equiv R\omega/c$ to  write our final expression for the polarizability in terms of the scatterer's radius, $R$. We may amend \cref{Equation: Static polarizability of uncoated cylinder perpendicular to axis without radiative loss} to introduce the effect of radiative loss (to maintain consistency with the optical theorem \cite{abajo2007collective, berman2006polarizability}), which gives us:
\begin{equation}
\frac{\varepsilon k^+/4-k^+/4}{\varepsilon/(\pi k^+)+1/(\pi k^+)-i\varepsilon^2(k^+)/4+i(k^+)/4}=\frac{\alpha \omega^2}{8c^2} \rightarrow\alpha= \frac{(\varepsilon-1)(2\pi R^2)}{(\varepsilon+1)-i(\varepsilon-1)\pi(k^+)^2/4}
\label{Equation: alphaxx quasistatic}
\end{equation}
This result is the same as Equation 17 in \cite{gomez2006extraordinary}. Note that if $\varepsilon\approx -1$, but has no dissipation, we get $\alpha \approx 8iR^2/(k^+)^2$, from which we immediately get $s\approx -1$. This is why if the real part of the permittivity/imaginary part of the conductivity is tuned to its CPA value (but $\varepsilon''=0, \sigma'=0$), we get a perfect phase shift in the subwavelength regime. We show how the fully retarded perpendicular polarizability differs from the quasistatic result in \cref{fig: supp figure 5}(a, b).

Now that we have shown that our formalism for computing the polarizability is indeed sound, we reiterate the relation between the dipolar scattering coefficient to the polarizability (perpendicular to the axis): 
\begin{equation}
\alpha = -\frac{4ic^2}{\omega^2}(s-1)
\label{Equation: polarizability to s}
\end{equation}

\subsection{Retarded polarizability of a coated cylinder (parallel to the cylinder's axis)}

For completeness, we also include the derivation of the polarizability along the cylinder axis. In contrast to the polarizability perpendicular to the cylinder axis, which is determined by the TM scattering coefficients, the polarizability parallel to the cylinder axis is determined by the TE scattering coefficients (we note that these are implemented in our code \cite{Coherent-Perfect-Absorption} by the function \texttt{e\_z\_cpa} with the angular momentum number set to unity). Besides this point, the derivation is more-or-less similar to the one given in the previous subsection. The scattered electric field is given by $E_{\text{scat}}(\mathbf{r})=(\omega/c)^2G_0(\mathbf{r}, \mathbf{r}_0)\alpha_{zz}E_{\text{inc}}(\mathbf{r}_0)$, where $E_{\text{inc}}(\mathbf{r})=2J_0(\omega|\mathbf{r}-\mathbf{r}_0|/c)$ (giving $E_{\text{inc}}(\mathbf{r}_0)=2$) and, from the Mie coefficients, $E_\text{scat}(\mathbf{r})=(s-1)H^{(1)}_0(k|\mathbf{r}-\mathbf{r}_0|)$ (note that both the incident and the scattered electric fields are polarized along the cylinder axis). From these three relations (and our expression for the Green's function given by \cref{Equation: Green's function}), we readily obtain the following relation: 
\begin{equation}
\begin{split}
    s-1=i(\omega/c)^2(\alpha_{zz}/2)\rightarrow \frac{J_0(k^-)J_0(k^+)'-\sqrt{\varepsilon}J_0(k^+)J'_0(k^-)}{-J_0(k^-)H_0^{(1)}(k^+)'+\sqrt{\varepsilon}H_0^{(1)}(k^+)J'_0(k^-)}=i(\omega/c)^2\alpha_{zz}/4\\\rightarrow \frac{-J_0(k^-)J_1(k^+)+\sqrt{\varepsilon}J_0(k^+)J_1(k^-)}{J_0(k^-)Y_1(k^+)-\sqrt{\varepsilon}Y_0(k^+)J_1(k^-)}\approx-(\omega/c)^2\alpha_{zz}/4\rightarrow \\\frac{-(k^+/2)+\varepsilon(k^+/2)}{-2/(\pi k^+)}=-\pi(\varepsilon-1)(k^+)^2/4=-(\omega/c)^2\alpha_{zz}/4\rightarrow\alpha_{zz}=\pi R^2(\varepsilon-1)
    \end{split}
    \label{Equation: exact alphazz}
\end{equation}
Note that this result could have been found without actually doing any math. If a static electric field is applied along the cylinder axis, there is no depolarization field, so the total dipole moment per unit length will simply be the area of the cylinder's cross section, $\pi R^2$, multiplied by the dielectric susceptibility, $\varepsilon-1$, of the constituent material of the cylinder. If we include radiative loss (which amounts to going one order higher in the expansion of the Bessel functions of the first kind in the denominator), this result changes to: 
\begin{equation}
\frac{(\varepsilon-1)(k^+/2)}{-2/(\pi k^+)-i(k^+/2)+i\varepsilon(k^+/2)}=-(\omega/c)^2\alpha_{zz}/4\rightarrow \frac{(\varepsilon-1)\pi R^2}{1-i(\varepsilon-1)\pi(k^+)^2/4} =\alpha_{zz}
\label{Equation: alphazz quasistatic}
\end{equation}
This result is the same as Equation 3 of \cite{gomez2006extraordinary}. We plot the real (imaginary) components of $\alpha_{zz}$ in \cref{fig: supp figure 5}(c) ((d)). There, we compare the quasistatic+radiative loss expression, given by \cref{Equation: alphazz quasistatic}, to the exact result given by \cref{Equation: exact alphazz}. 

\subsection{Polarizability and conductivity required for CPA}
\label{subsection: lattice cpa condition}

Although for general angles of incidence, it is more convenient to work with the magnetic field (since it is purely polarized along the cylinder axis while the electric field will have components along both the $\mathbf{e}_x$ and $\mathbf{e}_y$), for normal incidence it is just as convenient to work with the electric field (since it is purely polarized along the $x$ axis). In this case, we take all the cylinders to be identical dipoles along the $x$ axis. We write down a self consistency condition for each dipole moment, $\mathbf{p}_n$ (with $E_0$ denoting the incident electric field): 
\begin{equation}\mathbf{p}_n\cdot \mathbf{e}_x=\varepsilon_0\alpha\Bigg[E_0+(\omega^2\mu_0)\Bigg[1+\frac{c^2}{\omega^2}\partial_x^2\Bigg] \sum_{m\neq n}\frac{i}{4}H_0^{(1)}(\omega|\mathbf{r}_n-\mathbf{r}_m|/c)\mathbf{p}_m\cdot\mathbf{e}_x\Bigg]
\label{Equation: Self consistency for dipole moment 1}
\end{equation}
The first term on the right side of \cref{Equation: Self consistency for dipole moment 1}, $\varepsilon_0\alpha E_0$, is the contribution to the dipole moment by the external field. The remaining sum, which we will show is related to $G_{\text{lattice}}$, corresponds to the dipole moment induced by fields scattered by all of the other cylinders. At this point, we may move all the terms proportional to the dipole moment to the left side of the equation to write out. In doing so, we note that, for normal incidence, $\mathbf{p}_n=\mathbf{p}$ (since there is no change of phase from one lattice site to the next).  
\begin{equation}
\mathbf{e}_x\cdot\mathbf{p}\Bigg[1-\varepsilon_0\alpha(\omega^2\mu_0)\Bigg[1+\frac{c^2}{\omega^2}\partial_x^2\Bigg] \sum_{m\neq n}\frac{i}{4}H_0^{(1)}(\omega|\mathbf{r}_n-\mathbf{r}_m|/c)\Bigg]=\varepsilon_0\alpha E_0
\label{Equation: Self consistency for dipole moment 2}
\end{equation}
At this point, we are essentially done. \cref{Equation: Self consistency for dipole moment 2} uniquely determines the dipole moment given the external field. Once the dipole moment of each cylinder is determined, \cref{Equation: lattice sum result} uniquely determines the total scattered field (and, therefore, the reflection and transmission coefficients). As such, at this point we have all the necessary ingredients to determine the necessary perpendicular polarizability, $\alpha$, and thus the surface conductivity to achieve CPA. To write things a bit more explicitly, however, we make a few simplifications so that we may write out the CPA condition in a compact form. We start by simplifying \cref{Equation: Self consistency for dipole moment 2} by using the following identity: 
\begin{equation}
\Bigg[1+\frac{c^2}{\omega^2}\partial_x^2\Bigg] \sum_{m\neq n}\frac{i}{4}H_0^{(1)}(\omega|\mathbf{r}_n-\mathbf{r}_m|/c)=-\frac{c^2}{\omega^2}\partial_y^2\sum_{m\neq n}\frac{i}{4}H_0^{(1)}(\omega|\mathbf{r}_n-\mathbf{r}_m|/c),
\label{Equation: Greens function simplification from Helmholtz}
\end{equation}
which simply follows from the fact that the scalar Green's function satisfies the Helmholtz equation. The relation given by \cref{Equation: Greens function simplification from Helmholtz} allows us to write the self consistency equation for the dipole moment as follows: 
\begin{equation}
\mathbf{e}_x\cdot\mathbf{p}\Bigg[1+\alpha\partial_y^2\sum_{m\neq n}\frac{i}{4}H_0^{(1)}(\omega|\mathbf{r}_n-\mathbf{r}_m|/c)\Bigg]=\varepsilon_0\alpha E_0\rightarrow \mathbf{e}_x\cdot\mathbf{p}=\frac{\varepsilon_0\alpha E_0}{1+\alpha G_\text{lattice}}
\label{Equation: Self consistency for dipole moment 3}
\end{equation}
Note the appearance, as promised at the outset of this section, of $G_\text{lattice}$. From \cref{Equation: Self consistency for dipole moment 3}, we may easily obtain the scattered electric field, $\mathbf{E}_{\text{scat}}(\mathbf{r})\equiv E_{\text{scat}}(y)\mathbf{e}_x$, which is given by: 
\begin{equation} 
E_{\text{scat}}(y)= -\frac{\varepsilon_0\alpha E_0G_\text{far-field}}{1+\alpha G_\text{lattice}}=\frac{i(a\omega/c)}{2a^2}\frac{E_0e^{-i|y|\omega/c}}{1/\alpha -\sum_{m\neq 0} (i\omega/(4c|\mathbf{r}_m|))H_1^{(1)}(k|\mathbf{r}_m|)}
\end{equation}
From this, we may read off the reflection and transmission coefficients: 
\begin{equation} 
\begin{split}
&r(\omega) = \frac{i(a\omega/c)}{2a^2}\frac{1}{1/\alpha -\sum_{m\neq 0} (i\omega/(4c|\mathbf{r}_m|))H_1^{(1)}(\omega|\mathbf{r}_m|/c)}\\ &t(\omega) = 1+ \frac{i(a\omega/c)}{2a^2}\frac{1}{1/\alpha -\sum_{m\neq 0} (i\omega/(4c|\mathbf{r}_m|))H_1^{(1)}(\omega|\mathbf{r}_m|/c)}
\end{split}
\label{Equation: Reflection and transmission for dipole lattice}
\end{equation}
In order to get symmetric CPA, we require $r+t=0$ \cite{pu2012ultrathin}. Since $t=1+r$ from \cref{Equation: Reflection and transmission for dipole lattice}    , the symmetric CPA condition is equivalently given by $r(\omega)=-1/2$. We note that the antisymmetric CPA condition, $r-t=0$, is obviously not possible.

We now solve for the polarizability that will give us $r=-1/2$, which we denote by $\alpha_{CPA}$. From \cref{Equation: Reflection and transmission for dipole lattice}, we have the following closed-from solution for $\alpha_{CPA}$
\begin{equation}
  \frac{1}{\alpha_{CPA}}=-\frac{i\omega}{ac}+\sum_{m\neq 0} \frac{i\omega}{4a|m|c}H_1^{(1)}(\omega a|m|/c)
\end{equation}
Once $\alpha_{CPA}$ is calculated, the necessary scattering coefficient, $s$, follows from \cref{Equation: polarizability to s}. Lastly, we obtain the surface conductivity (also in closed form) from the value of $s$.

\section{Polarizability of a coated sphere}
\label{section: sphere polarizability from mie}

In this section, we calculate the retarded polarizability, $\alpha(\omega)$, of a coated sphere in terms of the TM scattering coefficients, $s_{l}$, which we derived in \cref{section: sphere exact cpa}. To do so, we consider an incident field with unit angular momentum ($l=1$) and, without loss of generality, azimuthal quantum number set to zero ($m=0$). That is, our incident electric and magnetic fields, $\mathbf{E}_\text{inc}(\mathbf{r})$, $\mathbf{H}_\text{inc}(\mathbf{r})$ are given by (using \cref{Equation: spherical bessel function approximations}):
\begin{equation}
\begin{split}
    &\mathbf{E}_{\text{inc}}(\mathbf{r})=\nabla\times(j_1(r\omega/c)\mathbf{r}\times \nabla \cos(\theta))=-\frac{2 j_1(r\omega/c)\cos(\theta)}{r}\mathbf{e}_\mathbf{r}+\frac{1}{r}\partial_r\Big(rj_1(r\omega/c)\Big)\sin(\theta)\mathbf{e}_\theta 
    \\ &\rightarrow \mathbf{E}_\text{inc}(0)\approx -\frac{2}{3}\frac{\omega}{c}\Bigg[\cos(\theta)\mathbf{e}_\mathbf{r}-\sin(\theta)\mathbf{e}_\theta\Bigg]=-\frac{2}{3}\frac{\omega}{c}\mathbf{e}_z, 
\\&\mathbf{H}_{\text{inc}}(\mathbf{r})=i\omega\varepsilon_0 j_1(r\omega/c)\sin(\theta)\mathbf{e}_\phi,
    \end{split}
\end{equation}
where $\mathbf{E}_{\text{inc}}(0)$ is the electric field at the location of the polarizable sphere ($\mathbf{r}=0$). The scattered magnetic field can by easily calculated by taking the curl of the free space Green's function and by noting that the sphere will be given by an oscillating electric dipole moment, $\mathbf{p}=\varepsilon_0\alpha\mathbf{E}_{\text{inc}}(0)=-\frac{2}{3}\varepsilon_0\alpha\frac{\omega}{c}\mathbf{e}_z$. Thus we obtain:
\begin{equation}
\begin{split}
    &\mathbf{H}_\text{scattered}(\mathbf{r})=\frac{2i}{3}\frac{\omega^2}{c}\varepsilon_0\alpha\nabla\times\frac{e^{i\omega r/c}}{4\pi r}\mathbf{e}_z=\frac{2i}{3}\frac{\omega^2}{c}\varepsilon_0\alpha\Bigg[\nabla\times \frac{e^{i\omega r/c}}{4\pi r}\Big(\cos(\theta)\mathbf{e}_\mathbf{r}-\sin(\theta)\mathbf{e}_\theta \Big)\Bigg]=
        \\& \frac{i}{6\pi}\frac{\omega^2}{c}\varepsilon_0\alpha\sin(\theta)\Bigg[1-ir\omega/c \Bigg]\frac{e^{i\omega r/c}}{r^2}\mathbf{e}_\phi=-\frac{1}{6\pi}\frac{\omega^4}{c^3}\varepsilon_0\alpha\sin(\theta)h_1^{(1)}(r\omega/c)\mathbf{e}_\phi
        \end{split}
\end{equation}
From this, we obtain the total magnetic field and, therefore, the polarizability in terms of the scattering coefficient for $l=1$ ($s\equiv s_1)$: 
\begin{equation}
\mathbf{H}_\text{total}=i\omega\varepsilon_0\sin(\theta)\Bigg[j_1(r\omega/c)+i\frac{\alpha}{6\pi}\Bigg(\frac{\omega}{c}\Bigg)^3h_1^{(1)}(r\omega/c) \Bigg]\mathbf{e}_\phi\rightarrow s-1=\frac{i\alpha}{3\pi }\Bigg(\frac{\omega}{c}\Bigg)^3
\end{equation}
This is the same result as in \cite{christensen2015localized, alu2011causality}. 

\section{Finite temperature effects on $E_{F, CPA}$ and $\omega_{CPA}$}
\label{section: Finite temperature effects}

In this section, we investigate the effect of finite temperature on the Fermi energy, $E_{F, CPA}$, and the angular frequency, $\omega_{CPA}$, at which we may achieve CPA at a given value of $k^+$ (and for a choice of $l/m$). Without loss of generality, we focus, in this section, on spheres (but we stress that the derivations in this section are not restricted, in any way, to the spherical geometry). At finite temperature, for graphene, we must amend the Drude conductivity as follows \cite{christensen2015localized}:
\begin{equation}
\sigma(\omega)=\frac{ie^2E_F}{\pi \hbar^2(\omega+i/\tau)}=\frac{4i\alpha c\varepsilon_0 E_F}{\hbar(\omega+i/\tau)}\rightarrow \frac{e^2}{\pi\hbar}\Bigg[\frac{2ik_BT}{\hbar(\omega+i/\tau)}\ln(2\cosh(E_F/2k_BT)) \Bigg],
\end{equation}
Clearly, the conductivity at finite temperature converges to the zero temperature result in the limit $E_F/k_BT\gg 1$ (for which $2\cosh(E_F/2k_BT)\rightarrow e^{E_F/2k_BT}$).
Since the finite temperature conductivity still has a Drude form, the CPA frequency, $\omega_{CPA}$, is not altered. That is, we still have the following, for any temperature:
\begin{equation}
\omega_{CPA}=\frac{1}{\tau}\Bigg[\frac{\sigma_{CPA}''(k^+, l)}{\sigma_{CPA}'(k^+, l)} \Bigg]
\end{equation}
We now find the Fermi energy corresponding to CPA at finite temperature, $T$, which has to satisfy the following equation:
\begin{equation}
\begin{split}
    &\sigma_{CPA}'(k^+, l)=\frac{8\alpha c\varepsilon_0 k_BT\tau \ln\Big[2\cosh(E_{F, CPA}/2k_BT) \Big]}{\hbar\Big[1+(\sigma_{CPA}''(k^+, l)/\sigma_{CPA}'(k^+, l))^2\Big]}\rightarrow \\ &\exp\Bigg[ \frac{E_{F, CPA}(k^+, T)}{2k_BT}\Bigg]+\exp\Bigg[-\frac{E_{F, CPA}(k^+, T)}{2k_BT}\Bigg] = \exp\Bigg[ \frac{E_{F, CPA}(k^+, T=0)}{2k_BT}\Bigg],
    \end{split}
\end{equation}
where $E_{F, CPA}(k^+, T=0)$ is the zero temperature value for the CPA Fermi energy given by Eq. (\equationefomega) of the main text (we have chosen to reintroduce the implicit $k^+$ dependence). This may be cast as a quadratic equation with the following solution:
\begin{equation} 
E_{F, CPA}(k^+, T)=2k_BT\ln \Bigg[\frac{f(k^+, T)\pm\sqrt{f(k^+, T)^2-4}}{2}\Bigg] = \pm 2k_BT\ln\Bigg[\frac{f(k^+, T)+\sqrt{f(k^+, T)^2-4}}{2}\Bigg] , 
\label{Equation: EF for finite temperature}
\end{equation}
where we have introduced the quantity, $f(k^+, T)$, which we define as follows:
\begin{equation}
f(k^+, T)\equiv \exp\Big[E_{F, CPA}(k^+, 0)/(2k_BT) \Big],
\end{equation}
We draw the reader's attention to the fact that we must require the following inequality to be satisfied:
 \begin{align*}
 T< \frac{E_{F, CPA}(k^+, 0)}{2 \ln(2)k_B}
 \end{align*}
 in order for \cref{Equation: EF for finite temperature} to yield a real valued (and, thus, sensible) solution for $E_{F, CPA}$. This sets an intrinsic upper bound for the temperature at which CPA may be achieved. 

\section{Critical coupling by isolated scatterers as multichannel CPA in the plane wave basis}
\label{Section: Single scatterer plane wave cpa}
In this section, we show that the absorption of a spherical/cylindrical wave by an isolated sphere/cylinder can be  thought of as absorption of a coherent superposition of incident plane waves. Our final results are given by \cref{Equation: Single scatterer plane wave CPA} (for cylinders) and \cref{Equation: Rayleigh expansion 4} (for spheres). 
\subsection{Isolated cylindrical scatterers as multichannel perfect absorbers in the plane wave basis}
Consider the TM polarized channels of an isolated cylinder. We excite the cylinder with the following magnetic field profile (oscillating at angular frequency $\omega$): 
\begin{equation}
    H_z(\mathbf{r})=\frac{(-i)^m}{2\pi}\int_0^{2\pi} e^{im\theta_k}e^{i\mathbf{k}\cdot\mathbf{r}}\text{d}\theta_k, 
    \label{Equation: Single scatterer plane wave CPA}
\end{equation}
with $\mathbf{k}=(\omega/c)(\cos(\theta_k), \sin(\theta_k))$ and $\mathbf{r}=r(\cos(\theta), \sin(\theta))$. We claim that the incident field given by \cref{Equation: Single scatterer plane wave CPA}  will be perfectly absorbed if $s_m=0$ \footnote{$s_m$ is the scattering coefficient for waves in the TM channel with angular momentum $m$}, leaving no outgoing flux.  This may be proven easily by using the Jacobi-Anger expansion: 
\begin{equation}
    \int_0^{2\pi} e^{im\theta_k}e^{i\mathbf{k}\cdot\mathbf{r}}\text{d}\theta_k=\int_0^{2\pi}e^{im\theta_k}e^{ikr\cos(\theta-\theta_k)}\text{d}\theta_k=\sum_{n=-\infty}^{\infty} \int_0^{2\pi}e^{im\theta_k}(i)^m e^{in(\theta-\theta_k)}J_n(kr)\text{d}\theta_k=2\pi (i)^m e^{im\theta}J_m(kr)
\end{equation}
This immediately proves that $H_{z}(\mathbf{r})$, given by \cref{Equation: Single scatterer plane wave CPA}, is just given by a Bessel function of the first kind of order $m$; i.e. $H_z(\mathbf{r})=J_m(kr)e^{im\theta}$. Thus, if $s_m=0$, we immediately see that the total field outside the scatterer will be given by an incoming Hankel function. Note that similar results for the TE channels of a cylinder follow analogously from the same arguments, just with the replacement $H_z\rightarrow E_z$. Furthermore, note that \cref{Equation: Single scatterer plane wave CPA} is a simple linear combination of plane waves propagating at all possible angles. The plane waves all have the same amplitude but different $\mathbf{k}$-dependent phases, $e^{im\theta_k}$. 

\subsection{Isolated spherical scatterers as multichannel perfect absorbers in the plane wave basis}
We now turn to the case of spheres. As in the previous subsection, we only consider the TM channel, since results for the TE channel may be derived immediately by replacing the magnetic field with the electric field. We start with the Rayleigh expansion of a (scalar) plane wave in three dimensions: 
\begin{equation}
e^{i\mathbf{k}\cdot\mathbf{r}}=4\pi\sum_{l=0}^\infty\sum_{m=-l}^{l}i^l j_l(kr)Y_{lm}(\hat{\mathbf{k}})Y^*_{lm}(\hat{\mathbf{r}}),
\label{Equation: Rayleigh expansion}
\end{equation}
where $\hat{\mathbf{k}}, \hat{\mathbf{r}}$ are unit vectors and $k=\omega/c$. From \cref{Equation: Rayleigh expansion} and the orthogonality of the spherical harmonics, we obtain: 
\begin{equation}
    \frac{i^l}{4\pi}\int e^{-i\mathbf{k}\cdot\mathbf{r}}Y_{lm}(\hat{\mathbf{k}})\text{d}\Omega_\mathbf{k}=j_l(kr)Y_{lm}(\hat{\mathbf{r}}),
    \label{Equation: Rayleigh expansion 2}
\end{equation}
where $d\Omega_\mathbf{k}=\sin(\theta_\mathbf{k})\text{d}\theta_\mathbf{k}\text{d}\phi_\mathbf{k}$ is the differential phase space element in momentum space. Up to now, we have been dealing with scalar quantities. We ultimately wish to obtain a plane wave expansion for the quantity $\mathbf{r}\times\nabla (j_l(kr)Y_{lm}(\hat{\mathbf{r}}))$ (since this is the form of a TM polarized magnetic field), which amounts to applying the angular momentum operator on \cref{Equation: Rayleigh expansion 2}: 
\begin{equation}
    -\frac{i^{l+1}}{4\pi}\int (\mathbf{r}\times\mathbf{k})e^{-i\mathbf{k}\cdot\mathbf{r}}Y_{lm}(\hat{\mathbf{k}})\text{d}\Omega_\mathbf{k}=\mathbf{r}\times\nabla (j_l(kr)Y_{lm}(\hat{\mathbf{r}}))
    \label{Equation: Rayleigh expansion 3}
\end{equation}
At this point, it may seem that we're done. However, note that \cref{Equation: Rayleigh expansion 3} has a pesky prefactor of $\mathbf{r}\times\mathbf{k}$, so it is not a true plane wave expansion. To obtain our desired plane wave expansion we write the cross product on the left side as a derivative: 
\begin{equation}
    \frac{i^{l+2}}{4\pi}\int Y_{lm}(\hat{\mathbf{k}}) (\mathbf{k}\times\nabla_\mathbf{k})e^{-i\mathbf{k}\cdot\mathbf{r}}\text{d}\Omega_\mathbf{k}=\mathbf{r}\times\nabla (j_l(kr)Y_{lm}(\hat{\mathbf{r}}))
\end{equation}
Writing $\nabla_\mathbf{k}\Big(Y_{lm}(\hat{\mathbf{k}})e^{-i\mathbf{k}\cdot\mathbf{r}} \Big)=Y_{lm}(\hat{\mathbf{k}})\nabla_\mathbf{k}e^{-i\mathbf{k}\cdot\mathbf{r}}+e^{-i\mathbf{k}\cdot\mathbf{r}}\nabla_\mathbf{k}Y_{lm}(\hat{\mathbf{k}})$ allows us to write a bonafide plane wave expansion (see also Equation 6.51 of \cite{stout2012spherical}): 
\begin{equation}
    \mathbf{r}\times\nabla (j_l(kr)Y_{lm}(\hat{\mathbf{r}}))=\frac{i^l}{4\pi}\int e^{-i\mathbf{k}\cdot\mathbf{r}}\mathbf{k}\times\nabla_\mathbf{k}Y_{lm}(\hat{\mathbf{k}})\text{d}\Omega_\mathbf{k}
    \label{Equation: Rayleigh expansion 4}
\end{equation}
Note that in deriving \cref{Equation: Rayleigh expansion 4} we used the following: 
\begin{equation}
    \int \mathbf{k}\times\nabla_\mathbf{k}(e^{-i\mathbf{k}\cdot\mathbf{r}}Y_{lm}(\hat{\mathbf{k}}))\text{d}\Omega_\mathbf{k}=\int \Big[e^{-i\mathbf{k}\cdot\mathbf{r}}Y_{lm}(\hat{\mathbf{k}})\nabla_\mathbf{k}\times\mathbf{k}-\nabla_\mathbf{k}\times\Big[ e^{-i\mathbf{k}\cdot\mathbf{r}}Y_{lm}(\hat{\mathbf{k}})\mathbf{k}\Big]\Big]\text{d}\Omega_\mathbf{k}=0,
\end{equation}

\section{Why is the bandwidth for subwavelength CPA in the TE channel small?}
\label{section: TE bandwidth}
In this section, we find the bandwidth of CPA in the TE polarized channel for a sphere coated with a material admitting a Dirac dispersion (such as graphene). We focus on $l=1$ and define $\sigma_{CPA}(k^+)\equiv \sigma_{CPA}(k^+, l)$ for brevity in the derivations below. To start, we first find $E_{F, CPA}$ and $\omega_{CPA}$. As we previously showed, subwavelength CPA in the TE polarized channel is only possible for $\sigma''(\omega_{CPA})<0$ (since $\sigma''_{CPA}(k^+)<0$ in the subwavelength regime). This is not possible in the intraband approximation of graphene, which includes only a Drude contribution (and, therefore, admits a strictly positive imaginary part of the conductivity). Therefore, from the outset, we need to include the interband contribution to the optical conductivity \cite{chen2011atomically}: 
\begin{equation}
\sigma_{CPA}(k^+)=\frac{ie^2}{\pi\hbar}\Bigg[ \frac{E_{F, CPA}(k^+)}{\hbar(\omega_{CPA}(k^+)+i\tau^{-1})}+\frac{1}{4}\ln\Bigg|\frac{2E_{F, CPA}(k^+)-\hbar\omega_{CPA}(k^+)}{2E_{F, CPA}(k^+)+\hbar\omega_{CPA}(k^+)} \Bigg|\Bigg]
\label{Equation: Interband graphene sigma}
\end{equation}
There is also a Heaviside function that is non-zero if $\hbar\omega_{CPA}>2E_{F, CPA}$ which contributes to the real part of the conductivity. However, we see from \cref{Equation: Interband graphene sigma} that we can have an arbitrarily negative conductivity as we approach $\hbar\omega=2E_F$ from below, and the results in this section do not depend on whether the frequency is larger than twice the Fermi energy or not. Thus, we implicitly take $\hbar\omega_{CPA}<2E_{F, CPA}$. As we will see below, we must also have $\hbar\omega\approx 2E_F$ (in order to have a sufficiently negative imaginary part of the conductivity), for which \cref{Equation: Interband graphene sigma} may be approximated as follows:
\begin{equation}
\sigma_{CPA}'(k^+)\approx\frac{ie^2}{\pi\hbar}\Bigg[\frac{E_{F, CPA}(2E_{F, CPA}/\hbar-i/\tau)}{\hbar(4E_{F, CPA}^2/\hbar^2+1/\tau^2)} +\frac{1}{4}\ln\Bigg|\frac{2E_{F, CPA}-\hbar\omega_{CPA}}{2E_{F, CPA}+\hbar\omega_{CPA}} \Bigg|\Bigg] 
\end{equation}
From this, we see that the real part of the conductivity actually fixes the value for the Fermi energy (if we fix the scattering time to e.g. $\tau\approx 6.4\times10^{-13}s$) as we did in the main text:
\begin{equation}
\frac{\sigma'_{CPA}(k^+)}{c\varepsilon_0}\approx \frac{e^2E_{F, CPA}/(c\varepsilon_0\tau)}{\pi(4E_{F, CPA}^2+\hbar^2/\tau^2)}=\frac{4\alpha E_{F, CPA}(\hbar/\tau)}{4E_{F, CPA}^2+(\hbar/\tau)^2} \end{equation}
Note that this is of the form $2\alpha x/(1+x^2)$, which has a maximum value of $\alpha$, the fine structure constant. So we already know that $\sigma'/c\varepsilon_0$ is restricted to be below the fine structure constant. For a sphere (dipolar mode), this immediately tells us that we require $k^+\lesssim 0.09$ and gives us a quadratic equation for the Fermi energy, which we solve as:
\begin{equation}
E_{F, CPA}(k^+)=\frac{\hbar}{\tau}\frac{\alpha\pm\sqrt{\alpha^2-\Big[\sigma_{CPA}'(k^+)/c\varepsilon_0\Big]^2}}{2\sigma_{CPA}'(k^+)/c\varepsilon_0}
\end{equation}
Once the Fermi energy is fixed, the only free parameter remaining is the frequency, which we obtain from the imaginary part of the conductivity: 
\begin{equation}
\frac{\sigma_{CPA}''(k^+)}{c\varepsilon_0}=4\alpha\Bigg[\frac{2E_{F, CPA}^2}{(4E_{F, CPA}^2+\hbar^2/\tau^2)} +\frac{1}{4}\ln\Bigg|\frac{2E_{F, CPA}-\hbar\omega_{CPA}}{2E_{F, CPA}+\hbar\omega_{CPA}} \Bigg|\Bigg],
\end{equation}
which has the following approximate solution:
\begin{equation}
\hbar\omega_{CPA}\approx 2E_{F, CPA}-4E_{F, CPA}\exp\Bigg[\Bigg[ \frac{\sigma''(k^+)}{c\varepsilon_0\alpha}-\frac{8E_{F, CPA}^2}{4E_{F, CPA}^2+(\hbar/\tau)^2} \Bigg]\Bigg]
\end{equation}
Note that since we know that $\sigma''(k^+)/c\varepsilon_0$ is negative (and of significant magnitude in the region $k^+\lesssim 0.09$), we self-consistently have that $\hbar\omega_{CPA}\approx 2E_{F, CPA}$. Now we calculate the bandwidth. We proceed as we did in the TM case, by finding $s(\omega_{CPA}+\delta)$. First we note that for $k^+<1$, we have:
\begin{equation}
    s(\omega_{CPA})\approx \frac{9/k^+-\sigma(\omega_{CPA})/(c\varepsilon_0)\Big[(k^+)^3 +3i\Big]}{9/k^++\sigma(\omega_{CPA})/(c\varepsilon_0)\Big[(k^+)^3 -3i\Big]}\rightarrow 
    s(\omega_{CPA}+\delta)\approx\frac{\delta \partial_\omega\sigma''(\omega_{CPA})/c\varepsilon_0}{\delta \partial_\omega\sigma''(\omega_{CPA})/c\varepsilon_0-2i(k^+)^2}
\end{equation}
Thus, the Lorentzian shape of the absorption spectrum and the half width at half maximum are given by: 
\begin{equation}
    |s(\omega_{CPA}+\delta)|^2=1-\frac{4(k^+)^4}{\delta^2| \partial_\omega\sigma''(\omega_{CPA})/c\varepsilon_0|^2+4(k^+)^4}\rightarrow 
    \delta_{hwhm}=\frac{2(k^+)^2}{|\partial_\omega\sigma''(\omega_{CPA})/c\varepsilon_0|}
\end{equation}
Note that $\partial_\omega\sigma''(\omega)/c\varepsilon_0\approx \hbar\alpha/(2E_{F, CPA}-\hbar\omega_{CPA})$. Therefore, we have: 
\begin{equation}
    \delta_{hwhm}\approx \frac{8(k^+)^2}{\hbar\alpha}E_{F, CPA}(k^+)\exp\Bigg[\Bigg[ \frac{\sigma''(k^+)}{c\varepsilon_0\alpha}-\frac{8E_{F, CPA}^2(k^+)}{4E_{F, CPA}(k^+)^2+(\hbar/\tau)^2} \Bigg]\Bigg]
\end{equation}
Note that the dominant behavior of the first exponential gives a $\exp(-3/(\alpha k^+))$ dependence for small $k^+$. Since $\alpha\approx 1/137$ is very small and $k^+\lesssim 0.09$, this is why the half width at half maximum is very small. 
\end{document}

%% file: setup.tex
\usepackage[utf8]{inputenc}
\usepackage[english]{babel}

\usepackage{microtype} 
\usepackage{xspace} 

\usepackage{txfonts}  
\usepackage{txfontsb} 

\usepackage{bm} 

\usepackage{xcolor}
\usepackage[]{graphicx}

\usepackage[]{booktabs}
\usepackage{array}
\usepackage{layouts}
\usepackage{multirow}

\usepackage{enumerate}
\usepackage[inline]{enumitem}

\usepackage{xr}
\makeatletter
\newcommand*{\addFileDependency}[1]{
  \typeout{(#1)}
  \@addtofilelist{#1}
  \IfFileExists{#1}{}{\typeout{No file #1.}}
}
\makeatother

\usepackage{hyperref}
\hypersetup{colorlinks,
	linkcolor={blue!75!black!80!yellow},
	citecolor={blue!75!black!80!yellow},
	urlcolor={blue!75!black!80!yellow}
}

\usepackage[capitalize,nameinlink]{cleveref}

\crefname{subequations}{Eqs.}{Eqs.} 
\Crefname{subequations}{Eqs.}{Eqs.}
\crefformat{subequations}{#2Eqs.~(#1)#3}
\Crefformat{subequations}{#2Eqs.~(#1)#3}
\crefname{page}{p.}{p.} 

\usepackage{placeins}

\usepackage{siunitx}
\DeclareSIUnit[number-unit-product = ]\percent{\char`\%} 

\usepackage[centering,hmargin=16mm,tmargin=30mm,bmargin=26mm]{geometry}

\usepackage{textcomp} 
\usepackage{xifthen}
\usepackage{etoolbox}
\newboolean{togglecomments}
\newboolean{toggletodos}
\newboolean{togglechanges}

\setboolean{togglecomments}{true}
\setboolean{toggletodos}{true}
\setboolean{togglechanges}{false} 

\newcommand{\textblacksquare}{$\blacksquare$}
\newcommand{\todo}[1]{\ifbool{toggletodos}%
	{\textcolor{orange!80!yellow!95!black}{\small\textsf{{}\textsuperscript{\textsc{\textsf{todo}}}}[\ignorespaces#1]}} 
	{}}     
\newcommand{\comment}[2]{\ifbool{togglecomments}%
		{\textcolor{blue!70!black}{\small\sf\textsuperscript{\textsc{\textsf{#1}}}[\ignorespaces#2]}} 
		{}}     
\newcommand{\remove}[1]{\ifbool{togglechanges}
	{}    
	{\textcolor{red!70!black}{\ignorespaces#1}}}
\newcommand{\inset}[1]{\ifbool{togglechanges}
	{#1}  
	{\textcolor{green!70!black}{#1}}}
\newcommand{\citeremind}[1]{%
	[\textcolor{blue!75!black!80!yellow}{\textblacksquare%
		\ifthenelse{\isempty{#1}}{}{\textsuperscript{\tiny\textsf{\ignorespaces#1}}}%
	}]\xspace}

\input{commands.tex}

\makeatletter
\newcommand{\raisemath}[1]{\mathpalette{\raisem@th{#1}}}
\newcommand{\raisem@th}[3]{\raisebox{#1}{$#2#3$}}
\makeatother

\renewcommand{\paragraph}[1]{\vskip 1ex\noindent\textbf{#1.}~}

\usepackage[eulergreek]{sansmath}
\makeatletter
\renewcommand\@make@capt@title[2]{%
    \@ifx@empty\float@link{\@firstofone}{\expandafter\href\expandafter{\float@link}}%
    \sansmath\sffamily\textbf{#1\@caption@fignum@sep}#2 
}%

\makeatother


%% file: commands.tex
\newcommand{\br}[2]{(\mathrm{#1}\kern.125ex | \kern.125ex#2)}

\newcommand{\appropto}{\mathrel{\vcenter{
			\offinterlineskip\halign{\hfil$##$\cr
				\propto\cr\noalign{\kern.2pt}\sim\cr\noalign{\kern-2.5pt}}}}}

\let\Re\relax 
\DeclareMathOperator{\Re}{Re}
\let\Im\relax 
\DeclareMathOperator{\Im}{Im}

\DeclareFontFamily{U}{mathx}{\hyphenchar\font45}
\DeclareFontShape{U}{mathx}{m}{n}{<5> <6> <7> <8> <9> <10>
                                  <10.95> <12> <14.4> <17.28> <20.74> <24.88>
                                  mathx10}{}
\DeclareSymbolFont{mathx}{U}{mathx}{m}{n}
\DeclareFontSubstitution{U}{mathx}{m}{n}